\documentclass[11pt,a4paper,english]{article}

\usepackage{graphicx}
\usepackage{amssymb,latexsym}
\usepackage{amsmath}
\usepackage{epsf}
\usepackage{pdfsync}
\usepackage{epsfig}
\usepackage[parfill]{parskip}
\usepackage[matrix,arrow,color]{xy}
\usepackage[usenames]{color}
\usepackage{hyperref}
\usepackage{bbm}

\input{xy}
\xyoption{all}

\input{epsf}

\makeatletter

\catcode`@=11
\renewcommand{\section}
{\@startsection{section}{1}{0pt}{\medskipamount}{\medskipamount}{\large\bf}}
\makeatletter\renewcommand{\subsection}
{\@startsection{subsection}{2}{\z@}{-3.25ex plus -1ex minus -.2ex}
{1.5ex plus .2ex}{\it }}

\numberwithin{equation}{section}
\catcode`@=12

\newcommand{\ba}{\begin{eqnarray*}}
\newcommand{\ea}{\end{eqnarray*}}
\newcommand{\ban}{\begin{eqnarray}}
\newcommand{\ean}{\end{eqnarray}}

\newcommand{\Tr}{{\rm Tr\,}}

\newcommand{\IZ}{\mathbb{Z}}
\newcommand{\IC}{\mathbb{C}}
\newcommand{\IP}{\mathbb{P}}

\newcommand{\IR}{\mathbb{R}}

\newcommand{\IF}{\mathbb{F}}

\newcommand{\frS}{\frak{S}}

\newcommand{\frE}{\frak{E}}
\newcommand{\frg}{\mathfrak{g}}

\newcommand{\frgl}{\mathfrak{gl}}

\renewcommand{\Re}{\ensuremath{\mathfrak{Re}}}

\newcommand{\cN}{{\cal N}}
\newcommand{\cM}{{\cal M}}
\newcommand{\cS}{{\cal S}}

\newcommand{\cH}{{\cal H}}

\newcommand{\cO}{{\cal O}}

\newcommand{\cR}{{\cal R}}

\newcommand{\cF}{{\cal F}}

\newcommand{\cU}{{\cal U}}

\newcommand{\sfP}{{\mathsf{P}}}

\newcommand{\sfR}{{\mathsf{R}}}

\newcommand{\sfh}{{\mathsf{h}}}
\newcommand{\sfg}{{\mathsf{g}}}
\newcommand{\sfH}{{\mathsf{H}}}

\newcommand{\unit}{\mathbbm{1}}   			

\newcommand{\mbf}[1]{{\boldsymbol {#1} }}
\newcommand{\torus}{{\mathbb T}}

\def\e{{\,\rm e}\,}

\def\dd{{\rm d}}

\newcommand{\Hom}{\mathrm{Hom}}

\newcommand{\End}{\mathrm{End}}

\def\beq{\begin{equation}}
\def\bee{\begin{equation}}
\def\eeq{\end{equation}}
\def\bea{\begin{eqnarray}}
\def\eea{\end{eqnarray}}
\def\bd{\begin{displaymath}}
\def\ed{\end{displaymath}}

\newcommand{\Cint}{\int\kern-10.5pt-\kern7pt}

\newcommand{\be}{\begin{equation}}
\newcommand{\ee}{\end{equation}}

\newcommand\fverbit{\egroup\item[\fbox{\unhbox\pippobox}]}
\newbox\pippobox

{

\def\be{\begin{equation}}
\def\ee{\end{equation}}
\def\bea{\begin{eqnarray}}
\def\eea{\end{eqnarray}}

\newtheorem{lemma}[equation]{Lemma}

\newtheorem{proposition}[equation]{Proposition}

\newcommand{\Proof}[1]{\noindent\underline{\textsf{Proof}}: #1 \hfill
  $\blacksquare$\\}

\begin{document}

\begin{titlepage}
\setcounter{page}{0}
\begin{flushright}
EMPG--16--11
\end{flushright}

\vskip 1.8cm

\begin{center}

{\Large\bf Chiral expansion and Macdonald deformation \\[2mm]
  of two-dimensional Yang-Mills theory}

\vspace{15mm}

{\large\bf Zolt\'an K\"ok\'enyesi$^{(a)}$, Annamaria Sinkovics$^{(a)}$ and
 Richard~J.~Szabo$^{(b)}$}
\\[6mm]
\noindent{\em $^{(a)}$ Institute of Theoretical Physics\\ MTA-ELTE
  Theoretical Research Group \\ E\"otv\"os Lor\'and University \\ 1117
  Budapest, P\'azm\'any s. 1/A, Hungary} \\ Email: \ {\tt
  kokenyesiz@caesar.elte.hu} \ , \ {\tt
  sinkovics@general.elte.hu}\\[4mm]
\noindent{\em $^{(b)}$ Department of Mathematics\\ Heriot-Watt
  University\\
Colin Maclaurin Building, Riccarton, Edinburgh EH14 4AS, UK\\ 
Maxwell Institute for Mathematical Sciences, Edinburgh, UK\\
The Higgs Centre for Theoretical Physics, Edinburgh, UK}\\
Email: \ {\tt
  R.J.Szabo@hw.ac.uk}

\vspace{20mm}

\begin{abstract}
\noindent
We derive the analog of the large $N$ Gross-Taylor holomorphic string
expansion for the refinement of $q$-deformed $U(N)$ Yang-Mills theory on a
compact oriented Riemann surface. The derivation combines Schur-Weyl duality for quantum groups with the Etingof-Kirillov theory of generalized quantum characters which are related to Macdonald polynomials. In the unrefined limit we reproduce the chiral expansion of $q$-deformed Yang-Mills theory derived by de~Haro, Ramgoolam and Torrielli. In the classical limit $q=1$, the expansion defines a new $\beta$-deformation of Hurwitz theory wherein the refined partition function is a generating function for certain parameterized Euler characters, which reduce in the unrefined limit $\beta=1$ to the orbifold Euler characteristics of Hurwitz spaces of holomorphic maps. We discuss the geometrical meaning of our expansions in relation to quantum spectral curves and $\beta$-ensembles of matrix models arising in refined topological string theory.
\end{abstract}

\end{center}


\end{titlepage}

\newpage


\tableofcontents

\bigskip

\section{Introduction and summary} \label{intro}

The OSV conjecture relates the partition function of a
four-dimensional BPS black hole in Type~IIA string theory compactified
on a Calabi-Yau threefold $X$ with the A-model topological string
amplitudes on $X$~\cite{OSV2004}. The black hole partition
function counts BPS microstates and can be used for calculation of
black hole entropy. On the topological string theory side, if the OSV
conjecture is valid, it may be used for a non-perturbative definition
of the topological string theory. In the case of a local Calabi-Yau
threefold $X$ which the total space of rank two holomorphic vector bundle
over a Riemann surface $\Sigma$, the counting of BPS states reduces to the
computation of the partition function of
$q$-deformed $U(N)$ Yang-Mills theory on $\Sigma$ (first introduced in~\cite{Klimcik1999}), and for
large $N$ it factorizes into a coupling of chiral and antichiral
partition functions of $q$-deformed $SU(N)$ Yang-Mills theory; the string theory dual of this two-dimensional gauge theory is topological string theory~\cite{Aganagic2005}. From a mathematical perspective, this duality states that the Gromov-Witten invariants of $X$ can be computed in terms of Hurwitz numbers of branched covers of $\Sigma$~\cite{Bryan2004} (see also~\cite{Caporaso2005a,Caporaso2005b,Caporaso2006,Szabo2010a}).

The refinement of the OSV conjecture states that the refined black
hole partition function has a large $N$ dual which is captured by
refined topological string amplitudes~\cite{Aganagic2012}. The refined
black hole partition function, which counts BPS states with spin,
reduces to the partition function of a refined version of two-dimensional $q$-deformed $U(N)$
Yang-Mills theory, and it factorizes into a coupling of chiral and
antichiral partition functions
of refined $q$-deformed $SU(N)$ Yang-Mills theory analogously to the
unrefined case. This refined $q$-deformed two-dimensional gauge theory
is called the $(q,t)$-deformation or Macdonald deformation of
Yang-Mills theory, and its string theory dual is refined topological
string theory. On open surfaces, the duals of boundary characters of
the $(q,t)$-deformed gauge theory on the refined topological string
theory side correspond to insertions of Lagrangian D3-branes in one of
the fibers which encircle the boundary; the special Wilson loop observables in Yang-Mills theory correspond to Lagrangian D-branes wrapping a one-cycle in the fiber~\cite{Aganagic2012}. In the present paper we develop a dual \emph{closed} string expansion of this gauge theory in the large $N$ limit.

The $(q,t)$-deformed Yang-Mills theory is also closely related to
various gauge theories in higher dimensions. It is related to refined
Chern-Simons theory on a circle bundle over the Riemann
surface $\Sigma$~\cite{Aganagic2011,Aganagic2012ref,Kokenyesi2013}. The
topological limit of the partition function was considered in
\cite{Gadde2011} where it was interpreted as the superconformal index
for a certain $\cN=2$ gauge theory of class $\cS$ in four
dimensions. These topological versions also arise in geometric
engineering of supersymmetric gauge theories in string theory. In
particular, the partition function of the two-dimensional topological
field theory version of the Macdonald deformation is equal to the
partition function of a five-dimensional $\cN=1$ gauge theory~\cite{Aganagic2012,Iqbal2015}.

The large $N$ phase structure of the Macdonald deformation on the sphere was studied in~\cite{Kokenyesi2013,Gorsky2016}, and related to refined black hole entropy and topological string theory. In this paper we derive the large $N$ expansion of $(q,t)$-deformed
Yang-Mills theory which is the analog of the Gross-Taylor holomorphic
string expansion of ordinary Yang-Mills theory. Ordinary
two-dimensional Yang-Mills theory can be solved exactly in the lattice
formulation~\cite{Migdal1975,Rusakov1990}. Using Schur-Weyl duality,
in~\cite{Gross1993,Gross1993a,Gross1993b} it was shown that the chiral
part of the partition function can be manipulated into a series consisting
of delta-functions over symmetric groups, which count equivalence
classes of branched covers of the Riemann surface $\Sigma$ in terms of Hurwitz
numbers, see e.~g.~\cite{Cordes1995,Lando2004}. It was further shown
in~\cite{Cordes1995,Cordes1994} that the chiral series is a generating
function for orbifold Euler characters of Hurwitz spaces of
holomorphic maps with fixed two-dimensional target space $\Sigma$; hence
two-dimensional Yang-Mills theory is dual to a closed two-dimensional
topological string theory with string coupling $g_{\rm
  str}=\frac1N$. This expansion was extended to the $q$-deformed case
by de~Haro, Ramgoolam and Torrielli in~\cite{deHaro2006}. In this
instance the $U(N)$ characters and dimensions are replaced by quantum
characters and quantum dimensions for the quantized universal
enveloping algebra of $U(N)$, and hence for large $N$ the quantum
version of Schur-Weyl duality~\cite{Jimbo1986} can be applied to the
chiral partition function to manipulate it into a series of delta-functions
on Hecke algebras of type~A~\cite{deHaro2006}. These expansions can also be applied to observables corresponding to open surfaces and Wilson loops.

The purpose of the present paper is to extend the large $N$ chiral
expansion of~\cite{deHaro2006} to the refined case. Our constructions
are based on the feature that, from a mathematical perspective, the
refinement uses the Etingof-Kirillov theory of characters for
intertwining operators~\cite{Etingof1993}, called ``generalized
characters'', which are related to Jack and Macdonald
polynomials in a similar way that ordinary characters are related to
Schur polynomials. Using the fact that Macdonald polynomials can be written
as vector-valued characters of the underlying quantum group, we can
apply quantum Schur-Weyl duality to these characters analogously to
the $q$-deformed case to obtain a more general and complex expansion into
delta-functions on Hecke algebras. While we borrow throughout from the wealth of
results already derived in~\cite{deHaro2006}, along the way we clarify
and extend these results in various directions. In particular, we
explicitly clarify the definitions of certain central elements
from~\cite{deHaro2006}, we define a new Fourier-type transformation on
quantum group characters to characters of central elements of the
Hecke algebra, and we prove a relation between refined and ordinary Littlewood-Richardson coefficients at large $N$.

Neither the $q$-deformed nor the refined expansions admit straightforward
worldsheet interpretations (much like the situation with refined
topological string theory). As such, our constructions may have some
independent mathematical interest in light of the property that the
undeformed chiral theory is the worldsheet field theory for classical Hurwitz spaces of branched
covers. To shed some light on this perspective, we shall study a
special classical limit of the $(q,t)$-deformed gauge theory in which $t=q^\beta$ and $q\rightarrow 1$
with fixed $\beta$. In this limit the Macdonald polynomials reduce to
the Jack symmetric functions, and the underlying Hecke algebra reduces
to the ordinary group algebra of the symmetric group. This limit of
the $(q,t)$-deformed partition function yields a new kind of deformation of the
usual Hurwitz theory of branched covering maps of Riemann surfaces,
which to the best of our knowledge has not appeared in
the literature before. In particular, we interpret the limiting
partition function as a generating function for certain
\emph{parameterized} Euler characters which at $\beta=1$ reduce to the usual orbifold
Euler characters of Hurwitz spaces.

Amongst the various speculations about the worldsheet interpretations
of these deformed two-dimensional gauge theories that are given
in~\cite{deHaro2006} and in the following, let us present one more to
close this introductory section. The enumeration of branched covers of
a Riemann surface $\Sigma$ can be obtained by computing the partition function
of a two-dimensional lattice gauge theory on a cellular decomposition
of $\Sigma$ with gauge group the symmetric
group~\cite{Kostov1997,Billo2001}; this model is related to chiral
two-dimensional Yang-Mills theory as discussed
by~\cite{Kostov1997}. It is then tempting to speculate that the
$q$-deformation of chiral two-dimensional Yang-Mills theory is obtained by
using instead a model whose gauge group is based on the corresponding
Hecke algebra of type~A.

The outline of the remainder of this paper is as follows. In
\S\ref{genasp} we review various aspects of the definition and
computation of the partition function of $(q,t)$-deformed
two-dimensional Yang-Mills theory, pointing out its various
geometrical incarnations within refined topological string theory in the related settings of M-theory, higher-dimensional
supersymmetric gauge theories, and deformed matrix models. In
\S\ref{sec:genqcharacters} we exploit the quantum version of
Schur-Weyl duality developed by~\cite{deHaro2006} to rewrite the
$(q,t)$-deformed dimension factors appearing in the partition function
in terms of generalized characters of the underlying Hecke algebras;
our final result is summarised in Proposition~\ref{eq:qtdim_big1}. In
\S\ref{sec:chiralpartfn} we compute the chiral expansion of the
refined topological partition function; our final result is summarised
by Proposition~\ref{eq:EulerExp}. We develop in particular the 
$\beta$-deformation of classical Hurwitz theory alluded to above; our
main result is summarised in Proposition~\ref{prop:parEuler}. In
\S\ref{sec:Observables} we extend these considerations to topological
partition functions on open Riemann surfaces and to Wilson loop
observables. Finally, four appendices at the end of the paper contain
technical details and definitions which are refered to throughout the
main text: Appendix~\ref{app:Quantumgroup} recalls the definition of
the pertinent quantum groups, Appendix~\ref{app:Hecke} describes the
corresponding Hecke algebras, Appendix~\ref{app:qt-traces} derives
some new identities amongst characters of symmetric groups and
supersymmetric Schur polynomials, and Appendix~\ref{sec:centralTrafo}
derives new Fourier-type transformations from quantum groups to
central elements of inductive limits of Hecke algebras.

\section{Macdonald deformation of Yang-Mills theory in two dimensions}  \label{genasp}

The Macdonald deformation of two-dimensional Yang-Mills theory is a two-parameter deformation of the usual two-dimensional gauge theory. It can be thought of as a refinement of the well-known $q$-deformation, or alternatively as a quantum deformation of the classical $\beta$-deformation which can be characterised in certain cases by $\beta$-ensembles of random matrix models. In this section we consider the partition function defining the gauge theory and its geometrical interpretations in the context of refined topological string amplitudes; more general amplitudes will be studied in~\S\ref{sec:Observables}.

\subsection{Combinatorial definition}

Let $\frg$ be the Lie algebra of a connected Lie group $G$ of rank
$N$. Let $\cR$ be the root system of $\frg$ and $\cR_+$ the system of
positive roots; similarly let $\Lambda\cong\IZ^N$ be the weight lattice of
$\frg$ with dominant weights
$\Lambda_+$. We fix an invariant bilinear form $(-,-)$
on $\frg$, usually the Killing form. Let
\bea
\rho=\frac12\, \sum_{\alpha\in \cR_+}\, \alpha
\eea
be the Weyl vector of $\frg$ labelling the trivial representation; we shall often assume that the rank $N$
is such that $\rho\in\IZ^N$, which in particular can be supposed in the large $N$ expansion that we consider in the following.

The partition function for the Macdonald deformation of 
Yang-Mills theory with gauge group $G$ on a closed oriented Riemann surface $\Sigma_h$ of
genus $h\geq0$ can be written as a
generalization of the Migdal-Rusakov heat kernel expansion given by~\cite{Aganagic2011,Aganagic2012ref,Aganagic2012}
\bea
Z_h(q,t;p) = \sum_{\lambda\in\Lambda_+} \, \frac{\dim_{q,t}(R_\lambda)^{2-2h}}{(g_\lambda)^{1-h}}\ q^{\frac p2\,
  (\lambda,\lambda)}\, t^{p\,(\rho,\lambda)} \ ,
\label{ZqtQpR}\eea
where the sum runs over all irreducible unitary representations
$R_\lambda$ of
$G$ which are parameterized by dominant weights
$\lambda\in\Lambda_+$. Here the degree $p\in\IZ$ and the deformation
parameters $q,t\in\IC^*$ satisfy $|q|<1$ and $|t|<1$ in order to ensure
that the series (\ref{ZqtQpR}) has a non-zero radius of convergence;
we shall sometimes assume $q,t\in(0,1)$ for convenience.
For simplicity of presentation, below we shall write some formulas for the
case when the refinement parameter
\bea
\beta=\frac{\log t}{\log q}
\eea
is a positive integer, and then extend our final results to arbitrary
$\beta\in\IC$ by analytic continuation.
The refined quantum dimension of the representation $R_\lambda$ is
\bea
\dim_{q,t}(R_\lambda)=\prod_{m=0}^{\beta-1} \ \prod_{\alpha\in \cR_+}\,
\frac{\big[(\lambda+\beta\,\rho,\alpha)+m\big]_q}{\big[(\beta\,\rho,\alpha)
  +m\big]_q}
\ ,
\label{qtdim}\eea
where
\bea
[x]_q=\frac{q^{x/2}-q^{-x/2}}{q^{1/2}-q^{-1/2}}
\label{qnumber}\eea
for $x\in\IR$ is a $q$-number. The Macdonald metric is given by
\bea \label{eq:MacdMetric}
g_\lambda= \frac1{N!} \ \prod_{m=0}^{\beta-1} \ \prod_{\alpha\in \cR_+}\,
\frac{\big[(\lambda+\beta\,\rho,\alpha)+m\big]_q}{\big[(\lambda+\beta\,
  \rho,\alpha)
  -m\big]_q} \ .
\eea

In this paper we shall specialize to the unitary gauge group $G=U(N)$. In this case there are convenient combinatorial expressions
available for the dimension and metric factors. The Weyl vector is $\rho=\big(\frac{N-1}2,\dots,-\frac{N-1}2\big)$ and the dominant weights
$\lambda\in \Lambda_+$ are parameterized by partitions with at most $N$ parts $\lambda=(\lambda_1,\dots,\lambda_N)$, $\lambda_1\geq\lambda_2\geq\cdots\geq\lambda_N\geq0$; they are in a one-to-one correspondence with Young diagrams $Y_\lambda\subset(\IZ_{>0})^2$ with at most $N$ rows. Then the refined quantum dimension and Macdonald metric have the equivalent forms
\bea
\dim_{q,t}(R_\lambda)&=& t^{\frac12\, (\|\lambda^{\rm t}\|-N\, |\lambda|)} \ \prod_{(i,j)\in Y_\lambda}\, \frac{1-t^{N-i+1}\, q^{j-1}}{1-t^{\lambda_j^{\rm t}-i+1}\, q^{\lambda_i-j}} \ , \nonumber \\[4pt]
g_\lambda&=& g_{\emptyset} \ \prod_{(i,j)\in Y_\lambda}\, \frac{1-t^{\lambda_j^{\rm t}-i}\, q^{\lambda_i-j+1}}{1-t^{\lambda_j^{\rm t}-i+1}\, q^{\lambda_i-j}} \, \frac{1-t^{N-i+1}\, q^{j-1}}{1-t^{N-i}\, q^j} \ ,
\label{eq:dimcomb}\eea
where $|\lambda|:=\sum_{i=1}^N\, \lambda_i$ and $\|\lambda\|:=\sum_{i=1}^N\, \lambda_i^2$, while
\be 
g_{\emptyset} = \frac1{N!} \ \prod_{m=0}^{\beta-1} \ \prod_{1\leq i<j\leq N}\,
\frac{\big[\beta\,(j-i) +m\big]_q}{\big[\beta\,(j-i)
  -m\big]_q} \ .
\ee
The products in (\ref{eq:dimcomb}) run over all boxes $(i,j)$ of the
Young diagram $Y_\lambda$ with $1\leq i\leq N$, $1\leq
j\leq\lambda_i$, and $\lambda^{\rm t}$ corresponds to the transposed
Young diagram, i.e. $\lambda_i^{\rm t}$ is the number of entries $\leq
i$ in~$Y_\lambda$.

\subsection{M-theory interpretation\label{sec:geometricint}}

The parameters of the Macdonald deformation are related to the equivariant parameters 
\bea
\epsilon_1=\frac1{\sqrt\beta} \ g_s \qquad \mbox{and} \qquad \epsilon_2=-\sqrt\beta \ g_s 
\label{eq:epsilonbetarel}\eea
of the $\Omega$-background~\cite{DijkgraafVafa2009} through
\bea
q=\e^{-\epsilon_1} \qquad \mbox{and} \qquad t=\e^{\epsilon_2} \ ,
\eea
where $g_s$ is the topological string coupling constant.
The refined A-model topological string theory that gives rise to the two-dimensional
gauge theory is defined on the non-compact Calabi-Yau threefold which
is the total space of the $\IC^2$-fibration
$\cO_{\Sigma_h}(p+2h-2)\oplus\cO_{\Sigma_h}(-p)$ over the Riemann
surface $\Sigma_h$. The complete geometrical picture of the $(q,t)$-deformed gauge theory involves M-theory on a particular 11-dimensional manifold~\cite{Aganagic2012}, which we now describe.

Let us start with five-dimensional $\cN=1$ supersymmetric gauge theory
on the total space of the flat affine bundle $\IC^2\times_\IZ\IR \to \torus$ over a circle
$\torus=S^1$ of radius $r$, where the $\IZ$-action is given by
\bea
(z,w,x) \ \longmapsto \ (q^n\,z,t^{-n}\, w,x+2\pi\, r\, n)
\label{eq:zwxaction}\eea
for $(z,w)\in\IC^2$, $x\in\IR$ and $n\in \IZ$. Here $(q,t)\in
(\IC^*)^2$ is regarded as an element of the maximal torus of the complex spin cover $Spin(4,\IC)=SL(2,\IC)\times SL(2,\IC)$ of the four-dimensional rotation group; the pair $(q,t^{-1})$ is the holonomy of a flat $Spin(4,\IC)$-connection on the bundle $\IC^2\times_\IZ\IR$. Correspondingly, one must also turn on an $SL(2,\IC)$ R-symmetry twist
\bea
T:=\begin{pmatrix}
\sqrt{q\, t^{-1}} & 0 \\ 0 & \sqrt{q^{-1}\, t}
\label{eq:TRsymmetry}\end{pmatrix}
\eea
in order to preserve the topological supersymmetry which is
broken when $\beta\neq1$. We can embed this gauge theory into the
M-theory compactification which is the total space of the bundle over
$\torus$ given by the quotient
\bea
{\rm
  Tot}\big(\cO_{\Sigma_h}(p+2h-2)\oplus\cO_{\Sigma_h}(-p)\big) \
\times_\IZ \ \IC^2 \ 
\times_\IZ \ \IR
\label{eq:Mtheorycomp}\eea
where $n\in\IZ$ acts on $\IC^2\times \IR$ as in
\eqref{eq:zwxaction} and rotates the fibre coordinates $(u,v)\in\IC^2$
of $\cO_{\Sigma_h}(p+2h-2)\oplus\cO_{\Sigma_h}(-p)$ by the matrix
$T^n$. The low energy effective action of this supersymmetric M-theory
background geometrically engineers the $\Omega$-deformed $\cN=1$
supersymmetric $U(1)$ gauge theory with $h$ adjoint hypermultiplets on
$\IC^2\times S^1$, where the integer $p$ corresponds to the level of
the five-dimensional Chern-Simons term; the corresponding partition
function is the generating function for the equivariant Hirzebruch
genus of a certain holomorphic vector bundle $\frE_{h,p}$ on the moduli space
of $U(1)$ instantons on $\IC^2$~\cite{Chuang2010}.

Let us now consider the M-theory background in which the $\IC^2$
factor in \eqref{eq:Mtheorycomp} is replaced with the hyper-K\"ahler
Taub-NUT space ${\sf TN}$, which is a local $S^1$-fibration over
$\IR^3$, where the $\IZ$-action on the local complex coordinates
$(z,w)\in\IC^2$ near the tip of ${\sf TN}$ is as in \eqref{eq:zwxaction}. We embed the
five-dimensional $\cN=1$ gauge theory into the $(2,0)$ superconformal
theory compactified on the fibre $\torus'=S^1$ of the Taub-NUT space,
which is the low energy limit of the theory on a single M5-brane with
worldvolume that is locally given by
\bea
\torus' \ \times \ {\rm
  Tot}\big(\cO_{\Sigma_h}(-p)\big) \
\times \ \torus \ .
\eea
After collapsing the M-theory circle $\torus'$ and equivariantly
localizing with respect to the $\IC^*$-scaling action induced by
\eqref{eq:TRsymmetry} along the fibres of the line bundle
$\cO_{\Sigma_h}(-p)$, we thereby construct
the $(q,t)$-deformed gauge theory on $\Sigma_h$; the corresponding
partition function is related to the generating function for the
Hirzebruch genus of the moduli space of $U(N)$ instantons on the local ruled surface ${\rm
  Tot}\big(\cO_{\Sigma_h}(-p)\big)$ over $\Sigma_h$~\cite{Aganagic2012,Kokenyesi2013}.

The $\Omega$-background symmetry
$(\epsilon_1,\epsilon_2)\mapsto(-\epsilon_2,-\epsilon_1)$ corresponds
to the inversion symmetry $\beta\mapsto\frac1\beta$ of the refinement
parameter together with the rank change $N\mapsto \beta\, (N-1) +1$. It acts on the Macdonald deformation parameters as $(q,t)\mapsto(t,q)$
which corresponds to the symmetry $p\mapsto 2-2h-p$ that exchanges the
two line bundle summands of the Calabi-Yau fibration over $\Sigma_h$.

\subsection{Quantum spectral curves and $\beta$-ensembles\label{sec:betaensemble}}

We can also define the $(2,0)$ theory by wrapping M5-branes on the
six-manifold $\Sigma_h\times \IC^2$ in \eqref{eq:Mtheorycomp}, equipped with a non-trivial fibration of $\IC^2$ over $\Sigma_h$ which specifies the $\Omega$-background~\cite{DijkgraafVafa2009}. In this case $\Sigma_h$ acquires an interpretation as the base of a branched covering by the Seiberg-Witten
curve of a four-dimensional $\cN=2$ gauge theory of class $\cS$, which
can in turn be regarded as the spectral curve of an associated Hitchin
system~\cite{Gaiotto2009,GaiottoMoore2009} that is quantized via a suitable deformation; the five-dimensional
gauge theories compactified on a circle of radius $r$ lead to a relativistic
($q$-deformed or difference) version of this Hitchin system. For $p=1$, the bound state of $N$ M5-branes is described by an $N$-sheeted branched covering of $\Sigma_h$ given by
\be 
\Sigma_{\rm SW}=\Big\{ (x,z)\in{\rm Tot}\big(\cO_{\Sigma_h}(-1)\big) \ \Big| \ x^N+\mbox{$\sum\limits_{j=2}^N$} \, {\rm t}_j(z)\, x^{N-j}=0\Big\} \ ,
\ee
where ${\rm t}_j$ is a $(j,0)$-differential on $\Sigma_h$.

Generally, the Seiberg-Witten curve is an affine curve
characterized by an algebraic relation of the form $P(x,y)=0$ for
$(x,y)\in\IC^2$. Turning on the $\Omega$-background lifts this
relation to a differential equation
$P(\,\widehat{x},\widehat{y}\,)\psi=0$ which quantizes the coordinate
algebra $\IC[x,y]$ to the Weyl algebra
$\IC[\hbar]\langle\,\widehat{x},\widehat{y}\,\rangle$ defined by the
commutation relations
\bea
[\,\widehat{x},\widehat{y}\,]=-\hbar \qquad \mbox{with} \quad \hbar=\sqrt\beta-\frac1{\sqrt\beta}= \frac{\epsilon_1+\epsilon_2}{g_s} \ .
\eea
We can represent $\widehat{x}$ as the multiplication operator by $x\in\IC$ and $\widehat{y}$ as the differential operator $\hbar \, \partial_x$. This differential equation is interpreted as a ``quantum curve''~\cite{Dijkgraaf2008}: The differential $\lambda_\hbar=\hbar\, \partial_x\log \psi(x,\hbar)\, \dd x$ is a ``quantum'' differential generating the ``quantum'' periods of the quantized Riemann surface which in the unrefined limit $\hbar=0$ coincides with the meromorphic differential $\lambda_0=y(x)\, \dd x$ of the original Seiberg-Witten curve (with $y=y(x)$ depending on $x$ through the equation $P(x,y)=0$). Thus refinement corresponds to a system of differential equations satisfied by the partition functions of the four-dimensional gauge theory. In the five-dimensional gauge theory, the quantum spectral curve is instead given by a difference equation $P(\,\widehat{X},\widehat{Y}\,)\psi=0$, where the difference operators $\widehat{X}=\e^{r\,\widehat{x}}$ and $\widehat{Y}=\e^{r\,\widehat{y}}$ obey $\widehat{X}\, \widehat{Y}=\underline{q}\, \widehat{Y}\, \widehat{X}$ with $\underline{q}=\e^{-r^2\,\hbar}$.

In order to understand this point from the perspective of two-dimensional gauge theory, following~\cite{Aganagic2012,Szabo2013,Kokenyesi2013} we define shifted weights $n_i=\lambda_i+\beta\, \rho_i$ for $i=1,\dots,N$ and rewrite the
partition function \eqref{ZqtQpR} for $G=U(N)$ in the form (up to overall normalization)
\bea
Z_h(q,t;p) = \sum_{n\in\IZ_0^N}\, \Delta_{q,t}\big(\e^{\epsilon_1\, n}\big)^{1-h}\, \Delta_{q,t}\big(\e^{-\epsilon_1\, n}\big)^{1-h} \, \e^{-\frac{p\, \epsilon_1}2\, (n,n)} \ ,
\eea
where $\IZ_0^N$ is the set of $N$-vectors of integers $n=(n_1,\dots,n_N)$ for which $n_i\neq n_j$ for all $i\neq j$, and the Macdonald measure $\Delta_{q,t}(x)$ is defined in Appendix~\ref{sec:centralTrafo}. Let us consider the case of genus $h=0$. Since $\Delta_{q,t}(1^N)=0$, we can then sum over all $n\in\IZ^N$ and following~\cite{Szabo2010} we rewrite the partition function on the sphere as
\be 
Z_0(q,t;p) = \prod_{i=1}^N \ \int_0^\infty\, \frac{\dd_q x_i}{
  x_i} \ \e^{-\frac{p}{2\epsilon_1}\, \log^2 x_i} \ \Delta_{q,t}(x)\,
  \Delta_{q,t}(x^{-1}) \ ,
\label{eq:qtmatrixmodel}\ee
where the multiple Jackson $q$-integral is defined by
\bea
\prod_{i=1}^N \ \int_{0}^\infty \, \frac{\dd_q x_i}{x_i} \ f(x) := (1-q)^N \, \sum_{n\in\IZ^N} \, f(q^n) 
\label{eq:Jackson}\eea
for a continuous function $f(z)$ on $(\IC^*)^N$ (provided the multiple series
is absolutely convergent). 

The rewriting \eqref{eq:qtmatrixmodel} demonstrates that the Macdonald
deformation of two-dimensional gauge theory can be described as a generalized
Gaussian matrix model in the $q$-deformed $\beta$-ensemble of
random matrix theory. In particular, for $p=1$ the indeterminancy of
the moment problem for the Stieltjes-Wigert distribution implies that
the discrete and continuous matrix models are
equivalent~\cite{Szabo2013}. In this case the geometrical
setup of \S\ref{sec:geometricint} greatly simplifies: The relevant
Calabi-Yau fibration is the conifold geometry ${\rm
  Tot}\big(\cO_{\IP^1}(-1)\oplus\cO_{\IP^1}(-1) \big)$, the
vector bundle $\frE_{0,1}$ is
trivial~\cite{Chuang2010,Iqbal2015}, while the surface ${\rm
  Tot}\big(\cO_{\IP^1}(-1)\big)$ is the blow-up of $\IC^2$ at two
points (with boundary the three-sphere $S^3$). 
This equivalence implies that we may replace the Jackson
integral in \eqref{eq:qtmatrixmodel} with an ordinary Riemann-Lebesgue integral, and by
setting $z_i:=\log x_i$ we may write
\bea
Z_0(q,t;1) = \prod_{i=1}^N \ \int_{-\infty}^\infty\, \dd z_i \
\e^{-\frac{1}{2\epsilon_1}\, z_i^2} \ \Delta_{q,t}\big(\e^z \big)\,
  \Delta_{q,t}\big(\e^{-z} \big) \ ,
\eea
which up to normalization coincides with the Stieltjes-Wigert matrix
model for refined Chern-Simons theory on
$S^3$~\cite{Aganagic2011}. However, for $p\neq1$ such an equivalence
ceases to hold and one must work directly with the discrete matrix model (\ref{eq:qtmatrixmodel}) for generic values of $p\in\IZ$.

Let us now consider the classical limit $q\to1$ defined by taking the limits
$\epsilon_1\to0$, $p\to\infty$ while keeping fixed the refinement
parameter $\beta$ (so that also $\epsilon_2\to0$) and the parameter
$a:=\epsilon_1\, p$; in this limit the $(q,t)$-deformed gauge theory
generally reduces to a $\beta$-deformation of ordinary Yang-Mills theory on a Riemann surface of area $a$. The right-hand side of \eqref{eq:Jackson} is a multiple
infinite Riemann sum, which for $q\to 1^-$ formally converges to
$\prod_{i=1}^N \, \int_0^\infty\, \frac{\dd x_i}{x_i} \, f(x)$. By rescaling $z_i\to\epsilon_1\, z_i$,
up to normalization we then find the partition function
\bea
\widetilde{Z}_0(a,\beta) &:=& \lim_{q\to1} \ \lim_{p\to\infty}\,
Z_0(q,t;p)\Big|_{\scriptstyle t=q^\beta \,,\,
    p=-\frac a{\log q}} \nonumber \\[4pt] &=&
\prod_{i=1}^N \ \int_{-\infty}^\infty\, \dd z_i \ \e^{-\frac a2\,
  z_i^2} \ \prod_{m=0}^{\beta-1} \ \prod_{i<j}\,
\big((z_i-z_j)^2-m^2\big) \ .
\label{eq:tildeZ0abeta}\eea
The planar limit is defined by taking
\bea
\tau_1=\epsilon_1\, N \qquad \mbox{and} \qquad \tau_2=\epsilon_2\, N
\eea
large but fixed for $N\to\infty$. In this limit the refinement
parameter $\beta=-\frac{\tau_2}{\tau_1}$ is finite, hence the product
in \eqref{eq:tildeZ0abeta} is finite and $m\,
\epsilon_1=\frac{m\,\tau_1}N\to 0$. By rescaling $z_i\to N\, z_i$ as
finite variables at large $N$, it follows that the planar limit
of the partition function \eqref{eq:tildeZ0abeta} is given by
\bea
\widetilde{Z}^{\rm pl}_0(\mu,\beta) =
\prod_{i=1}^N \ \int_{-\infty}^\infty\, \dd z_i \ \e^{-\frac \mu2\,
  z_i^2} \ \Delta(z)^{2\beta} \ ,
\label{eq:Mehtaintegral}\eea
where $\mu:=p\,\tau_1\, N$ and
\bea
\Delta(z)=\prod_{i<j} \, (z_i-z_j)
\eea
is the Vandermonde determinant.
Thus in this limit the weak coupling phase of the two-dimensional
gauge theory can be described by a Gaussian matrix model in the
classical $\beta$-ensembles of random matrix theory. In particular, $\log\widetilde{Z}^{\rm pl}_0(\mu,\beta)$ coincides with the partition function of two-dimensional $c=1$ string theory at radius $R=\beta$; this generalizes the usual identification of the conifold geometry with the $c=1$ string at the self-dual radius for $\beta=1$. In this formulation the symmetry $\beta\to\frac1\beta$ is manifest~\cite{DijkgraafVafa2009} and corresponds to T-duality invariance of the string theory. For later use and comparison, let us run through the details of this identification.

The matrix integral \eqref{eq:Mehtaintegral} is a special case of the Selberg integral which can be evaluated analytically in terms of Mehta's formula
\bea
\widetilde{Z}^{\rm pl}_0(\mu,\beta) = \Big(\, \frac{\sqrt{2\pi}}{\mu^{\frac12\,((N-1)\, \beta +1)}} \, \Big)^N \ \prod_{i=1}^N\, \frac{\Gamma(1+\beta\, i)}{\Gamma(1+\beta)} \ .
\label{eq:Mehtagamma}\eea
One can show that~\cite{Brini2010}
\bea
\prod_{i=1}^N\, \Gamma(1+\beta\, i) = \big(\, \sqrt{2\pi} \, \beta^{\frac12\,((N-1)\, \beta +1)}\, \big)^N \, \Gamma(1+N\, \beta)\, \Gamma(N)\, \Gamma_2\big(N;-\beta^{-1},-1\big) \ ,
\eea
where $\Gamma_2(\tau_2;\epsilon_1,\epsilon_2)$ is the Barnes double gamma-function defined via
\begin{eqnarray}
\log\Gamma_2(\tau_2;\epsilon_1,\epsilon_2) = -\left.\frac{\dd}{\dd s}\right|_{s=0} \ \frac1{\Gamma(s)} \ \int_0^\infty\, \frac{\dd t}{t} \ t^s\, \frac{\e^{-\tau_2\, t}}{\big(1-\e^{\epsilon_1\,t}\big)\, \big(1-\e^{\epsilon_2\, t}\big)}
\end{eqnarray}
which is the double zeta-function regularization of the infinite product $\prod_{m,n\geq0}\, (\tau_2-m\,\epsilon_1-n\,\epsilon_2)$. To obtain the large $N$ expansion of the partition function \eqref{eq:Mehtaintegral} we use the asymptotic expansion~\cite[Appendix~E]{Nakajima2003}
\begin{eqnarray}
\log\Gamma_2(\tau_2;\epsilon_1,\epsilon_2) &=& \frac1{\epsilon_1\, \epsilon_2}\, \big(\mbox{$\frac12$}\, \tau_2^2\,\log\tau_2- \mbox{$\frac34$}\, \tau_2^2\big)+\frac{\epsilon_1+\epsilon_2}{\epsilon_1\,\epsilon_2}\, \big(\tau_2\,\log\tau_2-\tau_2\big)\nonumber \\ && +\, \frac{\epsilon_1^2+\epsilon_2^2+3\, \epsilon_1\,\epsilon_2}{12\, \epsilon_1\, \epsilon_2}\, \log\tau_2- \sum_{n=3}^\infty\, \frac{d_n(\epsilon_1,\epsilon_2)\, \tau_2^{2-n}}{n\,(n-1)\, (n-2)} \ ,
\end{eqnarray}
where the series coefficients $d_n(\epsilon_1,\epsilon_2)$ are defined through the generating function
\begin{eqnarray}
\frac{t^2}{\big(1-\e^{\epsilon_1\,t}\big)\, \big(1-\e^{\epsilon_2\, t}\big)} =  \sum_{n=0}^\infty\, \frac1{n!}\, d_n(\epsilon_1,\epsilon_2)\, t^{n} \ .
\label{eq:dnepsilon}\end{eqnarray}
Introducing the Bernoulli numbers $B_m$ through the generating function
\begin{eqnarray}
\frac s{1-\e^s}=-\sum_{m=0}^\infty\, \frac1{m!}\, B_m\, s^{m}
\label{eq:Bernoulli}\end{eqnarray}
with $B_0 = 1$, $B_1 = \frac12 $, $B_2 = \frac16$ and $ B_k = 0$ for all $k >1$ odd, by comparing series expansions we find
\begin{eqnarray}
d_n(\epsilon_1,\epsilon_2) =g_s^{n-2} \ \sum_{k=0}^n\, {n\choose k}\, (-1)^{k-1}\, B_k\, B_{n-k}\, \beta^{k-\frac n2}
\end{eqnarray}
where we have used the relations \eqref{eq:epsilonbetarel}. By dropping overall prefactors, for the free energy $\widetilde{F}^{\rm pl}_0(\tau_2,\beta):= -\log\widetilde{Z}^{\rm pl}_0(\tau_2,\beta)=-\log\Gamma_2(\tau_2;\beta^{-1/2}\, g_s,\beta^{1/2}\, g_s) $ in the large $N$ limit this gives the asymptotic expansion
\begin{eqnarray}
\widetilde{F}^{\rm pl}_0(\tau_2,\beta) &=& \frac1{g_s^2}\, \big(\mbox{$\frac34$}\, \tau_2^2-\mbox{$\frac12$}\, \tau_2^2\,\log\tau_2\big)+\frac1{g_s}\, \big(\mbox{$\frac1{\sqrt\beta}$}-\sqrt\beta\, \big)\, \big(\tau_2-\tau_2\,\log\tau_2\big)\nonumber \\ && +\, \chi_0(\beta)\, \log\tau_2+ \sum_{n=1}^\infty\, \chi_n(\beta) \ \Big(\, \frac{g_s}{\tau_2}\, \Big)^n \ ,
\label{eq:freeenergybeta}\end{eqnarray}
where
\begin{eqnarray}
\chi_0(\beta)&=& -\frac14+\frac{\beta^{-1}}{12}+\frac{\beta}{12} \ , \nonumber \\[4pt] 
\chi_n(\beta)&=& (n-1)! \ \sum_{k=0}^{n+2}\, \frac{(-1)^{k-1}\, B_k\, B_{n+2-k}}{k!\, (n+2-k)!} \ \beta^{k-\frac n2-1} \qquad \mbox{for} \quad n\geq1 \ .
\label{eq:parEuler}\end{eqnarray}
Note that the expansion parameter is
\bea
\frac{g_s}{\tau_2}=\frac1{\sqrt\beta\, N} \ .
\eea

In the unrefined limit $\beta=1$, the identity
\begin{eqnarray}
\frac1{\big(1-\e^s\big)\, \big(1-\e^{-s}\big)} = -\frac{\dd}{\dd s}\frac1{1-\e^s}
\end{eqnarray}
together with \eqref{eq:Bernoulli} imply that $d_n(g_s,-g_s)=0$ for $n$ odd while $d_{2g}(g_s,-g_s)=g_s^{2g-2}\, (2g-1)\, B_{2g}$ for $g\geq1$, so that the non-vanishing coefficients
\begin{eqnarray}
\chi_0(1)=-\frac1{12} \qquad \mbox{and} \qquad
  \chi_{2g-2}(1)=\frac{B_{2g}}{2g\, (2g-2)} \quad (g>1)
\end{eqnarray}
coincide with the orbifold Euler characteristics $\chi_{\rm orb}(\cM_g)$ of the Riemann moduli spaces $\cM_g$ of genus $g\geq1$ complex curves, i.e. the Euler character of $\cM_g$ calculated by resolving its orbifold singularities. On the other hand, for $\beta=2$ one can use the identity
\begin{eqnarray}
\frac1{\big(1-\e^s\big)\, \big(1-\e^{-2s}\big)} = -\frac12\,\frac{\dd}{\dd s}\frac1{1-\e^s}+\frac12\, \frac{1}{1-\e^{-2s}} -\frac12\, \frac1{1-\e^{-s}}
\end{eqnarray}
together with \eqref{eq:Bernoulli} to infer that
\begin{eqnarray}
\chi_{2g-1}(2)= \sqrt2\ 2^{-g}\,
  \frac{\big(2^{2g-2}-\frac12 \big)\, B_{2g}}{2g\, (2g-1)} \qquad \mbox{and} \qquad \chi_{2g-2}(2)= 2^{-g}\, \chi_{\rm orb}(\cM_g)
\end{eqnarray}
for $g\geq1$, so that the coefficients $\chi_{2g-1}(2)$ are
proportional to the orbifold Euler
characteristics of the moduli spaces of certain real algebraic curves of
genus $g$~\cite{Goulden1999}. Thus in this case refinement corresponds
to the replacement of $\chi_{\rm orb}(\cM_g)$ with the \emph{parameterized}
Euler characters $\chi_n(\beta)$~\cite{Krefl2010,Krefl2013}, which
provide a geometric parameterization that interpolates between the orbifold
Euler characters of the moduli spaces of closed oriented Riemann
surfaces at $\beta=1$ and closed unoriented Riemann surfaces with
crosscap at $\beta=2$. In other words, the string theory at $\beta=2$
can be regarded as the orientifold of the string theory at $\beta=1$. From this perspective, it is natural to expect
that the generic $\beta$-deformed Euler characters $\chi_n(\beta)$ themselves describe characteristic classes of some related moduli spaces~\cite{Goulden1999}.

The expansion \eqref{eq:freeenergybeta} governs the leading order behaviour of the dual refined B-model topological string amplitudes on the mirror of the conifold geometry, which is the cotangent bundle $T^*S^3$. Generally, these local Calabi-Yau geometries are described by algebraic equations of the form
\begin{eqnarray}
u\,v+F(x,y)=0
\end{eqnarray}
in $\IC^4$, where the equation $F(x,y)=0$ describes an affine curve
$\Sigma$ in $\IC^2$ and a (local) function $y(x)$ which determines a
meromorphic differential $\lambda=y(x)\, \dd x$ giving the periods of
$\Sigma$. In the Gaussian matrix model \eqref{eq:Mehtaintegral} at
$\beta=1$, the Riemann surface $\Sigma$ is the corresponding rational spectral
curve which is given by a double cover of the $y$-plane
with $F(x,y)=x^2-y^2+m=0$, where $m$ is the K\"ahler parameter of the
resolved conifold. After a simple change of variables this spectral
curve can be regarded as the holomorphic curve $F(z,w)=z\,w-m=0$ in
$\IC^2$, which after refinement quantizes to the differential operator
$F(\,\widehat{z},\widehat{w}\,)=\hbar\,z\,\partial_z-m$~\cite{Eynard2008}; the quantum
curve in this case is the canonical example of a
D-module~\cite{Dijkgraaf2008} and it can be regarded as a differential
equation for certain correlators in the matrix model~\cite{DijkgraafVafa2009}. After
$q$-deformation, the quantum spectral curve for the conifold is
naturally described by a difference equation (rather than a
differential equation) with difference operator
\bea
F(\,\widehat{X},\widehat{Y}\,) = \big(\, 1-\underline{q}^{-1/2}\,
\widehat{X}\, \big)\, \widehat{Y}-\big(\, 1-\underline{Q}\
\underline{q}\, \widehat{X}\, \big) \ ,
\eea
where $\underline{Q}:=\e^{-r^2\, m}$; it can be thought of as a
differential equation for the partition functions of refined
topological string theory~\cite{Chen2013}. It is then natural to
expect that a similar quantum spectral curve governs the 
matrix model \eqref{eq:qtmatrixmodel} of the $q$-deformed
$\beta$-ensemble that represents the $(q,t)$-deformed gauge theory,
along the lines of~\cite{Zenkevich2015}. In the following, these lines
of reasoning will be applied to the closed string chiral
expansion of the $(q,t)$-deformed two-dimensional gauge theory to give
geometrical interpretations of the
Macdonald deformation in terms of contributions from deformed
characteristic classes associated to quantum Riemann surfaces.

\section{Generalized quantum characters as Hecke characters} \label{sec:genqcharacters}

In this section we develop a combinatorial
description of the dimension factors for the quantum universal enveloping algebra $\cU_q(\frgl_N)$ appearing in
(\ref{ZqtQpR}) in terms of
characters of the Hecke algebra $\sfH_q(\frS_n)$ of type
$A_{n-1}$; our final result is summarised in Proposition~\ref{eq:qtdim_big1}. For this, we shall use quantum
Schur-Weyl duality between $\cU_q(\frgl_N)$ and $\sfH_q(\frS_n)$. See Appendix~\ref{app:Quantumgroup} for relevant definitions and properties of quantum groups which are used throughout, and Appendix~\ref{app:Hecke} for those pertaining to Hecke algebras.

\subsection{Generalized characters and Macdonald polynomials}

The (unrefined) $q$-deformation of the standard two-dimensional
Yang-Mills theory can be obtained by replacing representations of $G$ with
quantum group modules, as in~\cite{deHaro2006,Szabo2013}. Refinement
then corresponds to the
$\beta$-deformation wherein (quantum) characters are replaced with
{generalized} characters; see e.g.~\cite[\S2]{Iqbal2011}
and~\cite[\S6.1]{Szabo2013}. If $V,W$ are finite-dimensional
representations of $\cU_q(\frgl_N)$, and $\Phi:V\to V\otimes W$ is a
non-zero intertwining operator for $\cU_q(\frgl_N)$, then the vector-valued function
\bea
\chi_\Phi(U)= \Tr_V\big(\Phi\, U \big)
\eea
on the maximal torus $T\subset G$ is called a \emph{generalized
  character}. Contrary to the classical case $q=1$, if the
representation $W$ is non-trivial then $\chi_\Phi(U)$ is not invariant
under the action of the Weyl group $\frS_N$ on $T$. Since the operator $\Phi$ preserves
weight, the vector $\chi_\Phi(U)$ actually takes values in the weight zero subspace
$W_0\subset W$.

To compute the generalized character explicitly, let $V^*$ denote the dual
$\cU_q(\frgl_N)$-module, and let $v_i,v^i$ be dual bases for
$V,V^*$. We can then identify $\Phi$ with an intertwiner $\Phi:
V^*\otimes V\to W$ and
\bea
\chi_\Phi(U)=\Phi\big(v^i\otimes U v_i\big) \ ,
\label{eq:gencharcompute}\eea
where throughout we use the Einstein summation convention for repeated upper and lower indices.
Since $v^i\otimes v_i=(\unit_{V^*}\otimes q^{-(\rho,H)})\, \mbf1_\IC$,
where $H=(H_1,\dots,H_N)$ are the Cartan generators of $\frgl_N$ and $\mbf1_\IC=\imath(1)$ with $\imath:\IC\to V^*\otimes V$ an
embedding of $\cU_q(\frgl_N)$-modules, we can also write the
generalized character as
\bea
\chi_\Phi(U) = \Phi\big((\unit_{V^*}\otimes q^{-(\rho,H)}\, U)
\mbf1_\IC \big) \ .
\eea
In the special instance where $W=\IC$ is the trivial representation of
$\cU_q(\frgl_N)$ and $\Phi:V\to V$ is the identity operator, so that
$\Phi:V^*\otimes V\to\IC$ is the canonical dual pairing, then
$\chi_{\unit_V}(U)=\chi_V(U)= \Tr_V(U)$ is the usual character of $U$ in the
representation $V$.

Now let $V=R_\lambda$ for fixed $\lambda\in\Lambda_+$ and $W=W_{\beta-1}$ for
fixed $\beta\in\IZ_{>0}$ where
\be \label{eq:defW}
W_{\beta-1}:= R_{\omega_1}^{\odot(\beta-1)\, N}\otimes(\det)^{-(\beta-1)}
\ee
is the $q$-deformation of
the traceless $(\beta-1)\, N$-th symmetric power of the first fundamental
representation $R_{\omega_1} =\IC^N$ of $G$, which is a finite-dimensional irreducible representation of $\cU_q(\frgl_N)$ with highest weight $(\beta-1)\,N\, \omega_1-(\beta-1)\,(1,\dots,1)=(\beta-1)\, (N-1,-1,\dots,-1)$. By~\cite[Lemma~1]{Etingof1993}, the space of intertwining operators
$\Hom_{\cU_q(\frgl_N)}\big(R_{\lambda'},R_{\lambda'} \otimes
W_{\beta-1}\big)$ for $\cU_q(\frgl_N)$ is one-dimensional if
$\lambda'=\lambda_\beta:= \lambda +(\beta-1)\, \rho$
for a highest weight $\lambda$ and zero otherwise; recall that
$\lambda$ is a dominant weight of $\frgl_N$ if and only if it is of
the form
\be
\lambda=a\, (1,\dots,1)+\sum_{i=1}^N\, n_i\, \omega_i
\ee
for some $n_i\in\IZ_{\geq 0}$ and $a\in\IC$, where $\omega_i=(1^i \
0^{N-i})$, $i=1,\dots,N$ are the fundamental weights of $\frgl_N$. It
follows that a non-zero $\cU_q(\frgl_N)$-homomorphism
$\Phi_{\lambda} :R_{\lambda_\beta} \to R_{\lambda_\beta} \otimes W_{\beta-1}$ is unique up
to normalization. As
the weight zero subspace $\big(W_{\beta-1}\big)_0$ is one-dimensional, the
corresponding {generalized character}
\bea
\chi_{\Phi_\lambda}(U):=
\Tr_{R_{\lambda_\beta}}\big(\Phi_\lambda\, U\big)
\label{eq:gencharacter}\eea
can be regarded as taking values in $\IC$. By~\cite[Theorem~1]{Etingof1993}, if $\lambda$ is a partition these generalized characters
are given in terms of the monic form $M_\lambda(x;q,t)$ of the
Macdonald polynomials at $t=q^\beta$, where $U=\e^{(z,H)}$ and $x=\e^z$. We
choose the normalization of $\Phi_\lambda$ and the identification
$\big(W_{\beta-1} \big)_{0}\cong\IC$ in such a way so that
\bea
\chi_{\Phi_\lambda}(U) = \frac{M_\lambda(x;q,t)}{\sqrt{g_\lambda}} \ .
\label{eq:gencharMacdonald}\eea
In the unrefined limit $\beta=1$, we have $g_\lambda=1$ and the
Macdonald polynomials reduce to the Schur polynomials
$M_\lambda(x;q,q)=s_\lambda(x)$ (independently of $q$), which coincide
with the ordinary characters $\chi_{R_\lambda}(U)=\Tr_{R_\lambda}(U)$
of the irreducible representation $R_\lambda$.

\subsection{Quantum Schur-Weyl duality}
\label{sec:qCWdual}

For $n\geq 2$, the
actions of $\cU_q(\frgl_N)$ and $\sfH_q(\frS_n)$ on $R_{\omega_1}^{\otimes n}$ are given respectively by the iterated coproduct $\Delta^{n-1}= (\Delta\otimes\unit^{\otimes(n-1)})\circ\cdots\circ(\Delta\otimes\unit)\circ
\Delta$ (see Appendix~\ref{app:Quantumgroup}) and by $q$-deformation of permutation of the factors (to be specified in~\S\ref{sec:refdim} below). These actions
commute, and so $R_{\omega_1}^{\otimes n}$ is a representation of the product
$\cU_q(\frgl_N)\times \sfH_q(\frS_n)$ which is completely reducible
to the form
\bea
R_{\omega_1}^{\otimes n}\cong \bigoplus_{\lambda\in\Lambda_+^{n}}\,
R_\lambda\otimes r_\lambda \ .
\label{eq:qschurweyl}\eea
Here $\Lambda_+^{n}\subset\Lambda_+$ is the set of partitions
$\lambda$ of $n$ with at most $N$ parts, i.e. $|\lambda|=n$, $R_\lambda$ is the corresponding irreducible representation of
$\cU_q(\frgl_N)$, and $r_\lambda$ is the representation
of $\sfH_q(\frS_n)$ associated with $\lambda$. Letting $P_\lambda$ denote
the quantum Young projector for the representation $r_\lambda$, one has
\bea
P_\lambda R_{\omega_1}^{\otimes n} \cong R_\lambda\otimes r_\lambda \ .
\label{eq:Youngprojector}\eea

Using quantum Schur-Weyl duality we can write the generalized
characters $\chi_{\Phi_\lambda}(U)$ for
$\lambda_\beta\in\Lambda_+^{n}$ as combinatorial expansions over
the symmetric group $\frS_n$ involving characters of the
Hecke algebra $\sfH_q(\frS_n)$. For this, we introduce a $\cU_q(\frgl_N)$-intertwiner
\bea
\Phi_n \, :\, R_{\omega_1}^{\otimes n} \ \longrightarrow \
R_{\omega_1}^{\otimes n}\otimes W_{\beta-1}
\label{eq:Phi1}\eea 
for each $n\geq0$, which can be defined in the following (non-canonical) way: As a $\cU_q(\frgl_N)$-module the vector space
$R_{\omega_1}^{\otimes n}$ decomposes into irreducible unitary
representations as
\bea \label{eq:PhiDefRep}
R_{\omega_1}^{\otimes n} = \bigoplus_{\lambda\in\Lambda_+^{n}}\,
R_\lambda^{\oplus d_\lambda(1)}
\eea
where $d_\lambda(1)= \dim (r_\lambda)$. We can use the projector property $\sum_{\lambda\in\Lambda_+^{n}}\, P_\lambda=\unit_{R_{\omega_1}^{\otimes n}}$ to write
\bea
\Phi_n = \sum_{\lambda,\mu\in\Lambda_+^{n}}\, \big(P_\mu\otimes \unit_{W_{\beta-1}} \big)\, \Phi_n\, P_\lambda \ ,
\eea
with
\bea \label{eq:PhiDefProj}
\big(P_\mu\otimes \unit_{W_{\beta-1}}\big)\, \Phi_n\, P_\lambda := \delta_{\lambda,\mu} \ \sum_{\lambda\in\Lambda_+^{n}}\, \Phi_{\lambda_{\beta-2}}\otimes\unit_{r_{\lambda}} \ ,
\eea
where we used (\ref{eq:PhiDefRep}) and $\Phi_{\lambda_{\beta-2}}\in\Hom_{\cU_q(\frgl_N)}\big(R_{\lambda},R_{\lambda} \otimes
W_{\beta-1}\big)$. In the large $N$ limit, if $\lambda$ is a dominant weight then so are $\lambda_\beta$ and $\lambda_{\beta-2}$, and thus $\Hom_{\cU_q(\frgl_N)}\big(R_{\lambda},R_{\lambda} \otimes W_{\beta-1}\big)$ is non-zero and one-dimensional if $\lambda\in\Lambda_+$.
This gives an identification of underlying linear transformations
\bea \label{eq:PhiDef}
\Phi_n = \bigoplus_{\lambda\in\Lambda_+^{n}}\, \Phi_{\lambda_{\beta-2}}\otimes\unit_{r_{\lambda}} \ .
\eea

By~\cite[eq.~(2.28)]{deHaro2006}, the
quantum Young projector can be written explicitly as the central element of the Hecke
algebra $\sfH_q(\frS_n)$ given by
\bea
P_\lambda = \frac{d_\lambda(q)}{q^{\frac{n\,(n-1)}4}\, [n]_q!} \ \sum_{\sigma\in\frS_n}\,
q^{-\ell(\sigma)}\, \chi_{r_\lambda}\big(\sfh(\sigma^{-1}) \big) \
\sfh(\sigma) \ ,
\label{eq:Youngprojexpand}\eea
where $\ell(\sigma)$ is the length of the permutation $\sigma\in\frS_n$, $\sfh(\sigma)\in\sfH_q(\frS_n)$ is the Hecke algebra element associated to $\sigma\in\frS_n$ and $\chi_{r_\lambda}\big(\sfh(\sigma^{-1}) \big)$ are the 
characters of the irreducible representation $r_\lambda$ of the Hecke algebra; here $d_\lambda(q)= \dim_q(r_\lambda)$ is a $q$-deformation of the dimension of the symmetric group representation $r_\lambda$, see Appendix~\ref{app:Hecke}.
Then we evaluate the trace $\Tr_{R_{\omega_1}^{\otimes n}} \big(\Phi_n \,
  U\, P_\lambda\big)$ in two different ways. Firstly, using
  (\ref{eq:Youngprojector}) and (\ref{eq:PhiDefRep}) along with (\ref{eq:gencharcompute}) we
  easily get
\be
\Tr_{R_{\omega_1}^{\otimes n}} \big(\Phi_n \,
  U\, P_\lambda\big) \, = \,
      \Tr_{R_{\lambda}}(\Phi_{\lambda_{\beta-2}} \, U)\, d_\lambda(1) \, = \, \chi_{\Phi_{\lambda_{\beta-2}}} (U) \, d_\lambda(1)  \ .  
\ee
Secondly, we substitute the explicit expansion (\ref{eq:Youngprojexpand}),
and hence for any weight $\lambda\in\Lambda_+^{n}$
we can express the vector-valued trace as
\bea
\Tr_{R_{\lambda}}(\Phi_{\lambda_{\beta-2}} \, U) \, = \, \frac{q^{-\frac{n\,(n-1)}4}}{ [n]_q!} \,
\frac{d_{\lambda}(q)}{d_{\lambda}(1)}\, \sum_{\sigma\in\frS_n}\,
q^{-\ell(\sigma)}\, \chi_{r_{\lambda}}\big(\sfh(\sigma^{-1})\big) \
\Tr_{R_{\omega_1}^{\otimes n}}\big(\Phi_n \, U\, \sfh(\sigma)\big) \ .
\label{eq:gencharHecke}\eea

It will prove useful later on to derive directly a formula for the inverse of this transformation of characters, generalizing \cite[eq.~(2.21)]{deHaro2006}.
\begin{lemma}
$\displaystyle{ \ \Tr_{R_{\omega_1}^{\otimes n}}\big(\Phi_n \, U\, \sfh(\sigma)\big)= \sum_{\lambda\in\Lambda_+^{n}}\, \chi_{r_{\lambda}}\big(\sfh(\sigma)\big) \ \chi_{\Phi_{\lambda_{\beta-2}}}(U)  \ . }$
\label{lem:schurweyl}\end{lemma}
\Proof{ 
Starting from the projector and centrality properties of $P_\lambda$ in the Hecke algebra along with (\ref{eq:PhiDefProj}) we compute
\bea
\Tr_{R_{\omega_1}^{\otimes n}}\big(\Phi_n \, U\, \sfh(\sigma)\big) 
&=& \sum_{\lambda,\mu\in\Lambda_+^{n}}\, \Tr_{R_{\omega_1}^{\otimes n}}\big((P_\mu \otimes \unit_{W_{\beta-1}})\,(P_\mu \otimes \unit_{W_{\beta-1}})\, \Phi_n \, P_\lambda \, U \, \sfh(\sigma) \, P_\lambda \big) \nonumber \\ [4pt]
&=& \sum_{\lambda\in\Lambda_+^{n}}\, \Tr_{R_{\lambda}\otimes r_{\lambda}}\Big( \big(\Phi_{\lambda_{\beta-2}}\otimes\unit_{r_{\lambda}}\big) \, \big(U\otimes \sfh(\sigma)\big) \Big) \\[4pt] &=& \sum_{\lambda\in\Lambda_+^{n}}\, \Tr_{R_{\lambda}} \big(\Phi_{\lambda_{\beta-2}}\, U\big)\, \Tr_{r_{\lambda}}\big(\sfh(\sigma)\big) \ = \ \sum_{\lambda\in\Lambda_+^{n}}\, \Tr_{R_{\lambda}} \big(\Phi_{\lambda_{\beta-2}}\, U\big) \, \chi_{r_{\lambda}}\big(\sfh(\sigma)\big) \nonumber 
\eea
as required.
}

By multiplying the left-hand side and the right-hand side of the character formula in Lemma~\ref{lem:schurweyl} with $q^{-\ell(\sigma)}\, \chi_{r_{\lambda'}}\big(\sfh(\sigma^{-1})\big)$, summing over all permutations $\sigma\in \frS_n$ and using the orthogonality relations for Hecke characters~\cite{deHaro2006}
\bea
\sum_{\sigma\in\frS_n}\, q^{-\ell(\sigma)}\,
\chi_{r_{\lambda}}\big(\sfh(\sigma)\big)\,
\chi_{r_{\lambda'}}\big(\sfh(\sigma^{-1})\big) =
\delta_{\lambda,\lambda'} \ q^{\frac{n\,(n-1)}4}\, [n]_q! \ \frac{d_{\lambda}(1)}{d_{\lambda}(q)} \ ,
\label{eq:orthogonality}\eea
we arrive at the expression (\ref{eq:gencharHecke}).

\subsection{Refined quantum dimensions\label{sec:refdim}}

We are finally ready to derive our Hecke character expansion for the
refined quantum dimensions (\ref{qtdim}). Firstly we note that the
refined quantum dimension (\ref{qtdim}) and the Macdonald metric
(\ref{eq:MacdMetric}) are both invariant under any shift of the
dominant weight $\lambda$ by the maximal partition $(1^N):=(1,\dots,1)$ of length $N$, i.e.
\bea \label{eq:dimshift}
\dim_{q,t}(R_{\lambda+a\,(1^N)}) = \dim_{q,t}(R_\lambda) \qquad \mbox{and} \qquad
g_{\lambda+a\, (1^N)} = g_\lambda \ .
\eea
We shall assume that $a$ is an integer. The refined quantum dimension is obtained by the
specialization $U=t^{(\rho,H)}$ in the generalized characters
(\ref{eq:gencharMacdonald}), i.e.
\be
\frac{\dim_{q,t}(R_\lambda)}{\sqrt{g_\lambda}} \, = \, \chi_{\Phi_\lambda}\big(q^{\beta\, (\rho,H)} \big)\, = \, \Tr_{R_{\lambda_\beta}}\big(\Phi_{\lambda} \, q^{\beta\, (\rho,H)}\big) \ .
\ee
We wish to substitute in the expansion
(\ref{eq:gencharHecke}), but the Hecke characters and dimensions are only defined for partitions, whereas $\lambda_\beta$ is not necessarily a partition. Hence we use the shift symmetry (\ref{eq:dimshift}) to get a partition $\lambda_\beta+a\, (1^N)$, which is true as long as $a\geq \frac{N-1}{2}\, (\beta-1)$. For definiteness we use the lowest value 
\be
a=\mbox{$\frac{N-1}{2}$}\, (\beta-1)
\ee
which for large $N$ can be regarded as integral. In the large $N$ expansion that we consider later on, we will typically also consider the limit $\beta\to1$ such that the quantity $a\,N$ is finite. Then we get
\be
\frac{\dim_{q,t}(R_\lambda)}{\sqrt{g_\lambda}} \, = \,
 \frac{q^{-\frac{n\,(n-1)}4}}{ [n]_q!} \,
\frac{d_{\lambda_\beta+a\, (1^N)}(q)}{d_{\lambda_\beta+a\, (1^N)}(1)}\, \sum_{\sigma\in\frS_n}\,
q^{- \ell(\sigma)}\, \chi_{r_{\lambda_\beta+a\,(1^N)}}\big(\sfh(\sigma^{-1})\big) \
\Tr_{R_{\omega_1}^{\otimes n}}\big(\Phi_n \, q^{\beta\, (\rho,H)} \, \sfh(\sigma)\big) \label{eq:qtdim}
\ee
for $\lambda_\beta+a\, (1^N)\in\Lambda_+^{n}$, i.e. $\lambda\in\Lambda_+^{n-a\, N}$. We can easily check that this
expression agrees with the anticipated formula for the quantum
dimension $\dim_q(R_\lambda)= \Tr_{R_\lambda}\big(q^{(\rho,H)}\big)$
in the unrefined limit: In the limit $\beta= 1$, we have
$\lambda_\beta= \lambda$, $a=0$, and the intertwiners
$\Phi_\lambda$ and $\Phi_n$ become identity operators on
finite-dimensional modules, so that (\ref{eq:qtdim}) coincides with the $q$-dimension formula from \cite[eq.~(2.33)]{deHaro2006}.

To manipulate the sum in (\ref{eq:qtdim}), let us first explicitly
specify how the Hecke operators act. From~\cite{King1992} we can
express $\sfh(\sigma)\in \sfH_q(\frS_n)$ as a product of generators
$\sfg_i$, $i=1,\dots,n-1$ in the same way that we express
$\sigma\in\frS_n$ in the form of a reduced word; we say that
$\sfh(\sigma)\in \sfH_q(\frS_n)$ is a reduced word if $\sigma\in\frS_n$
is a reduced word.
From \cite{deHaro2006}, $\sfg_i$ acts on $R_{\omega_1}^{\otimes n}$ as the braiding operator
\be
\sfg= q^{1/2} \ \check{\sfR}
\ee
on the tensor product $R_{\omega_1}\otimes R_{\omega_1}$ in the
$(i,i+1)$ slot of $R_{\omega_1}^{\otimes n}$, where
$\check{\sfR}=\sfP\,\sfR$ with $\sfP$ the flip operator $\sfP(v\otimes
w)=w\otimes v$; here $\sfR$ is the FRT quantum $R$-matrix~\cite{Faddeev1989}
\be
\sfR=q^{1/2}\, \sum_{i=1}^N \, H_{i}\otimes H_{i}+\sum_{i\neq j} \, H_{i}\otimes H_{j}+\big(q^{1/2}-q^{-1/2} \big)\, \sum_{i>j}\, E_{ij}\otimes E_{ji} \ ,
\label{eq:FRT}\ee
where $H_i=E_{ii}$ and $E_{ij}$ for $i,j=1,\dots,N$ act on the standard basis $\{e_k\} \subset R_{\omega_1}=\IC^N$ of the fundamental representation as
\be
E_{ij} e_k=\delta_{jk} \ e_i \ .
\label{eq:fundrep}\ee

Let us define the \emph{$(q,t)$-trace} of an element $x\in\sfH_q(\frS_n)$ by $\Tr_{q,t}(x):=\Tr_{R_{\omega_1}^{\otimes n}}\big(\Phi_n \, U \, x \big)$, where $U=t^{(\rho,H)}$ and $\Phi_n$ is the intertwiner (\ref{eq:Phi1}). This terminology is justified by
\begin{lemma}
 \ The $(q,t)$-trace is cyclic: $ \ \Tr_{q,t}(x\, y)=\Tr_{q,t}(y\, x)$ for all $x,y\in\sfH_q(\frS_n)$.
\label{lem:qtcyclic}\end{lemma}
\Proof{Since
the reduced words $\{\sfh(\sigma)\}_{\sigma\in\frS_n}$ form a basis of
$\sfH_q(\frS_n)$, we can express $\Tr_{q,t}(x\, y)$ as a linear
combination of $(q,t)$-traces $\Tr_{q,t} \big(\sfh(\sigma)\, \sfh(\tau)
\big)$, and therefore we only need to show that $\Tr_{q,t} \big(\sfh(\sigma)\,
\sfh(\tau) \big)=\Tr_{q,t} \big(\sfh(\tau)\, \sfh(\sigma) \big)$ for all
$\sigma,\tau\in\frS_n$. We first prove that $\Phi_n$ and the Hecke
algebra generators $\sfg_i$ commute. Let us consider a fixed element $f_j$ of a basis $\{f_i\}\subset W_{\beta-1}$, with corresponding dual basis $\{ f^i\}$. If we restrict the codomain of $\Phi_n$ to $f_j$, then
$
_{f_j}|\Phi_n \in \End_{\cU_q(\frgl_N)}\big(R_{\omega_1}^{\otimes n}\big)$ and using quantum Schur-Weyl duality we can decompose this restriction as
\be
_{f_j}\big|\Phi_n =\bigoplus_{\lambda\in\Lambda_+^{n}}\, \big(\, _{f_j}\big|\Phi_n \, \big)\big|_{R_\lambda}\otimes \unit_{r_\lambda} \ ,
\ee
where $(\, _{f_j} |\Phi_n \,) |_{R_\lambda}\in \End_{\cU_q(\frgl_N)}(R_\lambda)$. It follows that $_{f_j} |\Phi_n$ acts on the Hecke algebra representation $r_\lambda$ as the identity, and so $_{f_j}|\Phi_n$ and $\sfg_i$ commute. Note that $\sfg_i$ commutes with $U^{\otimes n}$, because both $\sfP$ and $\sfR\in \cU_q(\frgl_N)\otimes\cU_q(\frgl_N) $ commute with $t^{(\rho,H)}\otimes t^{(\rho,H)}$. Thus we get
\be
\Tr_{R_{\omega_1}^{\otimes n}}\big(\, _{f_j}\big|\Phi_n \, U\, x\, y\, \big)=\Tr_{R_{\omega_1}^{\otimes n}}\big(\, _{f_j}\big|\Phi_n \, U\, y\, x\, \big)
\ee
and hence
\bea
\Tr_{q,t}(x\, y)&=&\Tr_{R_{\omega_1}^{\otimes n}}\big(\Phi_n \, U\, x\,
  y\big) \nonumber \\[4pt] &=& f^j\big( \Tr_{R_{\omega_1}^{\otimes
    n}}(\,_{f_j}|\Phi_n \, U \, x\, y \, ) \big) \ f_j \nonumber\\[4pt]
&=& \Tr_{R_{\omega_1}^{\otimes n}}\big(\Phi_n \, U\, y\, x\big) \ = \
\Tr_{q,t}(y\, x) \ ,
\eea
as required.
}

We can use Lemma~\ref{lem:qtcyclic} to truncate the reduced words
$\sfh(\sigma)$ to minimal words in the sum (\ref{eq:qtdim}). A word is
said to be \emph{minimal} if it is both reduced and contains no
generators $\sfg_i$ more than once. Using the Hecke relations on $\sfg_i$
and cyclicity of the $(q,t)$-trace, we can then truncate the sum in
(\ref{eq:qtdim}) to conjugacy classes $T$ in $\frS_n$ and express the
$(q,t)$-trace $\Tr_{q,t} \big(\sfh(\sigma)\big)$ for any $\sigma\in T$
as the $(q,t)$-trace $\Tr_{q,t}\big(\sfh(m_T)\big) $ of the minimal word
$m_T\in T$ in the conjugacy class~\cite{King1992}. Hence following the derivation
of~\cite[eq.~(2.38)]{deHaro2006}, we can write the expansion
(\ref{eq:qtdim}) as
\be
\label{eq:qtdim_minW}
\frac{\dim_{q,t}(R_\lambda)}{\sqrt{g_\lambda}}\, =\, \frac{q^{-\frac{n\,(n-1)}4}}{ [n]_q!} \,
\frac{d_{\lambda_\beta+a\, (1^N)}(q)}{d_{\lambda_\beta+a\,(1^N)}(1)}\, \sum_{T\in\frS_n^\vee}\,
q^{- \ell(T)}\, \chi_{r_{\lambda_\beta+a\, (1^N)}}(C_T) \
\Tr_{R_{\omega_1}^{\otimes n}}\big(\Phi_n \, q^{\beta\, (\rho,H)} \,
\sfh(m_T)\big)
\ee
where $\ell(T)$ is the length of the permutation $m_T\in\frS_n$ and
$C_T$ are the same central elements of the Hecke
algebra $\sfH_q(\frS_n)$ as in~\cite{deHaro2006}. To write $C_T$
explicitly, we need to express an arbitrary character of the Hecke algebra
in terms of characters of minimal words~\cite{King1992} using
Lemma~\ref{lem:qtcyclic} and the Hecke relations (\ref{eq:HeckeRel})
from Appendix~\ref{app:Hecke} as
\be \label{eq:minwordsExpand}
\chi_{r_\lambda}\big(\sfh(\sigma)\big)  =  \sum_{T\in\frS_n^\vee}
\alpha_{T}(\sigma) \, \chi_{r_\lambda} \big(\sfh(m_T)\big) \ , 
\ee
where the expansion coefficients $\alpha_{T}(\sigma)$ do not depend
on the representation $r_\lambda$. Then the central element $C_T$ is defined by
\be \label{eq:CTdef}
C_T = q^{\ell(T)} \, \sum_{\sigma\in\frS_n} q^{-\ell(\sigma)} \, \alpha_{T}(\sigma^{-1}) \, \sfh(\sigma) \ .
\ee
The quantum Young projector (\ref{eq:Youngprojexpand}) can be rewritten as
\be \label{eq:QmYoungP_C}
P_\lambda = \frac{d_\lambda(q)}{q^{\frac{n\,(n-1)}4}\, [n]_q!} \ \sum_{T\in\frS_n^\vee}\,
q^{-\ell(T)} \, \chi_{r_\lambda}\big(\sfh(m_T) \big)\, C_T \ .
\ee
This transformation can be inverted to express $C_T$ in terms of the central elements $P_\lambda$, because the determinant of the transformation matrix is
non-zero. To see that the determinant of
$\chi_{r_\lambda}\big(\sfh(m_T) \big)$ is non-zero, we
use the orthogonality relation (\ref{eq:orthogonality}) and expand it into minimal words
using (\ref{eq:minwordsExpand}). Then we take the determinant of the
obtained expression and use multiplicativity of the determinant to
find that it is non-vanishing. Hence the centrality of the elements
$C_T$ follows from the centrality of the projectors.

\subsection{$(q,t)$-traces of minimal words}

We are left with the problem of evaluating the $(q,t)$-traces $\Tr_{R_{\omega_1}^{\otimes n}}\big(\Phi_n \, q^{\beta\, (\rho,H)} \,
\sfh(m_T)\big)$ of minimal words; we shall follow the strategy
of~\cite[Appendix~B]{deHaro2006}. For $n=1$, the $R$-matrix $\check{\sfR}$ acts trivially
while $Ue_i=t^{\rho_i}\, e_i$, where $\rho_i=\frac{N+1}2-i$. Using
\be
\dim_{q,t}(R_{\omega_1})=[N]_t = \Tr_{R_{\omega_1}}(U) \qquad
\mbox{and} \qquad g_{\omega_1} = g_\emptyset \
\frac{\big[ N\big]_t}{\big[\beta\, (N-1) +1\big]_q}
\label{eq:dimqtRfund}\ee
where generally
\be
[N]_{t^k}:= \frac{[k\, \beta\, N]_q}{[k\, \beta]_q} \ ,
\ee
we then easily find for the $(q,t)$-trace
\be 
\Tr_{R_{\omega_1}}\big(\Phi_1\, U\big) =
\frac{\dim_{q,t}(R_{\omega_1})}{\sqrt{g_{\omega_1}}} = \bigg(\,
\frac{\big[ N\big]_t\,\big[\beta\, (N-1)+1\big]_q}{g_{\emptyset} } \, \bigg)^{1/2} \ .
\ee
Note that here the intertwining operator $\Phi_1=\Phi_{\omega_1}
:R_{\omega_1}\to R_{\omega_1}\otimes W_{\beta-1}$ simply acts in the $(q,t)$-trace of $1$ to rescale the
normalization of the trace of $U$ by the Macdonald measure factor $(g_{\omega_1})^{-1/2}$.
In the unrefined limit $\beta=1$, this expression reduces as expected
to the quantum dimension of the fundamental representation
$\dim_q(R_{\omega_1})= [N]_q$. Below we shall also need the
generalization of the trace formula in (\ref{eq:dimqtRfund}) to arbitrary powers
$U^k$, $k\in\IZ_{>0}$, which is given by
\be
\Tr_{R_{\omega_1}}\big(U^k\big)= [ N]_{t^k} \ .
\label{eq:TrRfundUk}\ee

Next we set $n=2$. We can use the FRT formula (\ref{eq:FRT}) for the $R$-matrix to compute
\be 
\big(U^k \otimes \unit_{R_{\omega_1}} \big)\,\check{\sfR}(e_i\otimes e_j) = t^{k\, \rho_j}\, e_j\otimes e_i+t^{k\, \rho_j}\, \big(q^{1/2}-1 \big)\, \delta_{ij}\, e_i\otimes e_j+ t^{k\, \rho_i}\,\big(q^{1/2}-q^{-1/2} \big)\, \theta_{ij}\, e_i\otimes e_j
\ee
for any $k\in\IZ_{>0}$, where $\theta_{ij}:=1$ if $i<j$ and $\theta_{ij}:=0$ otherwise. From this expression one can easily derive the partial traces
\be 
\big(\Tr_{R_{\omega_1}}\otimes \unit_{R_{\omega_1}} \big)\Big( \big(U^k \otimes
\unit_{R_{\omega_1}} \big)\,\sfg_1 \Big) = t^{k\, \frac{N+1}2}\,
\frac{q-1}{t^k-1} \ \unit_{R_{\omega_1}}+\frac{t^k-q}{t^k-1} \
U^k
\label{eq:qt2trace}\ee
as operators in $\cU_q(\frgl_N)$ acting on the fundamental
representation $R_{\omega_1}$. In the unrefined limit $t=q$ at $k=1$, this operator reduces to $q^{\frac{N+1}2}\, \unit_{R_{\omega_1}}$ as in~\cite[eq.~(B.5)]{deHaro2006}; in the general case it is also diagonal but no longer proportional to the identity operator.

Let us decompose the representation $W_{\beta-1}$ into its one-dimensional weight subspaces $\big(W_{\beta-1}\big)_\alpha\cong \IC w^\alpha$; in particular, the isomorphism $\big(W_{\beta-1}\big)_0\cong \IC$ is given by mapping $w^0\mapsto 1$. Then one can find explicit formulas for the matrix elements of $\Phi_2:R_{\omega_1}^{\otimes2}\to R_{\omega_1}^{\otimes2}\otimes W_{\beta-1}$ in the following way: We write
\be 
\Phi_2(e_i\otimes e_j) = P_{ij}{}^{kl}{}_\alpha \ e_k\otimes e_l\otimes w^\alpha \ .
\ee
Then the condition that $\Phi_2$ is an intertwiner can be rewritten as a
system of linear equations for the expansion coefficients
$P_{ij}{}^{kl}{}_\alpha$. Since $\Phi_2$ is uniquely determined up to
the normalization in (\ref{eq:gencharMacdonald}), this linear system has
a unique solution which determines $P_{ij}{}^{kl}{}_\alpha$ as a
rational function in $q^{1/2}$ and $t^{1/2}=q^{\beta/2}$; with (\ref{eq:gencharMacdonald}) the
intertwining operators $\Phi_\lambda:R_{\lambda_\beta} \to
R_{\lambda_\beta} \otimes W_{\beta-1}$ are normalized in such a way that if
$v_{\lambda_\beta}$ is a highest weight vector of $R_{\lambda_\beta}$,
then the component of $\Phi_\lambda(v_{\lambda_\beta})$ in
$R_{\lambda_\beta}\otimes \big(W_{\beta-1}\big)_0$ is $(g_{\lambda_\beta})^{-1/2} \, v_{\lambda_\beta}\otimes w^0$. Setting $P_{ij}{}^{kl}:=P_{ij}{}^{kl}{}_0$, the $(q,t)$-trace of the generator $\sfg_1$ can then be written as
\bea
\Tr_{R_{\omega_1}^{\otimes2}}\big(\Phi_2\, U\, \sfg_1 \big) &:=& \big(\Tr_{R_{\omega_1}} \otimes \Tr_{R_{\omega_1}} \big)\big(\Phi_2\, (U\otimes U)\, \sfg_1\big) \nonumber \\[4pt] &=& q^{1/2}\, t^{N+1}\, \Big(\, \sum_{i,j=1}^N\, t^{-i-j}\, P_{ji}{}^{ij}+\big(q^{1/2}-1\big)\, \sum_{i=1}^N\, t^{-2i}\, P_{ii}{}^{ii} \nonumber \\ && \qquad \qquad \qquad +\, \big(q^{1/2}-q^{-1/2}\big)\, \sum_{i<j}\, t^{-i-j}\, P_{ij}{}^{ij}\, \Big) \ .
\label{eq:Trace2}\eea
In the unrefined limit $\beta=1$, we have $P_{ij}{}^{kl}=\delta_i{}^k\, \delta_j{}^l$ and it is easy to check that this expression reduces to $q^{\frac{N+1}2}\, [N]_q$ as in~\cite[eq.~(B.5)]{deHaro2006}. In the general case we have
\begin{lemma}
$\displaystyle{ \ P_{ij}{}^{kl}=(g_{\omega_1})^{-1} \ \delta_i{}^k\, \delta_j{}^l \ . }$
\label{lemma:P0}\end{lemma}
\Proof{
By $\cU_q(\frgl_N)$-equivariance we have the relations
\be 
\Phi_2\, \Delta(H_p)= \Delta^2(H_p)\, \Phi_2 \qquad \mbox{for} \quad p=1,\dots,N
\ee
as operators on $R_{\omega_1}^{\otimes2}\to
R_{\omega_1}^{\otimes2}\otimes W_{\beta-1}$, where $\Delta(H_p)=H_p\otimes
\unit+\unit\otimes H_p$ and $\Delta^2(H_p)=(\Delta\otimes\unit)
\Delta(H_p)$. We evaluate both sides of this equation on a generic basis
element $e_i\otimes e_j$ of $R_{\omega_1}\otimes R_{\omega_1}$, and
denote the action of the Cartan generators on the weight subspaces of
$W_{\beta-1}$ as $H_p w^\alpha=\alpha_p\, w^\alpha$ with $\alpha_p\in \IC$. Then the equality can be written as
\bea
&& \big(P_{pj}{}^{kl}{}_\alpha\, \delta_{pi}+P_{ip}{}^{kl}{}_\alpha\,
\delta_{pj}\big) \ e_k\otimes e_l\otimes w^\alpha \\ && \qquad \qquad \qquad\qquad \qquad \qquad \qquad \ = \  P_{ij}{}^{kl}{}_\alpha\, \big(\delta_{pk} \
e_p\otimes e_l+\delta_{pl} \ e_k\otimes e_p+\alpha_p \ e_k\otimes
e_l\big) \otimes w^\alpha \nonumber
\ .
\eea
In particular, for the weight $\alpha=0$ component we obtain the
constraints
\be
P_{ij}{}^{kl}\, \left(\delta_{pk}+\delta_{pl}-\delta_{pi}-\delta_{pj}\right)=0
\ee
for all $i, j,  k,  l,  p = 1,\ldots, N$. The tensor $P_{ij}{}^{kl}=
\delta_i{}^k\, \delta_j{}^l$ solves this equation, and it is the
unique solution up to normalization. The normalization is found as
above by observing that the intertwiner $\Phi_2$ acts in the $(q,t)$-trace as a
multiple of the identity operator $\unit_{R_{\omega_1}}\otimes \unit_{R_{\omega_1}}$,
with proportionality constant $(g_{\omega_1})^{-1}$.
}
Using Lemma~\ref{lemma:P0} we can straightforwardly express the $(q,t)$-trace (\ref{eq:Trace2}) in terms of $q$-numbers as
\be \label{eq:qtTraceg1}
\Tr_{R_{\omega_1}^{\otimes2}}\big(\Phi_2\, U\, \sfg_1 \big) =
\frac{\big[\beta\, (N-1)+1\big]_q}{g_\emptyset} \, \Big(\,
t^{\frac{N+1}2}\, \frac{q-1}{t-1}+ \frac{t-q}{t-1} \, \frac{[N]_{t^2}}{[N]_t}\, \Big) \ .
\ee

Next we have to generalise this expression to the $(q,t)$-trace of the
connected minimal word $\sfg_1\, \sfg_2 \cdots \sfg_{n-1}$ for~$n\geq2$. Defining
\be 
\Phi_n(e_{i_1}\otimes\cdots \otimes e_{i_n})=P_{i_1\cdots i_n}{}^{j_1\cdots j_n}{}_\alpha \ e_{j_1}\otimes\cdots \otimes e_{j_n}\otimes w^\alpha
\ee
and setting $P_{i_1\cdots i_n}{}^{j_1\cdots j_n}:=P_{i_1\cdots i_n}{}^{j_1\cdots j_n}{}_0$, by a completely analogous argument to that used in the proof of Lemma~\ref{lemma:P0} one obtains the constraints
\be
P_{i_1\cdots i_n}{}^{j_1\cdots j_n}\, \left(\delta_{p\, j_1}+\cdots+\delta_{p\,j_n}-\delta_{p\,i_1}-\cdots-\delta_{p\,i_n}\right)=0
\ee
for all $p,i_1,\dots,i_n,j_1,\dots,j_n=1,\dots,N$, and $P_{i_1\cdots i_n}{}^{j_1\cdots j_n}=\delta_{i_1}{}^{j_1}\cdots\delta_{i_n}{}^{j_n}$ solves this; thus again $\Phi_n$ acts in the $(q,t)$-trace as a multiple of the identity operator $\unit_{R_{\omega_1}}^{\otimes n}$ with the normalization determined as before to be $(g_{\omega_1})^{-n/2}$, and we find
\be \label{eq:PhiFundaId}
P_{i_1\cdots i_n}{}^{j_1\cdots j_n}=(g_{\omega_1})^{-n/2} \ \delta_{i_1}{}^{j_1}\cdots\delta_{i_n}{}^{j_n} \ .
\ee
We can use this result together with the partial trace formula
(\ref{eq:qt2trace}) and the traces (\ref{eq:TrRfundUk}) to calculate the $(q,t)$-trace
\bea \label{eq:qt3trace}
&& \Tr_{R_{\omega_1}^{\otimes n}} \big(\Phi_n \, U \, \sfg_{1} \,
\sfg_{2} \cdots \sfg_{n-1} \big) \ := \
\big(\Tr_{R_{\omega_1}}\big)^{\otimes n}\big(\Phi_{\omega_1}^{\otimes
  n} \, U^{\otimes n}\, \sfg_{1} \, \sfg_{2} \cdots \sfg_{n-1} \big)
\nonumber \\[4pt]
&& \qquad \ = \  (g_{\omega_1})^{-n/2} \ \Tr_{R_{\omega_1}} \Big( U\,
\big(\Tr_{R_{\omega_1}} \otimes \unit_{R_{\omega_1}}
\big)\big( (U\otimes \unit_{R_{\omega_1}} )\, \sfg_1 \big) 
\\ && \qquad \qquad \ \cdots \ \big(\Tr_{R_{\omega_1}}\otimes
\unit_{R_{\omega_1}}^{\otimes(n-2)} \big) \big( (U\otimes
\unit_{R_{\omega_1}} ^{\otimes(n-2)})\, \sfg_1\big)\, \big(\Tr_{R_{\omega_1}}\otimes
\unit_{R_{\omega_1}}^{\otimes(n-1)} \big) \big( (U\otimes
\unit_{R_{\omega_1}} ^{\otimes(n-1)})\, \sfg_1\big)\Big) \nonumber
\eea
recursively in $n$. To simplify the derivation, let us introduce the short-hand notation
\be
\xi_k:=t^{k\, \frac{N+1}{2}}\, \frac{q-1}{t^k-1} \qquad \mbox{and} \qquad \varphi_k:=\frac{t^k-q}{t^k-1} \ .
\ee
Using (\ref{eq:qt2trace}), we then compute the first partial trace from (\ref{eq:qt3trace}) as
\be
\mathcal{O}_1\, :=\,\big(\Tr_{R_{\omega_1}}\otimes
\unit_{R_{\omega_1}}^{\otimes(n-1)} \big) \big( (U\otimes
\unit_{R_{\omega_1}} ^{\otimes(n-1)})\, \sfg_1\big)=\xi_1\,\unit_{R_{\omega_1}}^{\otimes(n-1)}+\varphi_1 \,U\otimes \unit_{R_{\omega_1}}^{\otimes(n-2)} \ .
\ee
The factors of $U$ can be cyclically permuted in the partial traces, so for $0\leq m\leq n-1$ the $m$-th partial trace
\bea
\mathcal{O}_m &:=&\big(\Tr_{R_{\omega_1}} \otimes \unit_{R_{\omega_1}}^{\otimes(n-m)} 
\big)\big( (U\otimes \unit_{R_{\omega_1}}^{\otimes(n-m)} )\, \sfg_1 \big) 
\\ && \ \cdots \ \big(\Tr_{R_{\omega_1}}\otimes
\unit_{R_{\omega_1}}^{\otimes(n-2)} \big) \big( (U\otimes
\unit_{R_{\omega_1}} ^{\otimes(n-2)})\, \sfg_1\big)\, \big(\Tr_{R_{\omega_1}}\otimes
\unit_{R_{\omega_1}}^{\otimes(n-1)} \big) \big( (U\otimes
\unit_{R_{\omega_1}} ^{\otimes(n-1)})\, \sfg_1\big) \nonumber
\eea
can be written as
\be \label{eq:Omdef}
\mathcal{O}_m=\sum_{k=0}^m \, f_k^{(m)}[\xi,\varphi]\ U^{k}\otimes \unit_{R_{\omega_1}}^{\otimes (n-m-1)} \ ,
\ee
where $f_k^{(m)}[\xi,\varphi]$ for $0\leq k\leq m$ are polynomials in $\xi_l$ and $\varphi_l$ with $l\leq k $. We derive a recursion relation for $f_k^{(m)}$ inductively by writing
\be
\mathcal{O}_{m+1}=\sum_{k=0}^m \, f_k^{(m)} \ \big(\Tr_{R_{\omega_1}} \otimes \unit_{R_{\omega_1}}^{\otimes(n-m-1)} 
\big)\big( (U^{k+1}\otimes \unit_{R_{\omega_1}}^{\otimes(n-m-1)} )\, \sfg_1 \big)
\ee
and using (\ref{eq:qt2trace}) to get
\be
\mathcal{O}_{m+1}=\sum_{k=0}^m \, f_k^{(m)} \, \xi_{k+1} \, \unit_{R_{\omega_1}}^{\otimes (n-m-1)}+
\sum_{k=0}^m\, f_k^{(m)} \, \varphi_{k+1}\, U^{k+1}\otimes \unit_{R_{\omega_1}}^{\otimes (n-m-2)} \ .
\ee
Comparing this with the expansion (\ref{eq:Omdef}) for $\mathcal{O}_{m+1}$ yields the recursion relations
\bea
f_0^{(m+1)} &=&\sum_{k=0}^{m} \, f_k^{(m)} \, \xi_{k+1} \ ,\label{eq:re1} \\[4pt]
f_{k+1}^{(m+1)}&=&f_{k}^{(m)} \, \varphi_{k+1} \label{eq:rec2}
\eea
with initial condition $f^{(0)}_0=1$.

For $k>0$ we can use (\ref{eq:rec2}) to express $f^{(m)}_k$ entirely in terms of $f^{(m)}_0$ as
\be \label{eq:fmkphi}
f^{(m)}_k=\varphi_1\cdots\varphi_k \ f^{(m-k)}_0 = \frac{q^{k+1}}{q-1} \ \frac{\big(q^{-1};t\big)_{k+1} } {(t;t)_k} \ f^{(m-k)}_0 \ ,
\ee
where we introduced the $q$-Pochhammer symbols
\be \label{eq:Pochhammer}
(a;q)_k:=\prod_{l=0}^{k-1}\, \big(1-a \,
q^l\big) \qquad \mbox{for} \quad 0<k\leq \infty
\ee
and $(a;q)_0:=1$. Using (\ref{eq:re1}) we can express $f^{(m)}_0$ recursively as
\be
f^{(m)}_0=\sum_{k=0}^{m-1}\, f^{(k)}_0\, \phi_{m-k} \ ,
\ee
where we defined
\be \label{eq:phifunc}
\phi_k\, := \, \xi_k\, \varphi_1\cdots \varphi_{k-1} =-
t^{k\,\frac{N+1}{2}}\,q^k \,\frac{\big(q^{-1};t\big)_{k}}{(t;t)_{k}} \ .
\ee
It is easy to see that the solution to this recursion with
$f_0^{(0)}=1$ is given by an expansion into partitions of $m$ as
\be \label{eq:fmsum}
f^{(m)}_0 = \sum_{\lambda \in \Lambda_+^m} \, {L}_\lambda \, \phi_\lambda \ ,
\ee
where this formula should be understood in the large $N$ limit as it
involves a sum over \emph{all} partitions of $m$. Here
$\phi_\lambda:=\prod_{i=1}^{\ell(\lambda)}\, \phi_{\lambda_i}$ with $\ell(\lambda)$ the length of the partition $\lambda$, and the combinatorial weight
\be
{L}_\lambda=\frac{\ell(\lambda)!}{\big|{\rm Aut}(\lambda)\big|}
\label{eq:Llambda}\ee
is the number of distinguishable orderings of $\lambda$ (e.g.
${L}_{(2,1)}=2$ and ${L}_{(1,1)}=1$), where 
\be
\big|{\rm Aut}(\lambda)\big| = \prod\limits_{i=1}^{|\lambda|} \, m_i(\lambda)!
\ee
is the order of the automorphism group of $\lambda$ consisting of permutations $\sigma\in\frS_{\ell(\lambda)}$ such that $\lambda_{\sigma(i)}=\lambda_i$ for all $i$, and
$m_i(\lambda)$ is the number of parts of $\lambda$ equal to $i$. For example, the first four
terms are given by
\bea
  f^{(1)}_0&=& \phi_1 \ , \nonumber \\[4pt]
f^{(2)}_0&=&\phi_2+\phi_1^2 \ , \nonumber \\[4pt]
 f^{(3)}_0 &=&\phi_3+2\,\phi_1\,\phi_2+ \phi_1^3 \ , \nonumber \\[4pt]
f^{(4)}_0 &=&
 \phi_4+2\,\phi_1\,\phi_3+\phi_2^2+3\,\phi_1^2\,
 \phi_2+\phi_1^4 \ .
\eea

We can finally evaluate the $(q,t)$-trace of the connected minimal word using (\ref{eq:qt3trace}) and (\ref{eq:Omdef}) to write
\be
\Tr_{R_{\omega_1}^{\otimes n}} \big(\Phi_n \, U \, \sfg_{1} \,
\sfg_{2} \cdots \sfg_{n-1} \big) \, = \, (g_{\omega_1})^{-n/2} \ \Tr_{R_{\omega_1}} \big( U\, \mathcal{O}_{n-1}\big) \
 = \, (g_{\omega_1})^{-n/2} \ \sum_{k=0}^{n-1} \, f^{(n-1)}_k \ \Tr_{R_{\omega_1}} \big(U^{k+1}\big) \ ,
\ee
and using (\ref{eq:TrRfundUk}), (\ref{eq:fmkphi}) and (\ref{eq:fmsum}) we get
\bea
&& \Tr_{R_{\omega_1}^{\otimes n}} \big(\Phi_n \, U \, \sfg_{1} \,
\sfg_{2} \cdots \sfg_{n-1} \big) = (g_{\omega_1})^{-n/2} \,
 \frac{q^n}{q-1} \ 
\sum_{k=0}^{n-1} \, t^{(n-1-k)\, \frac{N+1}2} \,
 \frac{\big(q^{-1};t\big)_{k+1}}{ (t;t)_k}\, {\zeta}_{n-1-k} (q,t)
 \ [N ]_{t^{k+1}} \nonumber \\ && \label{eq:qtTraceformula}
\eea
where
\bea
\zeta_0(q,t)&:=& 1  \ , \nonumber
\\[4pt]
 \zeta_m(q,t)&:=&\sum_{\lambda\in\Lambda_+^m}\, (-1)^{\ell(\lambda)}\, L_\lambda \
\prod_{i=1}^{\ell(\lambda)}\,
\frac{\big(q^{-1};t\big)_{\lambda_i}}{
  (t;t)_{\lambda_i}} \qquad \mbox{for} \quad m>0 \ .
\label{eq:omegaMdef}\eea
It is straightforward to show that this expression reduces to
(\ref{eq:qtTraceg1}) for $n=2$, while for $n=3$ it reads as
\bea
\Tr_{R_{\omega_1}^{\otimes 3}} \big(\Phi_3 \, U \, \sfg_{1} \,
\sfg_{2} \big) &=& (g_{\omega_1})^{-3/2} \,\bigg(t^{N+1}\,
\frac{q-1}{t-1}\, \Big(\, \frac{q-1}{t-1}+\frac{t-q}{t^2-1}\, \Big)\, [ N]_t +t^{\frac{N+1}2}\,
\frac{(q-1)(t-q)}{(t-1)^2}\, [N]_{t^2} \nonumber\\ &&
\qquad \qquad \qquad +\, \frac{(t-q)\, \big(t^2-q\big)}{(t-1)\, \big(t^2-1\big)}\,
[ N]_{t^3} \bigg) \ .
\eea

Let us look at the unrefined limit $q=t$ of the expression
(\ref{eq:qtTraceformula}). In this case $(q^{-1};q)_{k+1}=0$ for $k>0$
from the definition (\ref{eq:Pochhammer}), so only the $k=0$ term in
(\ref{eq:qtTraceformula}) is non-zero and the sum
over partitions in (\ref{eq:omegaMdef}) receives a non-vanishing
contribution from only the maximal partition $\lambda=(1^m)$
with $\ell(\lambda)=m$ parts, and ${L}_{(1,\ldots,1)}=1$, so that
\be
\zeta_m(q,q)  =  \prod_{i=1}^{m} \, \frac{1}{q} = q^{-m} \qquad
\mbox{for} \quad m \geq 0 \ .
\ee
Then the $q=t$ limit of the $(q,t)$-trace formula
(\ref{eq:qtTraceformula}) becomes
\be
\Tr_{R_{\omega_1}^{\otimes n}} \big(\Phi_n \, U \, \sfg_{1} \,
\sfg_{2} \cdots \sfg_{n-1} \big)\Big|_{q=t}  \, = \, q^{(n-1)\, \frac{N+1}{2}} \, \left[ N \right]_q
\ee
which coincides with the unrefined quantum trace formula
of~\cite[eq.~(B.6)]{deHaro2006}. 
The ensuing simplicity of the unrefined limit as compared to the general
case (\ref{eq:qtTraceformula}) is explained in terms of the combinatorics of
symmetric functions in Appendix~\ref{app:qt-traces}.

We can now use (\ref{eq:qtTraceformula}) to evaluate
$\Tr_{R_{\omega_1}^{\otimes n}} \big(\Phi_n \, U \, \sfh(m_T)
\big)$. If the conjugacy class $T$ is composed of permutations which have $\mu_i(T)$ cycles of length $i$, then $n=\sum_i\, i\, \mu_i(T)$ and we get
\bea
&& \Tr_{R_{\omega_1}^{\otimes n}} \big(\Phi_n \, U \, \sfh(m_T) \big) 
 \\ && \qquad \ = \ (g_{\omega_1})^{-n/2} \ \frac{q^{n} \, t^{n\, \frac{N+1}{2}}}{(q-1)^{\sum_i \mu_i(T)}} \,
\prod_{i=1}^{n} \, \Big( \, \sum_{k=0}^{i-1} \, t^{-(k+1)\, \frac{N+1}2} \,
 \frac{\big(q^{-1};t \big)_{k+1}}{(t;t)_k}\, {\zeta}_{i-1-k} (q,t)
 \ [N]_{t^{k+1}}\, \Big)^{\mu_i(T)} \ .
 \nonumber 
\eea
Let us rewrite this formula in terms of the partitions $\mu=\mu(T)$ which parameterize the conjugacy classes $T=T_\mu$ as
\bea
\Tr_{R_{\omega_1}^{\otimes n}} \big(\Phi_n \, U \, \sfh(m_{T_\mu}) \big) 
=(g_{\omega_1})^{- n/2} \ \frac{q^{n} \, t^{n\, \frac{N+1}{2}}}{(q-1)^{\ell(\mu)}} \,
\prod_{i=1}^{\ell(\mu)} \, \Big( \,  \sum_{k=1}^{\mu_i} \, t^{-k\, \frac{N+1}2} \,
 \frac{\big(q^{-1};t \big)_{k}}{(t;t)_{k-1}}\, {\zeta}_{\mu_i-k} (q,t)
 \ [N]_{t^k}\, \Big)
\label{eq:minwordfinal}\eea
where $\ell(\mu)=\sum_i\,\mu_i(T)$ is the length of the partition $\mu$.

\subsection{Hecke character expansion}

We can finally substitute the formula \eqref{eq:minwordfinal} into (\ref{eq:qtdim_minW}) to get the main result of this section.
\begin{proposition}\label{eq:qtdim_big1}
 \ The refined quantum dimensions can be expressed as
\bea \nonumber
\frac{\dim_{q,t}(R_\lambda)}{\sqrt{g_\lambda}} &=& \frac{q^{-\frac{n\,(n-5)}4}\, t^{n\, \frac{N+1}{2}}}{(g_{\omega_1})^{n/2} \, [n]_q!} \,
\frac{d_{\lambda_\beta+a\, (1^N)}(q)}{d_{\lambda_\beta+a\, (1^N)}(1)}\, \sum_{\mu\in\Lambda_+^n}\,
\frac{q^{- \ell^*(\mu)}}{(q-1)^{\ell(\mu)}} \,
\chi_{r_{\lambda_\beta+a\, (1^N)}}(C_\mu) \\ && \qquad \qquad \qquad \qquad \qquad \qquad \times \
\prod_{i=1}^{\ell(\mu)} \, \Big( \, \sum_{k=1}^{\mu_i} \, t^{-k\, \frac{N+1}2} \,
 \frac{\big(q^{-1};t\big)_{k}}{(t;t)_{k-1}}\, {\zeta}_{\mu_i-k} (q,t)
 \ [N]_{t^k} \, \Big)
 \nonumber
\eea
for $\lambda\in\Lambda_+^{n-a\,N}$, where $\ell^*(\mu)=\sum_i\,
(i-1)\, \mu_i=n-\ell(\mu)$ is the colength of the partition $\mu$ (the
complement to its length) which coincides with the length of the
permutation (the minimal number of generators) that belongs to the
conjugacy class labelled by $\mu$, and the central Hecke algebra element
$C_\mu:=C_{T_\mu}$ is defined by \eqref{eq:CTdef}. The
coefficients $\zeta_m(q,t)$ are defined in \eqref{eq:omegaMdef} and \eqref{eq:Llambda}.
\end{proposition}
It easy to see that this refined quantum dimension formula reduces at $\beta=1$ to the quantum dimension formula from~\cite[eq.~(2.36)]{deHaro2006}.

\section{Chiral expansion of the partition function\label{sec:chiralpartfn}}

To explore the relations between the refinement of $q$-deformed Yang-Mills theory on
$\Sigma_h$ and a refined topological string theory, 
we consider the {topological} limit of the gauge theory which is the
limit of degree
$p=0$. In this section we will study the partition function which from (\ref{ZqtQpR})
is given by
\be
Z_h(q,t;0) = \sum_{n=0}^\infty \ \sum_{\lambda\in\Lambda^n_+} \,
\Big(\, \frac{\dim_{q,t}(R_\lambda)}{\sqrt{g_\lambda}}\, \Big)^{2-2h}
\ ,
\ee
analogously to \cite{deHaro2006}. The chiral expansion is the
asymptotic large $N$ expansion
defined by dropping the constraint $\ell(\lambda)\leq N$ on the
lengths of the partitions $\lambda$; our main results are summarised
in Propositions~\ref{eq:EulerExp} and~\ref{prop:parEuler}.

\subsection{Generalised quantum $\Omega$-factors}

Let us begin by rewriting the refined quantum dimension
from Proposition~\ref{eq:qtdim_big1} in a simpler condensed form. We define the element
\bea
\Omega_n(q,t) = \frac{t^{n\, \frac{N+1}{2}}}{\big([N]_t\big)^n} \,
\sum_{\mu\in\Lambda_+^n}\, \Big(\, \frac{q}{q-1} \, \Big)^{\ell(\mu)} \ \prod_{i=1}^{\ell(\mu)} \, \Big( \,  \sum_{k=1}^{\mu_i} \, t^{-k\, \frac{N+1}2} \,
 \frac{\big(q^{-1};t\big)_{k}}{(t;t)_{k-1}}\, {\zeta}_{\mu_i-k} (q,t)
 \ [N]_{t^k}\, \Big) \ 
  C_\mu \ .
\label{eq:Omegafactors}\eea
This is a sum of central elements $C_\mu$ of the Hecke algebra
$\sfH_q(\frS_n)$, so $\Omega_n(q,t)$ is also central. The normalization is chosen so
that the identity is the leading term at large $N$. For this, we note
that, under the assumptions $q,t \in(0,1)$, the largest terms of $\Omega_n(q,t)$ in the large $N$ expansion come from rectangular partitions of the form $\mu=(m,\ldots,m)$ which give the leading term
\be  \label{eq:leadinOmega}
\Big(\, \frac{q\, (q^{-1};t)_m}{(q-1)\, (t;t)_{m-1}}\, \Big)^{\ell(\mu)} \ .
\ee
For $\beta \geq 1$ and $k \in \IZ_{>0}$ we have
\be
\Big|\, \frac{1-q^{-1}\, t^k}{1-t^k} \, \Big|= \Big|\, \frac{1-t^{k-\frac{1}{\beta}}}{1-t^k} \, \Big|  < 1 \ .
\ee
This implies that the absolute value of (\ref{eq:leadinOmega}) is less
than $1$, unless $\mu=(1,\ldots,1)$ in which case it is equal to $1$. Hence the maximal partition $(1,\ldots,1)$ is the leading term which corresponds to the identity permutation, and we can write
\be \label{eq:omegaprimed}
\Omega_n(q,t) \, = \, 1 \, + \, \Omega_n'(q,t) \ ,
\ee
where $\Omega_n'(q,t)$ has the same form as $\Omega_n(q,t)$ except
that the sum runs over all non-maximal partitions $\mu$ of $n$.

The $q=t$ limit of \eqref{eq:Omegafactors} 
coincides with the unrefined element 
\be
\Omega_n(q,q)=\sum_{\mu\in\Lambda_+^n}\, q^{\frac{N-1}2\, \ell(\mu)} \
\big([N]_q\big)^{-\ell^*(\mu)} \ C_\mu
\label{eq:Omegaqq}\ee
from~\cite[eq.~(2.46)]{deHaro2006}. As we discuss further below, the
power $\big([N]_q\big)^{-\ell^*(\mu)}$ appearing here suggests an
interpretation of $\Omega_n(q,q)$ in terms of branch points on
$\Sigma_h$ in a topological string theory of worldsheet branched covers
of the target Riemann surface $\Sigma_h$, with string coupling $g_{\rm
  str}=\frac1{[N]_q}$.

With this new notation we can write the result of Proposition~\ref{eq:qtdim_big1} as
\be \label{eq:qtdim_small}
 \frac{\dim_{q,t}(R_\lambda)}{\sqrt{g_\lambda}} = \frac{q^{-\frac{n\,(n-1)}4}}{ [n]_q!} \,
\frac{d_{\lambda_\beta+a\,(1^N)}(q)}{d_{\lambda_\beta+a\, (1^N)}(1)}\, \Big(\, \frac{[ N]_{t}}{\sqrt{g_{\omega_1}}}\, \Big)^n \, \chi_{r_{\lambda_\beta+a\, (1^N)}}\big(\Omega_n(q,t)\big) \ .
\ee
This formula is very similar to the unrefined one from~\cite[eq.~(2.45)]{deHaro2006}, except that in our case the expansion parameter is 
\be
\frac{[N]_t}{\sqrt{g_{\omega_1}}} =\bigg(\,
\frac{\big[ N\big]_t\,\big[\beta\, (N-1)+1\big]_q}{g_{\emptyset} } \, \bigg)^{1/2}
\ee
and $g_\emptyset$ is of order $1$ in the large $N$ limit. This
expansion parameter respects the $\Omega$-background symmetry
$(\epsilon_1,\epsilon_2)\mapsto (-\epsilon_2,-\epsilon_1)$ described in
\S\ref{sec:geometricint}; however, fixing $p=0$ breaks this
symmetry of the topological partition function in the ensuing large
$N$ expansion.

\subsection{Chiral series}

We next use the following fact to evaluate the powers of the refined quantum dimensions: If $C$ is any central element of the Hecke 
algebra $\sfH_q(\frS_n)$ and $\sigma\in\frS_n$, then by~\cite[eq.~(2.44)]{deHaro2006} one has
\be \label{eq:central}
\chi_{r_{\lambda}}(C) \, \chi_{r_{\lambda}} \big(\sfh(\sigma)\big) \, = \, d_{\lambda} (1)\, \chi_{r_{\lambda}} \big(C \, \sfh(\sigma)\big) \ .
\ee
This implies
\bea
Z_h(q,t;0) &=& \sum_{n=a\,N}^\infty \
\sum_{\lambda\in\Lambda^{n-a\,N}_+} \ \Big(\,
\frac{q^{-\frac{n\,(n-1)}4}\, d_{\lambda_\beta+a\,(1^N)}(q)}{ [n]_q!}
\, \Big)^{2-2h} \, \frac{1}{d_{\lambda_\beta+a\,(1^N)}(1)} \, \Big(\,
\frac{[ N]_{t}}{\sqrt{g_{\omega_1}}} \,\Big)^{n\,(2-2h)} \nonumber \\ &&
\qquad \qquad  \qquad \qquad \qquad  \qquad \qquad \times \ \chi_{r_{\lambda_\beta+a\,(1^N)}}\big(\Omega_n(q,t)^{2-2h}\big) \ .
\eea
Because of our normalization (\ref{eq:omegaprimed}), the element $\Omega_n(q,t)$ is always formally invertible in the large~$N$ expansion.

By~\cite[Appendix~A.1]{deHaro2006} we have
\be \label{eq:PI}
\Big(\, \frac{[n]_q!}{ q^{-\frac{n\,(n-1)}4}\,
  d_{\lambda_\beta+a\,(1^N) }(q)} \, \Big)^2 \, = \,
\frac{1}{d_{\lambda_\beta +a\,(1^N)}(1)} \ \sum_{\sigma,\tau \in
  \frS_n} \, q^{-\ell(\sigma)-\ell(\tau )} \,
\chi_{r_{\lambda_\beta+a\,(1^N)}} \big(\sfh(\sigma)\,\sfh( \tau
)\,\sfh(\sigma^{-1})\,\sfh( \tau^{-1}) \big)
\ee
which yields
\bea
&&Z_h(q,t;0) \ = \ \sum_{n=a\,N}^\infty \
\sum_{\lambda\in\Lambda^{n-a\, N}_+} \
\frac{1}{d_{\lambda_\beta+a\,(1^N)}(1)}\, \Big(\,
\frac{q^{-\frac{n\,(n-1)}4}\, d_{\lambda_\beta +a\,(1^N)}(q)}{ [n]_q!}
\, \Big)^2\, \Big(\, \frac{[N]_t}{\sqrt{g_{\omega_1}}}
\,\Big)^{n\,(2-2h)} \nonumber \\ &&\qquad  
\times \ \sum_{\sigma_1,\tau_1,\dots ,\sigma_h,\tau_h\in \frS_n} \, q^{-\sum_i\,
  (\ell(\sigma_i)+\ell(\tau_i))} \,
\chi_{r_{\lambda_\beta+a\,(1^N)}}\Big(\Omega_n(q,t)^{2-2h} \ \prod_{i=1}^{h}\, \sfh(\sigma_i)\, \sfh(\tau_i)\, \sfh(\sigma_i^{-1})\, \sfh(\tau_i^{-1}) \Big)
\nonumber \\ &&
\eea
where we used (\ref{eq:central}) and the centrality property of~\cite[Appendix~A.2]{deHaro2006}. Following~\cite[Appendix~A.3]{deHaro2006} we introduce the element
\be
D_n \, = \, \frac{q^{-\frac{n\,(n-1)}4}}{[n]_q!} \, \sum_{k=0}^\infty
\, (-1)^k \ \sum_{\stackrel{\scriptstyle \sigma_1,\dots,\sigma_k\in
    \frS_n}{\scriptstyle \sigma_i\neq1}} \, q^{-\sum_i\, \ell (\sigma_i)} \ \prod_{j=1}^k\, \sfh(\sigma_j^{-1})\, \sfh(\sigma_j)
\label{eq:Dndef}\ee
in the completed Hecke algebra $\widehat{\sfH}_q(\frS_n)$. We can then express the $q$-dimension $d_{\lambda_\beta+a\,(1^N)}(q)$ in terms of the character of $D_n$ as
\be \label{eq:charD}
d_{\lambda_\beta+a\,(1^N)}(q) \, = \, \chi_{r_{\lambda_\beta+a\,(1^N)}} (D_n) \ .
\ee
Since $D_n$ is a series in central elements of $\sfH_q(\frS_n)$
by~\cite[Appendix~A.1]{deHaro2006}, it is central in
$\widehat{\sfH}_q(\frS_n)$ and using (\ref{eq:central}) we get
\bea \label{eq:PartBeforeDelta}
&& Z_h(q,t;0) \ = \ \sum_{n=a\, N}^\infty \, \frac{q^{-\frac{n\,(n-1)}2}}{ \big([n]_q!\big)^2} \, \Big(\, \frac{[N]_t}{\sqrt{g_{\omega_1}}} \, \Big)^{n\,(2-2h)} 
\ \sum_{\sigma_1,\tau_1,\dots,\sigma_h,\tau_h\in \frS_n} \, q^{-\sum_i\, (\ell(\sigma_i)+\ell(\tau_i))} \\ && 
\qquad \times \ \sum_{\lambda\in\Lambda_+^{n-a\, N}} \, d_{\lambda_\beta+a\,(1^N) }(q)\, \chi_{r_{\lambda_\beta+a\,(1^N)}}\Big(D_n\,\Omega_n(q,t)^{2-2h} \ \prod_{i=1}^{h}\, \sfh(\sigma_i)\, \sfh(\tau_i)\, \sfh(\sigma_i^{-1})\, \sfh(\tau_i^{-1}) \Big) \ .
\nonumber
\eea

We define a delta-function on Hecke algebras analogously to \cite{deHaro2006} by
\bea
\delta\big(\sfh(\sigma)\big)\,=\, \left\{ \begin{array}{l} 1 \qquad \text{if} \quad \sigma \, = \, 1 \ , \\[4pt]
0 \qquad \text{otherwise}  \ , \end{array} \right.
\eea 
and extended over $\sfH_q(\frS_n)$ by $\IC$-linearity. It can be expressed as the sum of characters of $\sfH_q(\frS_n)$ given by
\be\label{eq:deltaf}
 \delta\big(\sfh(\sigma)\big)=\frac{q^{-\frac{n\,(n-1)}4}}{[n]_q!} \ \sum_{\lambda\in \Lambda_+^n} \, d_\lambda (q) \, \chi_{r_\lambda} \big(\sfh(\sigma)\big) \ .
\ee
To write the partition function in terms of delta-functions as in the
unrefined case, we have to take the sum over all partitions of
$n$. There is a bijection between partitions $\alpha\in \Lambda_+^{n}$
such that $\alpha_i\geq(\beta-1)\, \rho_i+a=(\beta-1)\, (N-i)$ for $i=1,\dots, N$ and partitions in
$\Lambda_+^{n-a\, N}$. Thus we need to construct a step
function $\Theta_n(\beta)$ on partitions that cuts off the contributions
involving smaller partitions and allows us to sum over all
$\alpha\in\Lambda_+^{n}$; it is defined by the property
\bea
\chi_\alpha\big(\Theta_{n}(\beta)\big)\, =\, \left\{ \begin{array}{ll}
      d_\alpha(1) \quad & \mbox{if} \quad
      \alpha_i\geq (\beta-1)\,(N-i) \ , \\[4pt] 0 \quad 
                                                         & \mbox{otherwise} \ , \end{array}
    \right.
\eea
for $\alpha\in\Lambda_+^n$. The sum of quantum Young projectors (\ref{eq:Youngprojexpand}) given by
\be \label{eq:thetaDef}
\Theta_{n}(\beta) \, = \, \sum_{\substack{ \mu\in\Lambda_+^n \\ \mu_i\geq (\beta-1)\, (N-i)} } \, P_\mu
\ee
fulfills this criterion, 
and it is a central element of $\sfH_q(\frS_n)$ because the projectors are central.

We now insert $\big(d_{\lambda_\beta+a\,(1^N)}(1) \big)^{-1}\, \chi_{\lambda_\beta+a\,(1^N)}\big(\Theta_{n}(\beta)\big)=1$ in (\ref{eq:PartBeforeDelta}) and using (\ref{eq:central}) we get
\bea 
&& Z_h(q,t;0) \ = \ \sum_{n=a\, N}^\infty \, \frac{q^{-\frac{n\,(n-1)}2}}{ \big([n]_q!\big)^2} \, \Big(\, \frac{[N]_t}{\sqrt{g_{\omega_1}}} \, \Big)^{n\,(2-2h)} 
\ \sum_{\sigma_1,\tau_1,\dots,\sigma_h,\tau_h\in \frS_n} \, q^{-\sum_i\, (\ell(\sigma_i)+\ell(\tau_i))} \\ && 
\times \ \sum_{\lambda\in\Lambda_+^{n-a\, N}} \, d_{\lambda_\beta+a\,(1^N) }(q)\, \chi_{r_{\lambda_\beta+a\,(1^N)}}\Big(\Theta_n(\beta) \,D_n\, \Omega_n(q,t)^{2-2h} \ \prod_{i=1}^{h}\, \sfh(\sigma_i)\, \sfh(\tau_i)\, \sfh(\sigma_i^{-1})\, \sfh(\tau_i^{-1}) \Big) \ .
\nonumber
\eea
A partition $\alpha\in\Lambda_+^{n}$ satisfies
$\alpha_i\geq(\beta-1)\,(N-i)$ for $i=1,\dots,N$ if and only if it can be written as
$\alpha=\lambda_\beta+a\,(1^N)$ for some $\lambda\in\Lambda_+^{n-a\,
  N}$. The contributions involving
$\alpha_i<(\beta-1)\, (N-i)$ for some $i$ in the partition function
vanish because of the step function $\Theta_n(\beta)$. Hence we can shift
the summation range and sum over all $\alpha\in\Lambda_+^{n}$ to get
\bea 
Z_h(q,t;0) &=& \sum_{n=a\, N}^\infty \, \frac{q^{-\frac{n\,(n-1)}2}}{ \big([n]_q!\big)^2} \, \Big(\, \frac{[N]_t}{\sqrt{g_{\omega_1}}} \, \Big)^{n\,(2-2h)} 
\ \sum_{\sigma_1,\tau_1,\dots,\sigma_h,\tau_h\in \frS_n} \, q^{-\sum_i\,
  (\ell(\sigma_i)+\ell(\tau_i))} \\ && \qquad \quad
\times \ \sum_{\alpha\in\Lambda_+^{n}} \, d_{\alpha}(q)\, \chi_{r_{\alpha}}\Big(\Theta_n(\beta) \,D_n\, \Omega_n(q,t)^{2-2h} \ \prod_{i=1}^{h}\, \sfh(\sigma_i)\, \sfh(\tau_i)\, \sfh(\sigma_i^{-1})\, \sfh(\tau_i^{-1}) \Big) \ ,
\nonumber
\eea
and then using the expression for the delta-function from (\ref{eq:deltaf}) we arrive at
\bea \label{eq:expandedPartFu}
&& Z_h(q,t;0) \ = \ \sum_{n=\lceil a\, N\rceil}^\infty \,
\Big( \,
\frac{[N]_t}{\sqrt{g_{\omega_1}}} \, \Big)^{n\,(2-2h)} \ \frac{q^{-\frac{n\,(n-1)}4}}{ [n]_q!}  \\ && \qquad \times \ 
\sum_{\sigma_1,\tau_1,\dots,\sigma_h,\tau_h\in \frS_n} \, \delta\Big(\Theta_n(\beta) \,D_n\, \Omega_n(q,t)^{2-2h} \ \prod_{i=1}^{h}\, q^{-\ell(\sigma_i)- \ell(\tau_i)} \,  \sfh(\sigma_i)\, \sfh(\tau_i)\, \sfh(\sigma_i^{-1})\, \sfh(\tau_i^{-1}) \Big) \ .
\nonumber
\eea
Now we can analytically continue $\beta$ away from integer values,
because this expansion only depends on $\beta$ in the $q$-numbers and
in $a$, and we can choose a larger value for $a$; the smallest choice
is $a_0$ such that $a_0\, N=\lceil a\, N\rceil$, and $\lceil a\, N\rceil$ also vanishes in the unrefined limit. The restriction on the sum in the definition of the step function $\Theta_n(\beta)$ from (\ref{eq:thetaDef}) can also be continued in a straightforward way.
This expansion is the refined version of \cite[eq.~(2.56)]{deHaro2006}; it is a refined quantum deformation of the large $N$ chiral Gross-Taylor expansion. The Hecke element $\Theta_n(\beta)$ does not have any unrefined analog, as it reduces to $\Theta_n(\beta=1)=\sum_{\mu\in\Lambda_+^n}\, P_\mu=1$ by (\ref{eq:Youngprojexpand}) together with the expression for the delta-function from (\ref{eq:deltaf}) and we recover the unrefined expansion.

Finally, using (\ref{eq:omegaprimed}) we can expand the
$\Omega$-factors $\Omega_n(q,t)^{2-2h}$ in the completion $\widehat{\sfH}_q(\frS_n)$ via the power series
\be
\Omega_n(q,t)^{2-2h} \, = \, \sum_{L=0}^{\infty} \, d(2-2h,L) \ \Omega_n'(q,t)^L \ ,
\ee
where 
\be
d(m,L) \, = \, \frac{\Gamma(m+1)}{\Gamma(L+1)\,\Gamma(m-L+1)} \ .
\ee
As explained in \cite[\S7.1.2]{Cordes1995}, the binomial coefficient $d(2-2h,L)$ is the Euler characteristic $\chi(\Sigma_{h,L})$ of the configuration space
of $L$ points on the Riemann surface $\Sigma_h$, i.e. the $L$-tuples of distinct points on $\Sigma_h$ modulo the natural action of the permutation group $\frS_L$.
In this way we arrive at
\begin{proposition}\label{eq:EulerExp}
 \ The chiral series for the partition function of topological
$(q,t)$-deformed Yang-Mills theory on $\Sigma_h$ is given by
\bea \nonumber
&& Z_h(q,t;0) \ = \ \sum_{n=\lceil a\, N\rceil}^\infty \ \sum_{L=0}^{\infty}\,\big( g_{\omega_1} \big)^{n\,(h-1)} \ \big( [N]_t \big)^{n\,(2-2h-L)} \ \frac{\chi(\Sigma_{h,L})}{ [n]_q!} \ q^{-\frac{n\,(n-1)}4}\, t^{n\, L\, \frac{N+1}{2}} \\ \nonumber  &&  \qquad
\times \ \sum_{\sigma_1,\tau_1,\dots,\sigma_h,\tau_h\in \frS_n} \ \prod_{l=1}^{L} \
\sum_{\stackrel{\scriptstyle \mu^{l}\in\Lambda_+^n}{\scriptstyle \mu^l\neq(1^n)}} \, \Big(\, \frac{q}{q-1}\,\Big)^{\ell(\mu^{l})}
\ \prod_{i=1}^{\ell(\mu^l)} \, \Big( \,  \sum_{k=1}^{\mu_i^l} \, t^{-k\, \frac{N+1}2} \,
 \frac{\big(q^{-1};t\big)_{k}}{ (t;t)_{k-1}}\, {\zeta}_{\mu_i^l-k} (q,t)
 \ [ N]_{t^k}\, \Big) \\ && \qquad \qquad \qquad \qquad \qquad 
\times \ \delta\Big(\Theta_n(\beta) \,D_n\,
C_{\mu^{1}} \cdots C_{\mu^{L}} \ \prod_{j=1}^{h}\,
q^{-\ell(\sigma_j)-\ell(\tau_j)} \, \sfh(\sigma_j)\, \sfh(\tau_j)\, \sfh(\sigma_j^{-1})\, \sfh(\tau_j^{-1}) \Big) \ .
\nonumber
\eea
Here the central Hecke algebra elements $\Theta_n(\beta)$ and $D_n$ are defined in
\eqref{eq:thetaDef} and \eqref{eq:Dndef}.
\end{proposition}
This is a refined quantum deformation of the symmetric group
enumeration of covering maps of the Riemann surface $\Sigma_h$,
analogously to the description in terms of quantum spectral curves
discussed in \S\ref{sec:betaensemble}. In particular,
following~\cite{deHaro2006} it is tempting to suppose that this
expansion is captured by a balanced topological string
theory~\cite{Dijkgraaf1996} with target space the M-theory
compactification described in \S\ref{sec:geometricint}, which would
naturally compute Euler characters of certain moduli spaces of curves
in this background. We elaborate further on these points below.

In the unrefined limit $q=t$, the asymptotic expansion of Proposition~\ref{eq:EulerExp} becomes
\bea 
\nonumber
&& Z_h(q,q;0) \ = \ \sum_{n,L=0}^\infty \, \frac{\chi(\Sigma_{h,L})}{ [n]_q!} \  \sum_{\stackrel{\scriptstyle \mu^1,\dots,\mu^{L}\in\Lambda_+^n}{\scriptstyle \mu^l\neq(1^n)}} \,\big( [N]_q \big)^{n\,(2-2h-L)+\sum_{l} \ell(\mu^l)} \, q^{-\frac{n\,(n-3)}4+n\, L\, \frac{N+1}{2}-\frac{N-1}2\, \sum_l \ell(\mu^l)} \\ &&  \qquad
\times \ \sum_{\sigma_1,\tau_1,\dots,\sigma_h,\tau_h\in \frS_n} \,  \delta\Big(D_n\,
C_{\mu^{1}} \cdots C_{\mu^{L}} \ \prod_{j=1}^{h}\,
q^{-\ell(\sigma_j)-\ell(\tau_j)} \, \sfh(\sigma_j)\, \sfh(\tau_j)\,
\sfh(\sigma_j^{-1})\, \sfh(\tau_j^{-1}) \Big) \ ,
\label{eq:qEulerExp}\eea
independently of the parts of the partitions $\mu^1,\dots,\mu^L$. When $q$ is a power of a prime, as explained in~\cite{deHaro2006} it is
tempting to interpret this expansion in
terms of the enumeration of algebraic curves over the (finite) field
$\IF_q$, rather than $\IC$. As further evidence, we note that
$q$-deformation in this regard can be interpreted as replacing the
symmetric group $\frS_n$ with the general linear group $GL_n(\IF_q)$,
which sends a transposition $\sigma_i$ to a reflection in
$GL_n(\IF_q)$, i.e. the fixed space $(\IF_q^n)^{\sigma_i}$ is a
hyperplane in $\IF_q^n\cong\IF_{q^n}$ (as $\IF_q$-vector spaces). The
irreducible representations $u_\lambda$ of $GL_n(\IF_q)$ are also parameterized by
partitions $\lambda$ of $n$ and have dimensions $\dim(u_\lambda)=
(q;q)_n\, \dim_{q}(R_\lambda)$~\cite{Green1955}, while the character of a semisimple
reflection $\sigma$ in $GL_n(\IF_q)$ (characterised as having
$\det(\sigma)\in \IF_q^*\setminus\{1\}$) is given by
\bea
\chi_{u_\lambda}(\sigma) = \frac{\det(\sigma)}{[n]_q} \ \sum_{(i,j)\in
  Y_\lambda}\, q^{j-i} \ .
\eea
It would be interesting to understand if the corresponding refinements
can be likewise regarded in terms of generalised characters.

\subsection{$\beta$-deformed Hurwitz theory\label{sec:betaHurwitz}}

To understand better the geometrical effect of refinement as it occurs
in the expansion of Proposition~\ref{eq:EulerExp}, let us consider the classical limit $q\to1$ with $\beta$ fixed. In this limit the Macdonald polynomials reduce to the Jack polynomials which are ordinary generalized characters of irreducible $U(N)$ representations~\cite{Etingof1993}, and the Hecke algebra reduces to the ordinary group algebra of the symmetric group $\IC[\frS_n]$. It is straightforward to show that the $\Omega$-factors reduce to
\be
\lim_{q\rightarrow 1} \, \Omega_n(q,t) \big|_{t=q^\beta}\,=\, \sum_{\mu\in \Lambda_+^n} \, \frac{{\mit\Delta}_\mu(\beta)}{N^{\ell^*(\mu)}} \ C_\mu  \ ,
\label{eq:Omegabeta}\ee
where $C_\mu=\sum_{\sigma\in T_\mu}\, \sigma$ and we have defined 
\be
{\mit\Delta}_\mu(\beta) \, = \, \prod_{i=1}^{\ell(\mu)} \, \Big(\, \sum_{k=1}^{\mu_i}\, \gamma_k(\beta) \ \sum_{\lambda\in\Lambda_+^{\mu_i-k}} \, \beta^{-\ell(\lambda)}\, \frac{\ell(\lambda)!}{z_\lambda} \ \prod_{j=1}^{\ell(\lambda)}\, \gamma_{\lambda_j}(\beta) \, \Big) \ ,
\label{eq:mitDelta}\ee
with
\bea
\gamma_1(\beta)&=&1 \ , \nonumber \\[4pt]
\gamma_k(\beta) &= & \prod_{l=1}^{k-1}\, \frac{\beta\, l - 1}{\beta\, l} \ = \ \frac{\Gamma(k-\frac 1 \beta)}{\Gamma(1-\frac 1 \beta)\,\Gamma(k)} \qquad \mbox{for} \quad k>1 \ .
\label{eq:gammakbeta}\eea
The integer
\be
z_\lambda=\prod_{i=1}^{|\lambda|}\, i^{m_i(\lambda)}\, m_i(\lambda)!
\ee
is the order of the stabilizer, under conjugation, of any element of the conjugacy class $T_\lambda$.
In the unrefined limit, ${\mit\Delta}_\mu(\beta)\to1$ as $\beta\rightarrow 1$ and \eqref{eq:Omegabeta} coincides with the unrefined $\Omega$-factor from \cite[eq.~(6.5)]{Cordes1995}, in which case the weights of the sum depend only on the colengths $\ell^*(\mu)=n-\ell(\mu)$ of the partitions $\mu\in\Lambda_+^n$ (the same is true of the unrefined $q$-deformed $\Omega$-factors \eqref{eq:Omegaqq}). In marked contrast, for $\beta\neq1$ the weights depend explicitly on the parts of the partition $\mu$ through the combinatorial coefficients ${\mit\Delta}_\mu(\beta)$.

The expansion into Euler characters given in Proposition~\ref{eq:EulerExp} reduces to
\bea
\widetilde{Z}_h(\beta) &:= & \lim_{q\to1}\, Z_h(q,t;0)\big|_{t=q^\beta} \nonumber \\[4pt]
&=& \sum_{n=\lceil a\,N \rceil}^{\infty} \, \tilde{g}(\beta)^{n\,(h-1)} \ \sum_{L=0}^\infty \, \frac{\chi(\Sigma_{h,L})}{n!} \ \prod_{l=1}^L \ \sum_{\stackrel{\scriptstyle \mu^{l}\in\Lambda_+^n}{\scriptstyle \mu^l\neq(1^n)}} \, {\mit\Delta}_{\mu^l}(\beta) \ N^{n\,(2-2h)-\sum_{l=1}^L\, \ell^*(\mu^l) } \nonumber \\
&& \qquad \qquad \times \ \sum_{\lambda\in\Lambda_+^n} \,
\omega_\lambda (\beta) \
\sum_{\sigma_1,\tau_1,\dots,\sigma_h,\tau_h\in \frS_n} \
\delta\Big(C_\lambda \, C_{\mu^1} \cdots C_{\mu^L} \ \prod_{i=1}^h \,
[\sigma_i , \tau_i] \Big) \ ,
\eea
where 
$[\sigma,\tau]=\sigma\, \tau\, \sigma^{-1}\, \tau^{-1}$ denotes the
\emph{group} commutator and
\be 
\tilde g(\beta):= \frac N{\beta\, (N-1)+1} \ \prod_{m=0}^{\beta-1} \ \prod_{1\leq i < j \leq N} \, \frac{\beta\, (j-i)+m}{\beta\, (j-i)-m} \ .
\label{eq:tildegbeta}\ee
We used (\ref{eq:QmYoungP_C}) and (\ref{eq:dimqtRfund}), and defined a new deformation weight
\be
\omega_\lambda (\beta) \, = \, \frac{1}{n!}\, \sum_{\substack{ \mu\in\Lambda_+^n \\ \mu_i\geq (\beta-1)\,(N-i)} } \, d_\mu \, \chi_{r_\mu}(m_{T_\lambda}) \ ,
\label{eq:omegalambda}\ee
which reduces to $\delta(C_\lambda)=\delta_{\lambda , (1^n)}$ in the
$\beta=1$ limit. Rewriting this expansion entirely as sums over
elements of the symmetric group $\frS_n$ reveals that it is a 
$\beta$-deformation of the ordinary chiral Gross-Taylor series
in~\cite[eq.~(7.3)]{Cordes1995}, containing an extra class sum, and
with extra deformation weights ${\mit\Delta}_\mu(\beta)$ and
$\omega_\lambda(\beta)$. In this expansion the weights depend
explicitly on the parts of the partitions and not only on their colengths, although they are decoupled according to distinct partitions. 

To make contact with Hurwitz theory, we shall collect terms with a fixed value of the integer
\be 
d=\sum_{l=1}^L\, \ell^*(\mu^l)
\label{eq:degree}\ee
and set
\be 
2g-2 = n\, (2h-2)+d \ .
\label{eq:RHformula}\ee
For each $n\geq\lceil a\,N \rceil$, let us first consider the contribution from the maximal partition $\lambda=(1^n)$, for which $C_{(1^n)}=1$ and
\be
\omega_{(1^n)} (\beta) \, = \, \frac{1}{n!}\, \sum_{\substack{ \mu\in\Lambda_+^n \\ \mu_i\geq (\beta-1)\, (N-i)} } \, (d_\mu)^2 \ .
\label{eq:omega1n}\ee
Let $H_{h,n}(\mu^1,\dots,\mu^L)$ be the number, weighted by $\frac1{n!}$,
of $(2h{+}L)$-tuples consisting of permutations
$\sigma_i,\tau_i\in\frS_n$, $i=1,\dots,h$ and central elements
$C_{\mu^l}$ of $\frS_n$ sitting in fixed conjugacy classes
$T_{\mu^l}$, $l=1,\dots,L$ such that $C_{\mu^1} \cdots C_{\mu^L} \,
\prod_{i=1}^h \, [\sigma_i , \tau_i]= 1$; this quantity is called a (combinatorial) Hurwitz number, and it has an explicit combinatorial expression given by the Frobenius-Schur formula~\cite[Appendix~A]{Lando2004}
\be 
H_{h,n}\big(\mu^1,\dots,\mu^L\big) = (n!)^{2h}\,
\sum_{\lambda\in\Lambda_+^n}\, \frac1{(d_\lambda)^{L+2h-2}} \
\prod_{l=1}^L\, \frac{\chi_{r_\lambda}\big(m_{T_{\mu^l}} \big)}{z_{\mu^l}} \ .
\ee
Geometrically, $H_{h,n}(\mu^1,\dots,\mu^L)$ counts the number of
$n$-sheeted branched covers $f:\Sigma_g\to\Sigma_h$ with $L$ fixed
branch points $z^1,\dots,z^L\in\Sigma_h$ whose ramification profiles
are specified by partitions $\mu^1,\dots,\mu^L\in\Lambda_+^n$ such
that $\mu^l= \big(\mu^l_1,\dots,\mu^l_{\ell(\mu^l)} \big)$ are the ramification indices of the preimages $f^{-1}(z^l)$ (encoding how many sheets of the cover $f:\Sigma_g\to\Sigma_h$ come together above the point $z^l\in\Sigma_h$, see e.g.~\cite[\S5]{Cordes1995}), weighted by $\frac1{|{\rm Aut}(f)|}$ where ${\rm Aut}(f)$ is the automorphism group of the covering map consisting of automorphisms $\alpha:\Sigma_g\to\Sigma_g$ such that $f\circ\alpha=f$; the degree $d$ of such a holomorphic map over $\Sigma_h\setminus\{z^1,\dots,z^L\}$ is given by (\ref{eq:degree}), while the genus $g$ of the covering surface is given by the Riemann-Hurwitz formula \eqref{eq:RHformula}. Note that $H_{h,n}(\mu^1,\dots,\mu^L)$ is independent of the branch point positions $z^1,\dots,z^L\in\Sigma_h$ and also of the choice of (fixed) complex structure on $\Sigma_h$. Incorporating the contributions from all partitions $\lambda\in\Lambda_+^n$, we then have
\bea
\widetilde{Z}_h(\beta) &=& \sum_{n=\lceil a\,N \rceil}^{\infty} \,
\tilde{g}(\beta)^{n\,(h-1)} \ \sum_{\lambda\in\Lambda_+^n} \,
\omega_\lambda (\beta) \ \sum_{d=0}^\infty\, \Big(\, \frac1N\,
\Big)^{2g-2} \ \sum_{L=0}^d \, \chi(\Sigma_{h,L}) \nonumber \\ &&
\qquad \qquad \times \ \sum_{\stackrel{\scriptstyle \mu^{1},\dots,\mu^L \in\Lambda_+^n}{\scriptstyle \mu^l\neq(1^n)\,,\, \sum_{l=1}^L \ell^*(\mu^l)=d}} \, {\mit\Delta}_{\mu^1}(\beta)\cdots {\mit\Delta}_{\mu^L}(\beta) \ H_{h,n}\big(\lambda, \mu^1,\dots,\mu^L\big) \ .
\label{eq:ZbetaHurwitz}\eea
From this expression we can infer at least four novel aspects of the closed
string expansion of the $\beta$-deformation of two-dimensional
Yang-Mills theory, interpreted from the geometric point of view of
Hurwitz theory:

\underline{\bf 1.} \ Branched covers of index $n<\lceil a\,N \rceil$ do not
contribute to the string expansion. This feature has important
ramifications for the planar limit of the gauge theory which we
discuss below.

\underline{\bf 2.} \ The refinement introduces an additional weighting by the
quantity \eqref{eq:tildegbeta} such that the expansion parameter does
not simply distinguish the genera of the covering worldsheets $\Sigma_g$. Below we shall replace this weight with its leading term $\tilde g(\beta)=\frac1\beta$ in the large $N$ limit.

\underline{\bf 3.} \ The string expansion \eqref{eq:ZbetaHurwitz} generically
involves deformations of the enumeration of branched covers
$f:\Sigma_g\to\Sigma_h$ in terms of Hurwitz numbers $H_{h,n}(\mu^1,\dots,\mu^L)$ which include an additional marked point with holonomy in the representation \eqref{eq:defW} of $U(N)$; the inclusion of such marked points is the earmark of refinement and is captured by the intertwining operators defining the generalised characters~\cite{Iqbal2011,Szabo2013}. Accordingly, the Hurwitz numbers $H_{h,n}\big(\lambda, \mu^1,\dots,\mu^L\big)$ account for additional branching over this marked point with ramification profile $\lambda\in\Lambda_+^n$. Due to the deformation weights $\omega_\lambda (\beta)$, for $\beta\neq1$ their contributions are strongly suppressed in the large $N$ limit.

\underline{\bf 4.} \ The expansion \eqref{eq:ZbetaHurwitz} involves
\emph{weighted} sums of Hurwitz numbers, with deformation weights
${\mit\Delta}_{\mu^l}(\beta)$ and $\omega_\lambda (\beta)$ depending explicitly
on the parts of the partitions $\mu^l$ and $\lambda$ which label the winding numbers of closed strings around the branch points in the target space $\Sigma_h$. This
deformation obstructs a rewriting of the partition function as a
generating function of characters of Hurwitz  spaces of
holomorphic maps $f:\Sigma_g\to\Sigma_h$, as occurs in the unrefined
case~\cite{Cordes1995}, and as such an interpretation as a balanced topological
string theory~\cite{Dijkgraaf1996} with string coupling $g_{\rm str}= \frac1N$ and two-dimensional
target space $\Sigma_h$ is not immediately evident. In fact, this
weighting suggests that the string expansion involves contributions
from \emph{marked} covers $f^{\rm m}:\Sigma_g\to\Sigma_h$; a marking
of a branched cover $f:\Sigma_g\to\Sigma_h$ consists of a marking of
each of its branch points $z^l$ for $l=1,\dots,L$, i.e. a choice of
labelling $\big\{w^l_1,\dots,w^l_{\ell(\mu^l)} \big\}=f^{-1}(z^l)$ such that $\mu^l_i$ is the ramification index at $w^l_i$. An automorphism $\alpha:\Sigma_g\to\Sigma_g$ of a marked cover preserves the labels $w^l_i$, and we denote the corresponding marked cover counts by $H_{h,n}^{\rm m}(\mu^1,\dots,\mu^L)$. The action of the automorphism group ${\rm Aut}(f)$ on the labels of $f^{\rm m}$ gives a group homomorphism
\be 
{\rm Aut}(f) \ \longrightarrow \ \prod_{l=1}^L\, {\rm Aut}\big(\mu^l\big)
\ee
whose kernel is ${\rm Aut}(f^{\rm m})$ and whose image has index given
by the number $m$ of markings of $f$ (up to isomorphism). It follows that $|{\rm Aut}(f)|\, m = |{\rm Aut}(f^{\rm m})| \, \prod_{l=1}^L\, |{\rm Aut}(\mu^l)|$, and hence the combinatorial expansion (\ref{eq:ZbetaHurwitz}) can be written in terms of marked Hurwitz numbers as
\bea
\widetilde{Z}_h(\beta) &=& \sum_{n=\lceil a\,N \rceil}^{\infty} \,
\beta^{-n\,(h-1)} \ \sum_{\lambda\in\Lambda_+^n} \,
\frac{\omega_\lambda (\beta)}{\big|{\rm Aut}(\lambda)\big|} \ \sum_{d=0}^\infty\, \Big(\, \frac1N\,
\Big)^{2g-2} \ \sum_{L=0}^d \, \chi(\Sigma_{h,L}) \label{eq:ZbetaHurwitzmarked} \\ &&
\qquad \qquad \times \ \sum_{\stackrel{\scriptstyle
    \mu^{1},\dots,\mu^L \in\Lambda_+^n}{\scriptstyle
    \mu^l\neq(1^n)\,,\, \sum_{l=1}^L \ell^*(\mu^l)=d}} \,
\frac{{\mit\Delta}_{\mu^1}(\beta)}{\big|{\rm Aut}\big(\mu^1\big)\big|}
\cdots \frac{{\mit\Delta}_{\mu^L}(\beta)}{\big|{\rm
    Aut}\big(\mu^L\big)\big|}  \ H^{\rm m}_{h,n}\big(\lambda,
\mu^1,\dots,\mu^L\big) \ . \nonumber
\eea

While the refined weights (\ref{eq:omegalambda}) have a natural
meaning as deformations of the identity (see \eqref{eq:omega1n}), it
would be interesting to understand better the geometrical significance
of the combinatorial weights \eqref{eq:mitDelta} in terms of orbifold
Euler characteristics of moduli spaces of Riemann surfaces, as
suggested by the appearence of the binomial-type coefficients
\eqref{eq:gammakbeta}. We can give further insight into this
perspective following the geometric interpretation of refinement from
\S\ref{sec:betaensemble}. Let $\cH_{n,d,h,L}$ denote the Hurwitz
 space of isomorphism classes of $n$-sheeted branched covers
$f:\Sigma_g\to\Sigma_h$ of degree $d$ with $L$ branch points. It has
the structure of a discrete fibration
\bea
\pi_{n,d,h,L}\,:\, \cH_{n,d,h,L} \ \longrightarrow \ \Sigma_{h,L}
\label{eq:Psicover}\eea
over the configuration space of $L$ indistinguishable points on
$\Sigma_h$, which sends the class of a holomorphic map $f:\Sigma_g\to\Sigma_h$ to the branch locus of $f$. There is also a natural map
\bea
\cH_{n,d,h,L} \ \longrightarrow \ \cM_g
\label{eq:mapMg}\eea
which sends the class of the cover $f:\Sigma_g\to\Sigma_h$ to the
class of the curve $\Sigma_g$; the image of $\cH_{n,d,h,L}$ under this
map is a subvariety of the moduli space $\cM_g$ of genus $g$
curves. Recall from \S\ref{sec:betaensemble} that, in the planar limit
of the gauge theory on the sphere, refinement can be interpreted
geometrically as replacing the orbifold Euler characters $\chi_{\rm
  orb}(\cM_g)$ with the parameterized Euler characters
\eqref{eq:parEuler}. It is natural to think of pulling back the
corresponding characteristic classes under the map \eqref{eq:mapMg}, and for fixed $N$ we define the \emph{parameterized Euler character}
\bea 
\chi_{n,d,h,L}(\beta) &:=& \sum_{\lambda\in\Lambda_+^n} \,
\frac{\omega_\lambda (\beta)}{\beta^{n\,(h-1)}} \ \sum_{\stackrel{\scriptstyle
    \mu^{1},\dots,\mu^L \in\Lambda_+^n}{\scriptstyle
    \mu^l\neq(1^n)\,,\, \sum_{l=1}^L \ell^*(\mu^l)=d}} \,
{\mit\Delta}_{\mu^1}(\beta)\cdots {\mit\Delta}_{\mu^L}(\beta)
\label{eq:parEulerbeta}\\ && \qquad \qquad \qquad \qquad \qquad \qquad \qquad \qquad \times \ \chi(\Sigma_{h,L}) \, H_{h,n}\big(\lambda, \mu^1,\dots,\mu^L\big)
\nonumber\eea
for $n\geq\lceil a\,N \rceil$. Via the fibration \eqref{eq:Psicover}, in the unrefined limit it reduces to the orbifold Euler character
\bea
\chi_{n,d,h,L}(1) = \chi_{\rm orb}(\cH_{n,d,h,L})
\eea
of the singular variety $\cH_{n,d,h,L}$~\cite{Cordes1995}. Then we can
rewrite \eqref{eq:ZbetaHurwitz} in the more suggestive form
\begin{proposition}
 \ The chiral series for the partition function of topological $\beta$-deformed Yang-Mills theory on $\Sigma_h$ is the generating function
\bea \nonumber
\widetilde{Z}_h(\beta) = \sum_{n=\lceil a\,N \rceil}^{\infty} \ \sum_{d=0}^\infty\, \Big(\, \frac1N\,
\Big)^{2g-2} \ \sum_{L=0}^d \, \chi_{n,d,h,L}(\beta)
\eea
for the parameterized Euler characters \eqref{eq:parEulerbeta}, where $g$ is determined from $n$, $h$ and $d$ by the Riemann-Hurwitz formula \eqref{eq:RHformula}.
\label{prop:parEuler}\end{proposition}
This generalizes the string theory interpretation of the unrefined case~\cite{Cordes1995}, wherein the orbifold Euler characters of Hurwitz spaces $\chi_{\rm orb}(\cH_{n,d,h,L})$ are replaced under refinement by the parameterized Euler characters $\chi_{n,d,h,L}(\beta)$. As in \S\ref{sec:betaensemble}, it is natural to expect that these $\beta$-deformed characters are themselves associated to characteristic classes of some related moduli spaces; in particular, for $\beta=2$ the deformation weights are given by
\bea
{\mit\Delta}_{\mu}(2)= 2^{\ell^*(\mu)}\, \prod_{i=1}^{\ell(\mu)} \, \Big(\, \sum_{k=1}^{\mu_i}\, \frac{(2k-3)!!}{(k-1)!} \ \sum_{\lambda\in\Lambda_+^{\mu_i-k}} \,  \frac{\ell(\lambda)!}{z_\lambda} \ \prod_{j=1}^{\ell(\lambda)}\, \frac{(2\lambda_j-3)!!}{(\lambda_j-1)!}\, \Big) \ .
\eea
However, in the present case the characters are non-polynomial
functions of $\beta$; below we will compare their forms explicitly
with the parameterized Euler characters \eqref{eq:parEuler}.

For the case of a spherical target space $\Sigma_0=\IP^1$, certain classes of Hurwitz numbers can be expressed as integrals of psi-classes and Hodge classes over
the Deligne-Mumford moduli spaces of punctured curves $\overline{\cM}_{g,n}$~\cite{Ekedahl2001}. The corresponding partition functions are annihilated by the differential operator of a quantum curve, see e.g.~\cite{Do2015} for a review; it would be interesting to see if there is a similar quantum spectral curve underlying the partition function $\widetilde{Z}_0(\beta)$. On the other hand, orbifold Hurwitz numbers lead to partition functions which are annihilated by the difference operator of a quantum curve~\cite{Do2015}, and it would be interesting to understand the general $(q,t)$-deformed partition function $Z_0(q,t;0)$ also in this context.

\subsection{Planar limit}

In the planar limit of \S\ref{sec:betaensemble}, the pertinent
generalised Selberg integrals can also be expressed in terms of Jack
symmetric functions~\cite{Goulden1999}, which gives a geometrical
meaning to the refinement parameter $\beta$ as a combinatorial
invariant of non-orientability for maps of graphs into surfaces. Let
us now compare the leading term of the partition function
(\ref{eq:expandedPartFu}) for $h\geq 2$ in the classical limit $q=1$
with the parameterized Euler characteristics \eqref{eq:parEuler}. This amounts to setting the $\Omega$-factors
$\Omega_n(q,t)$ to $1$ and keeping only the $n=a\, N$ term of the sum
in (\ref{eq:expandedPartFu}), which yields
\be
\label{eq:Zlead}
Z_h^{\mathrm{pl}}(q,t;0)\, = \, \frac{q^{-\frac{a\, N\,(a\, N-1)}4}}{
  [a\, N]_q!} \, \big( d_{(\beta-1)\, \rho+a\, (1^N)}(q)\big)^{2-2h}\,
\Big(\, \frac{[N]_t}{\sqrt{g_{\omega_1}}}\, \Big)^{a\, N\, (2-2h)} \ .
\ee
In the classical limit this becomes
\bea
\widetilde Z_h^{\rm pl}(\beta) = \frac{1}{(a\,N)!} \,  \big(
d_{(\beta-1)\, \rho+a\, (1^N)}\big)^{2-2h} \, \Big(\, \frac{N^4}{\tilde g(\beta)}\, \Big)^{a\, N\, (1-h)} \ ,
\eea
where we used (\ref{eq:dimqtRfund}).
We can rewrite \eqref{eq:tildegbeta} in the form
\bea
\tilde g(\beta) = \frac N{\beta\, (N-1)+1} \
\frac{\prod\limits_{i=1}^N\, \frac{\Gamma(\beta\, i)}{\Gamma(\beta)}}{\prod\limits_{m=0}^{\beta-1} \ \prod\limits_{i=1}^{N-1} \, \frac{\beta^i\, \Gamma\big(i+1-\frac m\beta\big)}{\Gamma\big(1-\frac m\beta\big)}} \ ,
\eea
and using the dimension formula \eqref{eq:dlambda1} from Appendix~\ref{app:Hecke}
we can write the dimension of the symmetric group representation corresponding to the partition $(\beta-1)\,\rho+a\,(1^N)$ as
\be
 d_{(\beta-1)\, \rho+a\, (1^N)} = \frac{\beta^{\frac{N\,(N-1)}{2}}\, \Gamma(a\,N+1)\, G(N+1)}{\prod\limits_{i=1}^{N-1} \, \Gamma(1+\beta\, i)} \ ,
\ee
where $G(z)$ is the Barnes $G$-function with the property that
$G(N+1)=\prod_{i=1}^N\, \Gamma(i)$. The appearence of this Barnes
function suggests, following~\cite{Ooguri2002}, that our asymptotic
expansion could be related to refined topological closed string theory on the resolved conifold geometry.

The corresponding free energy
$\widetilde{F}_h^{\rm pl}(\beta):= -\log \widetilde Z_h^{\rm
  pl}(\beta)$ can be expanded as a power series in $\beta$, whose
coefficients are combinations of Bernoulli numbers, in much the same way
that we dealt with the partition function \eqref{eq:Mehtagamma}. The resulting expansion is somewhat complicated, so we content ourselves with an integral representation from which the expansion is straightforwardly extracted. For this, we use the integral formula
for the gamma-function~\cite[eq.~(3.6)]{Krefl2013}
\be
\log\Gamma(z)=\int_0^\infty \, \frac{\dd x}{x} \ \frac{1}{\e^x-1}\,
\big((z-1)\, (1-\e^{-x})+\e^{-x\, (z-1)}-1\big) \ ,
\ee
which holds for $\Re(z) > 0$. After some calculation, one infers the free energy
\bea
\widetilde{F}_h^{\rm pl}(\beta) &=& (h-1) \, \big( N\,
    (N-1) \, \log\beta \,- \, \mbox{$\frac{1}{2}$}\,a\,N^2\, (N-1)\, 
  \beta \, \log\beta \, - \, 3\, a\, N \, \log N \nonumber
  \\ && \qquad \qquad - \, a\, N \, \log(\beta\, (N-1)+1 )
   \big)
  \, + \, \int_0^\infty\, \frac{\dd x}{x} \ \frac{1}{\e^x-1} \, \mathcal{F}_h^{\beta,N}(x) \ , 
\eea
where we have defined 
\bea
 \mathcal{F}_h^{\beta,N}(x) &=& a\, N\, \big(
 1+\mbox{$\frac\beta2$}\, N\, (N-1)\, (1-h)\big) \, \big(1-\e^{-x} \big) \nonumber \\ 
  && + \, (2h-2)\, \Big(\, \frac{1-\e^{-\beta\, (N-1)\,
      x}}{1-\e^{\beta\, x}}\, -\, \frac{1-\e^{-(N-1)\, x}}{1-\e^{x}}
  \, \Big) + (2h-1)\, \left( \e^{-a\, N\, x}-1\right) \nonumber \\ 
    && + \, a\, N\, (h-1)\, \Big(\, \frac{\e^{-\beta\, (N-1)\, x} +
      \e^{x}-\e^{-(\beta \, N-1)\, x}-1}{1-\e^{\beta\, x}}-N+1 \\ 
		&& \qquad \qquad \qquad \qquad + \, N\e^{-(\beta-1)\,
                  x}+\mbox{$\frac\beta2$}\, N\, (N-1) \e^{-\beta\, x} \,
                 \big(\e^x-1
                  \big) \, \Big) \ . \nonumber
\eea
In the $\beta=1$ limit (which also induces $a=0$) the planar free
energy vanishes, as we expect from~(\ref{eq:Zlead}).

The power series
expansion in $\beta$ can now be obtained by using the
generating function \eqref{eq:Bernoulli} to expand the denominators
$(1-\e^{\beta\,x})^{-1}$ and the integral identities
of~\cite[Appendix~A]{Krefl2013}.
For example, we can readily compute the contribution
\bea
-\int_0^\infty\, \frac{\dd x}{x} \ \frac{1}{\e^x-1}\,\frac{1-\e^{-\beta\, (N-1)\,
      x}}{1-\e^{\beta\, x}} \, =\, \sum_{n=0}^\infty\, \cF_n^N \ \beta^n
\eea
where
\bea
\cF_0^N&=&(N-1)\, \big(\,\mbox{$\frac1\varepsilon+\frac12\, \log\varepsilon+\frac12\,(\gamma-\log2\pi)$}\, \big) \ , \nonumber \\[4pt]
\cF_n^N&=& \zeta(n)\, (n-1)! \, \sum_{k=0}^n\, (N-1)^{k+1}\, \frac{B_{n-k}}{(k+1)!\, (n-k)!} \qquad \mbox{for} \quad n\geq1 \ .
\eea
Here $\varepsilon\to0^+$ gives the leading one-loop linear and
logarithmic divergences, $\gamma$ is the Euler-Mascheroni constant,
and $\zeta(z)$ is the Riemann zeta-function. These formulas explicitly
illustrate
the analytic dependence of the parameterized Euler characteristics
\eqref{eq:parEulerbeta} on the refinement parameter $\beta$, as
compared to the polynomial characters \eqref{eq:parEuler}.

\section{Chiral expansions of defect observables\label{sec:Observables}}

Two-dimensional Yang-Mills theory also involves observables corresponding to insertions in the partition function of operators supported on real codimension one defects in $\Sigma_h$. In this final section we extend the chiral expansion of~\S\ref{sec:chiralpartfn} to these observables.

\subsection{Boundaries}

We first describe the large $N$ expansion of the refinement of $q$-deformed Yang-Mills theory on open Riemann surfaces of genus $h$ with $b$ boundaries; via suitable gluing rules they give the building blocks for all Yang-Mills amplitudes. The holonomies along the boundaries are specified by generalized quantum characters of elements $U_i\in T$, $i=1,\dots,b$, and the partition function is given by
\be 
Z_{h,b}(q,t;p; U_1,\dots,U_b) = \sum_{\lambda\in\Lambda_+} \, \frac{\dim_{q,t}(R_\lambda)^{2-2h-b}}{(g_\lambda)^{1-h-b/2}} \ q^{\frac p2\,
  (\lambda,\lambda)}\, t^{p\,(\rho,\lambda)} \ \prod_{i=1}^b\, \chi_{\Phi_\lambda} (U_i) \ .
\ee
Again we consider only the topological gauge theory and study
\be \label{eq:ZBoundary}
Z_{h,b}(q,t;0;U_1,\ldots,U_b) \,= \sum_{n=0}^\infty \ \sum_{\lambda\in\Lambda^n_+} \, \Big(\, \frac{\dim_{q,t}(R_\lambda)}{\sqrt{g_\lambda}}\,\Big)^{2-2h-b} \ \prod_{i=1}^b\, \chi_{\Phi_\lambda} (U_i) \ .
\ee
This partition function was also considered in~\cite{Aganagic2012} but with a different normalization for the boundary characters.

We begin by using the transformation from Appendix~\ref{sec:centralTrafo} to change to a basis of central elements $C_i$ of $\sfH_q(\frS_\infty)$ and set
\bea \label{eq:fourier}
&& Z_{h,b}(q,t;0;C_1,\ldots,C_b) \\ \nonumber
 &&  \ := \ \int_{T^b} \ \prod_{i=1}^{b}\, [\dd U_i]_{q,t} \ \sum_{n_i=a\,
   N}^\infty\, \frac{q^{-\frac{n_i\, (n_i-1)}{4}}}{[n_i]_q!} \
 \Tr_{R_{\omega_1}^{\otimes n_i}} \big(\, \Phi_{n_i}\,\Theta_{n_i}(\beta)\, C_{i}\,U_i^\dagger\,\big) \
 Z_{h,b}(q,t;0; U_1,\ldots,U_b) \ ,
\eea
where the intertwining operator $\Phi_{n_i}$ is
defined in~\S\ref{sec:qCWdual}, the step function $\Theta_{n_i}(\beta)$
is defined in (\ref{eq:thetaDef}), and the integration
measure $[\dd U]_{q,t}$ on the maximal torus $T\subset G$ given by (\ref{eq:Macintmeasure})
defines the Macdonald inner product of generalized characters as an integral over
holonomies~\cite{Etingof1993}. Since the commutant of the
representation of $\cU_q(\frgl_N)$ on $R_{\omega_1}^{\otimes n}$ is
the Hecke algebra $\sfH_q(\frS_n)$, the map (\ref{eq:fourier}) may be
regarded as the refined version of the quantum Fourier transformation
of the boundary holonomy amplitudes. The
  generalized characters are orthonormal with respect to the Macdonald inner product, i.e.
\be \label{eq:GenCarORt}
\int_T \, [\dd U]_{q,t} \ \chi_{\Phi_\lambda} (U)\, \chi_{\Phi_{\lambda'}} (U^\dagger) \, = \, \delta_{\lambda, \lambda'}
\ee
for $\lambda,\lambda'\in \Lambda_+$. Using Lemma~\ref{lem:schurweyl} we can write
\bea 
\Tr_{R_{\omega_1}^{\otimes n}} \big(\, \Phi_n\,\Theta_{n}(\beta)\,
C_{i}\,U_i^\dagger\, \big)&=&\sum_{\lambda\in \Lambda_+^{n}} \, \chi_{r_{\lambda}}\big(\Theta_{n}(\beta) \, C_i \big) \, \chi_{\Phi_{\lambda_{\beta-2}}}(U_i^\dagger) \nonumber \\[4pt] &=& \sum_{\mu\in \Lambda_+^{n-a\, N}} \, \chi_{r_{\mu_{\beta}+a\,(1^N)}}(C_i) \, \chi_{\Phi_{\mu+a\,(1^N)}}(U_i^\dag) \ ,
\label{eq:FourierExpand}\eea
which yields
\bea
Z_{h,b}(q,t;0;C_1,\ldots,C_b) &=&  \sum_{n=a\,N}^\infty\, \Big(\, \frac{q^{-\frac{n\, (n-1)}{4}}}{[n]_q!}\, \Big)^b \ \sum_{n'=0}^\infty \ \sum_{\lambda\in\Lambda_+^{n'}}\, \Big(\, \frac{\dim_{q,t}(R_\lambda)}{\sqrt{g_\lambda}}\, \Big)^{2-2h-b} \nonumber \\
&& \qquad \qquad \qquad \times \ \sum_{\mu_1,\dots,\mu_b \in\Lambda_+^{n-a\, N}} \
\prod_{i=1}^b\, \chi_{r_{\mu_{i\beta}+a\,(1^N)}}(C_i) \
\delta_{\lambda ,  \mu_i+a\,(1^N)} \label{eq:ZcentralBasis} \\[4pt]  
& = &  \sum_{n=a\, N}^\infty\, \Big(\, \frac{q^{-\frac{n\, (n-1)}{4}}}{[n]_q!}\, \Big)^b \,  \sum_{\mu\in\Lambda_+^{n-a\, N}} \, \Big( \, \frac{\dim_{q,t}(R_\mu)}{\sqrt{g_\mu}}\, \Big)^{2-2h-b} \, \prod_{i=1}^b\, \chi_{r_{\mu_{\beta}+a\,(1^N)}}(C_i) \nonumber
\eea
where we used (\ref{eq:GenCarORt}) and (\ref{eq:dimshift}).

We can now use (\ref{eq:qtdim_small}) to expand the refined quantum dimensions and from (\ref{eq:central}) we get
\bea 
&& Z_{h,b}(q,t;0;C_1,\ldots,C_b) \ = \ \sum_{n=\lceil a\, N\rceil}^\infty\, \Big(\, \frac{[ N]_t}{\sqrt{g_{\omega_1}}} \,\Big)^{n\, (2-2h-b)} \ \Big(\, \frac{q^{-\frac{n\,(n-1)}4}}{ [n]_q!} \, \Big)^b
\nonumber \label{eq:Zhbfinal} \\ && \qquad \qquad \qquad 
\times \ \sum_{\sigma_1,\tau_1,\dots,\sigma_h,\tau_h\in\frS_n} \,
\delta\Big(\Theta_n(\beta) \,(E_n)^{b-1}\, \Omega_n(q,t)^{2-2h-b}
\\ && \qquad \qquad \qquad \qquad \qquad \qquad \qquad \qquad \times \ \prod_{i=1}^{h}\, q^{-\ell(\sigma_i)-\ell(\tau_i)} \, \sfh(\sigma_i)\, \sfh(\tau_i)\, \sfh(\sigma_i^{-1})\, \sfh(\tau_i^{-1}) \ \prod_{j=1}^b\, C_j \Big) \ ,
\nonumber
\eea
where the central element $E_n$ is defined by~\cite[Appendix~A.3]{deHaro2006}
\be
E_n \, := \,  \sum_{\sigma\in\frS_n} \, q^{-\ell(\sigma)}\, \sfh(\sigma^{-1})\,\sfh(\sigma)
\label{eq:Encentral}\ee
with the properties
\be
E_n^{-1} \, = \,\frac{q^{-\frac{n\,(n-1)}4}}{ [n]_q!} \ D_n \qquad
\mbox{in} \quad \widehat{\sfH}_q(\frS_n)
\ee
and
\be
\chi_{r_\lambda}(E_n)\,= \, q^{\frac{n\,(n-1)}4} \ [n]_q! \ \frac{d_{\lambda}(1)^2 }{d_{\lambda}(q)} \qquad \text{for} \quad \lambda\in\Lambda_+^n \ .
\ee
For fixed $n$ this expression reduces to \cite[eq.~(3.6)]{deHaro2006}
in the unrefined limit, whereas our derivation gives the full
partition function summed over all indices $n$. In particular, this
partition function is a refined quantum deformation of the counting of
holomorphic maps with specified monodromies $C_j$ at the boundaries~\cite{Ramgoolam1995}; by expanding the $\Omega$-factors, in the classical limit $q=1$ it can be expressed in terms of parameterized Euler characters as in \S\ref{sec:betaHurwitz}.

Let us look at some of the basic amplitudes which are the building
blocks for the entire $(q,t)$-deformed gauge theory. The topological disk amplitude (with puncture of holonomy in the representation \eqref{eq:defW}) is the case $h=0,b=1$ in \eqref{eq:ZBoundary} which evaluates to
\be 
Z_{0,1}(q,t;0;U) \,= \sum_{n=0}^\infty \ \sum_{\lambda\in\Lambda^n_+} \, \frac{\dim_{q,t}(R_\lambda)}{\sqrt{g_\lambda}} \ \chi_{\Phi_\lambda} (U) = \delta_{q,t}\big(U,q^{\beta\,(\rho,H)} \big) \ ,
\ee
where $\delta_{q,t}$ is the delta-function in the measure $[\dd U]_{q,t}$. This shows that the wavefunction $\Psi(U)$ for a disk in the topological $(q,t)$-deformed gauge theory is supported on generalized quantum group holonomies of flat connections on a disk, generalising the unrefined case of \cite[eq.~(3.7)]{deHaro2006} wherein $\delta_{q,q}$ is the delta-function in the Haar measure for $U(N)$. Dually, we can represent the disk partition function in a form that depends solely on Hecke algebra quantities by using \eqref{eq:Zhbfinal} to write
\be 
Z_{0,1}(q,t;0;C) \,= \sum_{n=\lceil a\, N\rceil}^\infty\, \Big(\, \frac{[ N]_t}{\sqrt{g_{\omega_1}}} \,\Big)^{n} \ \frac{q^{-\frac{n\,(n-1)}4}}{ [n]_q!} \ \delta\big(\Theta_n(\beta)\, \Omega_n(q,t)\, C\big) \ ,
\ee
independently of the central elements \eqref{eq:Encentral}. Similarly, the punctured topological cylinder amplitude is obtained from \eqref{eq:ZBoundary} with $h=0,b=2$, giving
\be 
Z_{0,2}(q,t;0;U_1,U_2) \,= \sum_{n=0}^\infty \ \sum_{\lambda\in\Lambda^n_+} \, \chi_{\Phi_\lambda} (U_1)\, \chi_{\Phi_\lambda} (U_2) = \delta_{q,t}(U_1,U_2)
\ee
with the dual formulation
\be 
Z_{0,2}(q,t;0;C_1,C_2) \,= \sum_{n=\lceil a\, N\rceil}^\infty\, \Big(\, \frac{q^{-\frac{n\,(n-1)}4}}{ [n]_q!} \,\Big)^{2} \ \delta\big(\Theta_n(\beta)\, E_n\, C_1\, C_2\big)
\ee
independently of the $\Omega$-factors \eqref{eq:Omegafactors}.

\subsection{Wilson loops}

The natural closed defect observables of the gauge theory are of
course the Wilson loops which correspond to simple closed curves on
the surface $\Sigma_h$. For definiteness, let us consider the large
$N$ expansion of a single Wilson loop in the representation
$R_\lambda$ on a simple oriented closed curve which divides the
Riemann surface $\Sigma_h$ into two faces of genera $h_1$ and $h_2$
with $h=h_1+h_2$. The expectation value of the Wilson loop operator is given by~\cite{Szabo2013}
\bea
W_\lambda(q,t;p;h_1,h_2) &=& \sum_{\mu,\nu\in\Lambda_+} \ \int_T \, [\dd U]_{q,t} \  \frac{\dim_{q,t}(R_\mu)^{1-2h_1}}{(g_\mu)^{\frac12-h_1}} \, \frac{\dim_{q,t}(R_\nu)^{1-2h_2}}{(g_\nu)^{\frac12-h_2}} \ \ q^{\frac p2\,
  (\lambda,\lambda)}\, t^{p\,(\rho,\lambda)} \nonumber \\ && \qquad
\qquad \qquad \qquad \qquad \qquad \qquad \times \ \chi_{\Phi_\mu} (U)\, \chi_{\Phi_\lambda} (U)\, \chi_{\Phi_\nu} (U^\dag) \ .
\eea
In the topological limit $p=0$ we can use the orthonormality relation (\ref{eq:GenCarORt}) to obtain
\be \label{eq:Wilson}
W_\lambda(q,t;0;h_1,h_2) \,= \sum_{\mu,\nu\in\Lambda_+} \, \Big(\, \frac{\dim_{q,t}(R_\mu)}{\sqrt{g_\mu}}\,\Big)^{1-2h_1} \, \Big(\, \frac{\dim_{q,t}(R_\nu)}{\sqrt{g_\nu}}\, \Big)^{1-2h_2} \, \widetilde{N}_{\mu \, \lambda}^\nu \ ,
\ee
where $\widetilde{N}_{\mu \, \lambda}^\nu $ are refined fusion coefficients defined by the relation
\be \label{eq:NtildeDef}
\chi_{\Phi_\mu}(U)\, \chi_{\Phi_\lambda}(U) \, = \, \sum_{\nu\in\Lambda_+} \, \widetilde{N}_{\mu \, \lambda}^\nu \ \chi_{\Phi_\nu}(U)
\ee
expressing the completeness of the Macdonald polynomials $M_\lambda(x;q,t)$ in the ring of symmetric functions.
Below we compare them to the Littlewood-Richardson coefficients $N_{\mu\, \lambda}^\nu\in\IZ_{\geq0}$ which give the multiplicities in the decomposition of tensor products of irreducible $U(N)$-modules as
\be \label{NDef}
R_{\mu}\otimes R_\lambda = \bigoplus_{\nu\in \Lambda_+}\, R_\nu^{\oplus N_{\mu\, \lambda}^\nu} \ ,
\ee
and the same decomposition is true as $\cU_q(\frgl_N)$-modules. To suitably express $\widetilde{N}_{\mu \, \lambda}^\nu$ and expand the Wilson loops we need a couple of preliminary lemmata.
\begin{lemma}
\ If $\lambda$, $\mu$, $\nu$, $\lambda_\beta$, $\mu_\beta$ and
$\nu_\beta$ are all partitions, then for large $N$ one has
\[
\widetilde{N}_{\mu \, \lambda}^\nu \,= \, N_{\mu_\beta\, \lambda_\beta}^{\nu_\beta}
\]
in $\cU_q(\frgl_N)$.
\label{lem:tildeN}
\end{lemma}
\Proof{
Let us consider $\Tr_{R_{\omega_1}^{\otimes n}} (\Phi_n \,
P_{\lambda_\beta}\, U)$ for $\lambda\in\Lambda_+^n$. The trace takes
values in the weight zero subspace of $W_{\beta-1}$ from
(\ref{eq:defW}), and in this subspace the intertwiner $\Phi_n$ acts proportionally to the identity on $R_{\omega_1}^{\otimes n}$ via (\ref{eq:PhiFundaId}). This yields
\be
\Tr_{R_{\omega_1}^{\otimes n}} (\Phi_n \, P_{\lambda_\beta} \, U) \, = \, (g_{\omega_1})^{-n/2} \ \Tr_{R_{\omega_1}^{\otimes n}} (P_{\lambda_\beta} \, U) \ ,
\ee
where the second trace is an ordinary $\IC$-valued trace. We can use quantum Schur-Weyl duality and the definition of the quantum Young projectors from (\ref{eq:Youngprojector}) together with the definition of $\Phi_n$ in (\ref{eq:PhiDef}) to get
\be
d_{\lambda_\beta}(1)\,\chi_{\Phi_\lambda} (U) \, = \,d_{\lambda_\beta}(1)\, (g_{\omega_1})^{-n/2} \ \Tr_{R_{\lambda_\beta}} (U) \ .
\ee
It follows that the generalized character and the trace of $U$ differ only by a factor as
\be
\Tr_{R_{\lambda_\beta}} (U) \, = \, (g_{\omega_1})^{n/2} \, \chi_{\Phi_\lambda} (U) \ .
\ee
Using in addition the definitions of the coefficients (\ref{NDef}) and (\ref{eq:NtildeDef}), we then get
\bea
 \chi_{\Phi_\mu}(U) \,\chi_{\Phi_\lambda}(U) & =
 &(g_{\omega_1})^{-(|\mu|+|\lambda|)/2} \ \Tr_{R_{\mu_\beta}} (U) \ \Tr_{R_{\lambda_\beta}} (U) \nonumber \\[4pt] &=& (g_{\omega_1})^{-(|\mu|+|\lambda|)/2}\, \sum_{\nu\in\Lambda_+} \, N_{\mu_\beta \, \lambda_\beta}^\nu  \, \Tr_{R_{\nu}} (U)  \ = \ \sum_{\nu\in\Lambda_+} \, N_{\mu_\beta \, \lambda_\beta}^\nu  \,  \chi_{\Phi_{\nu_{\beta-2}}} (U) 
\eea
and
\bea 
\chi_{\Phi_\mu}(U) \,\chi_{\Phi_\lambda}(U) \,=\, \sum_{\nu\in\Lambda_+} \, \widetilde{N}_{\mu \, \lambda}^{\nu} \, \chi_{\Phi_{\nu}} (U) \ .
\eea
The result now follows by taking the inner product
(\ref{eq:GenCarORt}) of $\chi_{\Phi_\mu}(U) \,\chi_{\Phi_\lambda}(U)$
with $\chi_{\Phi_{\nu}}(U)$ in each of these expressions and comparing the two results.
}
\begin{lemma} \ $
\widetilde{N}_{\mu \, \lambda}^\nu\, = \, \widetilde{N}_{\mu+a\,(1^N) \ \lambda+a\,(1^N)}^{\nu+2a\,(1^N)} \ . $
\label{lem:Nshift}
\end{lemma}
\Proof{
We use the shift property of the Macdonald polynomials from~\cite[\S
IV, \ eq.~(4.17)]{Macdonald1995} which reads
\be
M_{\lambda+a\,(1^N)}(x;q,t)=x^a \, M_{\lambda}(x;q,t)
\ee
where $x^a:=(x_1\cdots x_N)^a$. Together with (\ref{eq:dimshift}) this implies 
\be \label{eq:CharSift}
\chi_{\Phi_{\lambda+a\,(1^N)}}(U)=x^a \, \chi_{\Phi_{\lambda}}(U) \ ,
\ee
where $x=\e^z$ and $U=\e^{(z,H)}$. We then obtain
\bea
\chi_{\Phi_{\mu+a\,(1^N)}} (U) \, \chi_{\Phi_{\lambda+a\,(1^N)}} (U)
\,=\, x^{2a}\, \chi_{\Phi_{\mu}} (U) \, \chi_{\Phi_{\lambda}} (U) \, =
\, \sum_{\nu \in\Lambda_+}\,  \widetilde{N}_{\mu\,\lambda}^{\nu}
\, \chi_{\Phi_{\nu +2a\,(1^N)}}(U)
\eea
and
\bea
\chi_{\Phi_{\mu+a\,(1^N)}} (U) \, \chi_{\Phi_{\lambda+a\,(1^N)}} (U) \,=\,\sum_{\nu\in\Lambda_+}\, \widetilde{N}_{\mu+a\,(1^N) \ \lambda+a\,(1^N)}^{\nu} \, \chi_{\Phi_{\nu}}(U) \ .
\eea
The result now follows by taking the inner product
(\ref{eq:GenCarORt}) of $\chi_{\Phi_{\mu+a\,(1^N)}} (U) \,
\chi_{\Phi_{\lambda+a\,(1^N)}} (U)$ with the generalized character
$\chi_{\Phi_{\nu+2a\,(1^N)}}(U)$ in each of these expressions and comparing the two results.
}

To work out the large $N$ expansion of the Wilson loop (\ref{eq:Wilson}), we use the expansion of the Littlewood-Richardson coefficients in terms of Hecke characters given by~\cite[eq.~(4.11)]{deHaro2006}
\bea 
N_{\mu \, \lambda}^\nu &=& \frac{q^{-\frac{n_1\,(n_1-1)}4}}{[n_1]_q!}
\, \frac{q^{-\frac{n_2\,(n_2-1)}4}}{[n_2]_q!} \,
\frac{d_{\mu}(q)}{d_{\mu}(1)} \, \frac{d_{\lambda}(q)}{d_{\lambda}(1)}
\ \sum_{\sigma_1\in\frS_{n_1}} \ \sum_{ \sigma_2\in\frS_{n_2}}\,
q^{-\ell(\sigma_1)-\ell(\sigma_2)} \label{eq:Nexp} \\ && \qquad \qquad
\qquad \qquad \qquad \qquad \qquad \times \
\chi_{r_{\mu}}\big(\sfh(\sigma_1^{-1})\big) \,
\chi_{r_{\lambda}}\big(\sfh(\sigma_2^{-1})\big) \,
\chi_{r_{\nu}}\big(\sfh(\sigma_1)\cdot \sfh(\sigma_2)\big) \ , \nonumber
\eea
where $|\mu|=n_1$, $|\lambda|=n_2$, $|\nu|=n_1+n_2=: n$, and $\sfh(\sigma_1)\cdot \sfh(\sigma_2)$ acts on $\sfH_q(\frS_{n})$ via $\sfg_1,\ldots, \sfg_{n_1-1}\in\sfH_q(\frS_{n_1})\subset\sfH_q(\frS_{n})$ and $\sfg_{n_1+1},\ldots ,\sfg_{n_1+n_2-1}\in\sfH_q(\frS_{n_2})\subset\sfH_q(\frS_{n})$.
We rewrite the expectation value of the Wilson loop (\ref{eq:Wilson}) using (\ref{eq:dimshift}), Lemma~\ref{lem:Nshift} and Lemma~\ref{lem:tildeN} to get
\bea
W_\lambda(q,t;0;h_1,h_2) &=& \sum_{n_1=a\, N}^\infty \ \sum_{n=2a\,
  N}^\infty \ \sum_{\mu\in\Lambda_+^{n_1-a\, N}} \
\sum_{\nu\in\Lambda_+^{n-2a\, N}} \, N_{\mu_\beta+a\,(1^N) \
  \lambda_\beta+a\,(1^N)}^{\nu_\beta+2a\,(1^N)} \\ && \qquad \qquad \times \ \Big(\,
\frac{\dim_{q,t}(R_{\mu+a\,(1^N)})}{\sqrt{g_{\mu+a\,(1^N)}}}\,\Big)^{1-2h_1}
\, \Big(\,
\frac{\dim_{q,t}(R_{\nu+2a\,(1^N)})}{\sqrt{g_{\nu+2a\,(1^N)}}}\,
\Big)^{1-2h_2} \ , \nonumber
\eea
for $|\lambda|=n_2-a\, N$. Again we expand a transformed version of the Wilson loop given by
\be
W(q,t;0;h_1,h_2;C) \, = \, \frac{q^{-\frac{n_2\, (n_2-1)}{4}}}{[n_2]_q!} \ \sum_{\lambda\in\Lambda_+^{n_2-a\, N}} \,
\chi_{r_{\lambda_\beta+a\,(1^N)}}(C) \ W_\lambda(q,t;0; h_1,h_2) \ ,
\ee
where $C$ is an arbitrary central element of the Hecke algebra
$\sfH_q(\frS_{n_2})$. Using now the expansion of the
Littlewood-Richardson coefficients $N_{\mu \, \lambda}^\nu$ from
(\ref{eq:Nexp}), the expansion of the refined quantum dimensions from
(\ref{eq:qtdim_small}), the character of the central element $D_n$
from (\ref{eq:charD}), the definition of the step function
$\Theta_n(\beta)$ from (\ref{eq:thetaDef}), the properties
(\ref{eq:PI}) and (\ref{eq:central}), and the definition of the
delta-functions on Hecke algebras from (\ref{eq:deltaf}) we finally
arrive at the chiral series for Wilson loop observables given by
\bea 
&& W(q,t;0;h_1,h_2;C) \nonumber \\ && \ = \  \sum_{n_1=\lceil a\, N\rceil}^\infty \ \sum_{n=\lceil 2a\, N\rceil}^\infty \, \Big(\, \frac{[N]_t}{\sqrt{g_{\omega_1}}} \, \Big)^{ n_1\,(1-2h_1)+ n\,(1-2h_2)} \ \delta_{n_1+n_2 , n} \, \frac{q^{-\frac{n_1\,(n_1-1)}4}}{ [n_1]_q!} \, \frac{q^{-\frac{n_2\,(n_2-1)}4}}{ [n_2]_q!} \nonumber \\ && \times \ 
\sum_{\sigma_1\in\frS_{n_1}} \ \sum_{\sigma_2\in\frS_{n_2}} \,
q^{-\ell(\sigma_1)-\ell(\sigma_2)} \, \delta\Big(\Theta_{n_2}(\beta)
\, C\,\sfh\big(\sigma_2^{-1}\big)\Big) \, \delta\Big(\Theta_{n_1}(\beta) \,D_{n_1} \,\Omega_{n_1}(q,t)^{1-2h_1}\, \Pi_{n_1}^{(h_1)}\, \sfh\big(\sigma_1^{-1}\big)\Big) \nonumber  \\
&& \qquad \qquad \qquad \qquad \qquad \qquad \qquad \qquad \times \ \delta\Big(
\Theta^2_{n}(\beta)\, \Omega_{n}(q,t)^{1-2h_2}\,
\Pi_{n}^{(h_2)} \, \big(\sfh(\sigma_1)\cdot\sfh(\sigma_2)\big) \Big) \ ,
\eea
where we have defined
\be
\Theta^2_{n}(\beta) \, = \, \sum_{\substack{
    \mu\in\Lambda_+^n \\ \mu_i\geq (\beta-1)\, \rho_i+2a} } \,  P_\mu
\ee
and
\be
\Pi_n^{(h)} \, = \,\sum_{\sigma_1,\tau_1,\dots,\sigma_h,\tau_h\in \frS_n} \ \prod_{i=1}^{h} \, q^{-\ell(\sigma_i)-\ell(\tau_i)} \, \sfh(\sigma_i)\, \sfh(\tau_i)\, \sfh(\sigma_i^{-1})\, \sfh(\tau_i^{-1}) \ .
\ee
This expression is the refined version
of~\cite[eq.~(4.18)]{deHaro2006}. It is a refined quantum deformation
of the counting of covering worldsheets with boundary that maps to the
corresponding Wilson graph on $\Sigma_h$ according to the specified
monodromy $C$~\cite{Cordes1994,Ramgoolam1995}; the expansion into parameterized orbifold Euler characters in the classical limit $q=1$ proceeds as in \S\ref{sec:betaHurwitz}.

\section*{Acknowledgments}

We thank Sanjaye Ramgoolam and Alessandro Torrielli for helpful
discussions and correspondence.
R.J.S. thanks the Alfred Renyi Institute of Mathematics
in Budapest for support and its staff for the warm hospitality during finishing
stages of this work. This work was supported in part by the Actions
MP1210 ``The String Theory Universe'' and MP1405 ``Quantum Structure of Spacetime'' funded by the European
Cooperation in Science and Technology (COST), by the Advanced Grant LDTBUD from the European Research Council, and by the
Consolidated Grant ST/L000334/1 from the UK Science and Technology
Facilities Council.

\appendix

\section{Quantum group $\cU_q(\frgl_N)$\label{app:Quantumgroup}}

For a generic value of $q$, let $\cU_q(\frgl_N)$ be the associative algebra over $\IC$ with generators $E_i,F_i$ for $i=1,\dots,N-1$ and $q^{\pm\,H_i/2}$ for $i=1,\dots,N$ obeying the relations
\bea
q^{H_i/2}\, E_i\, q^{-H_i/2} &=& q^{1/2}\, E_i \ , \nonumber \\[4pt]
q^{H_i/2}\, E_{i-1}\, q^{-H_i/2} &=& q^{-1/2}\, E_{i-1} \ , \nonumber \\[4pt]
q^{H_i/2}\, F_i\, q^{-H_i/2} &=& q^{-1/2}\, F_i \ , \nonumber \\[4pt]
q^{H_i/2}\, F_{i-1}\, q^{-H_i/2} &=& q^{1/2}\, F_{i-1} \ , \nonumber \\[4pt]
\big[q^{H_i/2},E_j\big] \ = \ \big[q^{H_i/2},F_j\big] &=&0 \qquad \mbox{for} \quad j\neq i,i-1 \ , \nonumber \\[4pt]
[E_i,F_j]&=&\delta_{ij}\, \frac{q^{(H_i-H_{i+1})/2}-q^{-(H_i-H_{i+1})/2}}{q^{1/2}-q^{-1/2}} \ , \nonumber \\ [4pt]
[E_i,E_j]\ = \ [F_i,F_j] &=& 0 \qquad \mbox{for} \quad |i-j|>1 \ , \nonumber \\[4pt]
E_i^2\, E_j-\big(q^{1/2}+q^{-1/2}\big)\, E_i\, E_j\, E_i+E_j\, E_i^2 &=&0 \qquad \mbox{for} \quad |i-j|=1 \ , \nonumber \\[4pt] 
F_i^2\, F_j-\big(q^{1/2}+q^{-1/2}\big)\, F_i\, F_j\, F_i+F_j\, F_i^2 &=&0 \qquad \mbox{for} \quad |i-j|=1 \ .
\eea
In the fundamental representation (\ref{eq:fundrep}), we have $H_i=E_{ii}$, $E_i=E_{i\, i+1}$ and $F_i=E_{i+1\, i}$. The coproduct on $\cU_q(\frgl_N)$ is defined by
\bea
\Delta(E_i)&=& E_i\otimes q^{-(H_i-H_{i+1})/2}+q^{(H_i-H_{i+1})/2}\otimes E_i \ , \nonumber \\[4pt]
\Delta(F_i)&=& F_i\otimes q^{-(H_i-H_{i+1})/2}+q^{(H_i-H_{i+1})/2}\otimes F_i \ , \nonumber \\[4pt]
\Delta\big(q^{H_i/2}\big) &=& q^{H_i/2}\otimes q^{H_i/2} \ .
\eea

\section{Hecke algebra of type $A_{n-1}$\label{app:Hecke}}

The symmetric group $\frS_n$ of degree $n\geq2$ is generated by the elementary transpositions $\sigma_i=(i \ i+1)$ for $i=1,\dots,n-1$ satisfying the relations
\be 
\sigma_i\, \sigma_{i+1}\, \sigma_i = \sigma_{i+1}\, \sigma_i\, \sigma_{i+1} \ , \qquad \sigma_i\, \sigma_j=\sigma_j\, \sigma_i \quad \mbox{for} \quad |i-j|>1 \qquad \mbox{and} \qquad \sigma_i^2=1 \ .
\ee
The \emph{length} $\ell(\sigma)$ of a permutation $\sigma\in\frS_n$ is the smallest integer $r$ such that there exists $i_1,\dots,i_r$ with $\sigma= \sigma_{i_1}\cdots \sigma_{i_r}$; such an expression is called a decomposition of $\sigma$ into a \emph{reduced word}. Note that decompositions into reduced words are not unique.

The \emph{Hecke algebra} $\sfH_q(\frS_n)$ of $\frS_n$ for $n\geq2$ is the
algebra over $\sfH_q(\frS_0)=\sfH_q(\frS_1):=\IC[q,q^{-1}]$ generated by $\sfg_i$ for $i=1,\dots,n-1$ with the relations
\be \label{eq:HeckeRel}
\sfg_i\, \sfg_{i+1}\, \sfg_i = \sfg_{i+1}\, \sfg_i\, \sfg_{i+1} \ , \qquad \sfg_i\, \sfg_j=\sfg_j\, \sfg_i \quad \mbox{for} \quad |i-j|>1\qquad \mbox{and} \qquad (\sfg_i-q)\, (\sfg_i+1)=0 \ .
\ee
The inverse of the generator $\sfg_i$ is
\be
\sfg_i^{-1}=q^{-1}\, \sfg_i+\big(q^{-1}-1 \big) \ .
\ee
If $\sigma=\sigma_{i_1}\cdots \sigma_{i_r}$ is a decomposition of $\sigma\in\frS_n$ in the form of a reduced word, then we set $\sfh(\sigma)=\sfg_{i_1}\cdots \sfg_{i_r}\in \sfH_q(\frS_n)$. One can show that $\sfh(\sigma)$ is independent of the decomposition of $\sigma$ into a reduced word and that $\big\{\sfh(\sigma)\big\}_{\sigma\in\frS_n}$ is a $\IC[q,q^{-1}]$-basis of the free $\IC[q,q^{-1}]$-module $\sfH_q(\frS_n)$. Given $\sigma,\tau\in\frS_n$ with $\ell(\sigma\,\tau)= \ell(\sigma)+\ell(\tau)$, one has $\sfh(\sigma)\,\sfh(\tau)= \sfh(\sigma\,\tau)$. The algebra $\sfH_q(\frS_n)$ is a $q$-deformation of the group algebra $\IC[\frS_n]$; in the classical limit $q=1$ the element $\sfh(\sigma)$ becomes the permutation~$\sigma$.
The combinatorial identity
\be 
\sum_{\sigma\in\frS_n}\, q^{\ell(\sigma)} = q^{\frac{n\, (n-1)}4}\, [n]_q!
\ee
expresses a $q$-deformation of the order of $\frS_n$, where we defined the $q$-factorial $[n]_q!:= [1]_q\cdots [n]_q$. 

The irreducible representations $r_\lambda$ of the symmetric group $\frS_n$ are in one-to-one correspondence with partitions $\lambda$ of $n$. In particular, the sign representation $\det= \bigwedge^nR_{\omega_1}^{\otimes n}$ corresponds to the trivial partition $\lambda=(n)$ while the trivial representation corresponds to the maximal partition $\lambda=(1^n)$ with $n$ parts. The splitting (\ref{eq:qschurweyl}) then gives the decomposition of $R_{\omega_1}^{\otimes n}$ into subrepresentations corresponding to its $\lambda$-isotopical components.
The $q$-deformation of the dimension of $r_\lambda$ is given by
\be 
d_\lambda(q) = \frac{\prod\limits_{i<j}\,
  \big(q^{\ell_i}-q^{\ell_j}\big)}{\prod\limits_{i=1}^{\ell(\lambda)} \, \big(q-1\big)\, \big(q^2-1 \big)\cdots\big(q^{\ell_i}-1\big)}\, \frac{\big(q-1\big)\, \big(q^2-1 \big)\cdots\big(q^{n}-1\big)}{q^{\frac{\ell(\lambda)\, (\ell(\lambda)-1)\, (\ell(\lambda)-2)}6}} \ ,
\ee
where $\ell(\lambda)$ is the length of the partition $\lambda$ (the number of non-zero $\lambda_i$) and
$\ell_i=\lambda_i+\ell(\lambda)-i$. In the classical limit $q\to1$ this expression reduces to the usual dimension formula
\be 
d_\lambda (1)=d_\lambda:=\chi_{r_\lambda}(1) =\frac{n!}{\prod\limits_{i=1}^{\ell(\lambda)} \, \ell_i!} \ \prod_{1\leq i< j\leq \ell(\lambda)} \, (\ell_i-\ell_j) \ .
\label{eq:dlambda1}\ee

\section{$(q,t)$-traces and symmetric functions} \label{app:qt-traces}

In this paper we are interested in the large $N$ expansion of refined $U(N)$
Yang-Mills amplitudes. The computation of the traces
$\Tr_{R_{\omega_1}^{\otimes n}}\big(\Phi_n \, U \,
\sfh(m_T)\big)$ in this limit can be related to some combinatorial
identities involving symmetric functions. For this, we shall say that the
minimal word $\sfh(m_T)$ of $\sfH_q(\frS_n)$ has \emph{connectivity
  class} $\mu(T)=\big(\mu_1(T),\dots,\mu_n(T) \big)$ if the conjugacy class
$T\in\frS_n^\vee$ is parameterized by the partition $\mu(T)$ of $n$,
i.e. any element of $T$ is composed of reduced words with $\mu_i(T)$ cycles of length
$i$. The minimal word $m_T$ in the conjugacy class $T$ has length
\be
\ell^*\big(\mu(T)\big)=\sum_{i=1}^n \, (i-1)\, \mu_i(T) \ ,
\ee
and $\ell(\mu(T)) =\sum_{i=1}^n\, \mu_i(T)$ is the total number of cycles in the cycle decomposition of $T$.

We interpret
Lemma~\ref{lem:schurweyl} as an
expansion in (normalized) Macdonald polynomials as
\bea
\Tr_{R_{\omega_1}^{\otimes n}}\big(\Phi_n \, U\, \sfh(\sigma)\big)= \sum_{\lambda_\beta\in\Lambda_+^{n}}\, \chi_{r_{\lambda_\beta}}\big(\sfh(\sigma)\big) \ \frac{M_\lambda(x;q,t)}{\sqrt{g_\lambda}} \ .
\label{eq:Macdonsum1}\eea
Let us consider the unrefined limit $\beta=1$, wherein the normalized Macdonald polynomials in (\ref{eq:Macdonsum1}) reduce to Schur polynomials $s_\lambda(x)$. We can then apply \cite[Theorem~1 and Definition~1]{King1992} to get
\be
\Tr_{R_{\omega_1}^{\otimes n}}\big(U\, \sfh(m_T)\big)=\prod_{i=1}^n\, p_{\mu_i(T)}(q;x) \ ,
\label{eq:Kingthm}\ee
where
\be
p_r(q;x):=\sum_{\stackrel{\scriptstyle a,b=0}{\scriptstyle a+b=r-1}}^{r-1}\, (-1)^b \, q^a \, s_{(a+1 \ 1^b)}(x) \ .
\label{eq:prqxdef}\ee
By \cite[Lemma~1]{King1992} we can equivalently write (\ref{eq:prqxdef}) as
\be
p_r(q;x)=\frac{1}{q-1} \ \sum_{\lambda\in\Lambda_+^r} \, s_\lambda(q|-1) \, s_\lambda(x) \ ,
\label{eq:prSUSYSchur}\ee
where $s_\lambda(q|-1)$ is a supersymmetric Schur
function~\cite[Section~4.4]{Szabo2013}. The sum in
(\ref{eq:prSUSYSchur}) 
can be evaluated by using the Cauchy-Binet identity for supersymmetric Schur functions~\cite[eq.~(4.27)]{Szabo2013}
\begin{eqnarray}
\sum_{\lambda \in\Lambda_+}\,s_{\lambda }(x|z)\,s_{\lambda }(y|w)
= \prod_{i,j=1}^N \ \frac{(1+x_i\, w_j)\, (1+y_i\, z_j)}{(1-x_i\,
  y_j)\, (1-z_i\, w_j)} \ ,
\label{ScSUSY}\end{eqnarray}
which at the specializations $z=(0,\dots,0)$, $y=(q,0,\dots,0)$ and
$w=(-1,0,\dots,0)$ yields
\be
\sum_{\lambda\in\Lambda_+} \, s_\lambda(q|-1) \, s_\lambda(x) =
\prod_{i=1}^N \,
\frac{1-x_i}{1-q\,x_i} \ .
\label{eq:CBspec}\ee
The sum over partitions of $r$ can in this way be computed by using
the homogeneity property $s_\lambda(\alpha\,x)= \alpha^{|\lambda|}\,
s_\lambda(x)$ of Schur polynomials and the generating function
\be
\sum_{r=1}^\infty\, \alpha^r \ \sum_{\lambda\in\Lambda_+^r} \, s_\lambda(q|-1) \, s_\lambda(x) 
=
\sum_{\lambda\in\Lambda_+} \, s_\lambda(q|-1) \, s_\lambda(\alpha\, x) 
\ee
for $\alpha\in\IC$. Using (\ref{eq:CBspec}) we then find
\be
p_r(q;x)=\left. \frac1{q-1}\, \frac1{r!}\,
  \frac{\partial^r}{\partial\alpha^r}\right|_{\alpha=0} \ \prod_{i=1}^N \,
\frac{1-\alpha\, x_i}{1-\alpha\, q\,x_i} \ .
\label{prderiv}\ee
In particular, the connected minimal word $\sfh(m_T)=\sfg_1\, \sfg_2\cdots
\sfg_{n-1}$ belongs to the connectivity class $\mu(T)=(n)$ and the
corresponding trace gives exactly $p_n(q;x)$, so that
\be
\Tr_{R_{\omega_1}^{\otimes n}}\big( U\, \sfg_1\, \sfg_2\cdots
\sfg_{n-1}\big) = \frac{1}{1-q} \ \sum_{\lambda\in\Lambda_+^n} \, s_\lambda(q|-1)
\, s_\lambda(x) \ .
\ee
At the specialization $U=q^{ (\rho,H)}$ we can compare this formula with the explicit computation of the trace from~\cite[eq.~(B.6)]{deHaro2006} to arrive at the combinatorial identity
\begin{proposition}
$\displaystyle{ \ \sum_{\lambda\in\Lambda_+^n} \, s_\lambda(q|-1) \, s_\lambda(q^\rho) = (q-1)\, q^{(n-1)\, \frac{N+1}2}\, [N]_q \ . }$
\end{proposition}
This identity can be compared explicitly with the formula
(\ref{prderiv}), in which case the specialization of the product in
(\ref{eq:CBspec}) to $x_i=\alpha\, q^{\frac{N+1}2-i}$ yields the
generating function
\be
\sum_{\lambda\in\Lambda_+} \, s_\lambda(q|-1) \, s_\lambda(\alpha\,
q^\rho) = \frac{1-\alpha\, q^{-\frac{N-1}2}}{1-\alpha\, q^{\frac{N+1}2}} \ .
\ee
Substituting now $p_r(q;q^\rho)= q^{(r-1)\, \frac{N+1}2}\, [N]_q$ into (\ref{eq:Kingthm}) we get
\be 
\Tr_{R_{\omega_1}^{\otimes n}}\big(q^{ (\rho,H)} \, \sfh(m_T)\big)=\prod_{i=1}^n\, \big(q^{(i-1)\, \frac{N+1}2}\, [N]_q \big)^{\mu_i(T)} = q^{\frac{N+1}2\, \ell^*(\mu(T))}\, \big([N]_q\big)^{\ell(\mu(T))}
\ee
as in~\cite[eq.~(B.7)]{deHaro2006}. We are not aware of any analogous simplifying identities for generating functions of Macdonald polynomials which could aid in simplifying the $(q,t)$-traces for $\beta\neq1$.

\section{Center of $\sfH_q(\frS_{\infty})$}\label{sec:centralTrafo}

There is a natural embedding of Hecke algebras $\sfH_q(\frS_{n}) \hookrightarrow \sfH_q(\frS_{n+1})$, and so the inductive limit of $\sfH_q(\frS_{n})$ as $n\to\infty$ exists~\cite{King1992}; we write this inductive limit as $\sfH_q(\frS_{\infty})$. We want to find an inductive limit of central elements of the Hecke algebras as well. Firstly we need an embedding of central elements given by a monomorphism
\be
\xymatrix{
\varphi_n\, : \, \widetilde{Z}\big(\sfH_q(\frS_{n})\big) \ \ar@{^{(}->}[r] \ & \  \widetilde{Z}\big(\sfH_q(\frS_{n+1})\big)
} \ ,
\ee
where $\widetilde{Z}\big(\sfH_q(\frS_n)\big)$ is a linear subspace of the center $Z\big(\sfH_q(\frS_n)\big)$ of the algebra $\sfH_q(\frS_n)$ such that
\be \label{eq:CindlimitTr}
\sum_i\, \Tr_{R_{\omega_1}^{\otimes n}}\big(\Phi_n \, x\,C_i^{(n)}\, U\big) \, = \, \sum_i\, \Tr_{R_{\omega_1}^{\otimes n}} \big(\Phi_{n}\, x\,C_i^{(n+1)}\, U\big) \ ,
\ee
for $U\in T$ and $x\in\sfH_q(\frS_n)$, where $C_i^{(n)}$ span a linear basis of $\widetilde{Z}\big(\sfH_q(\frS_{n})\big)$.

According to~\cite[Theorem~2.14]{Dipper1987}, the center $Z\big(\sfH_q(\frS_n)\big)$ is the algebra of symmetric polynomials in the \emph{Murphy operators} $L_i$, $i=1,\dots, n$, which are defined as
\bea
L_1&=& \sfh(1) \ = \ 1 \ , \nonumber \\[4pt] L_i&=&q^{-(i-1)}\, \sum_{j=1}^{i-1}\, q^{j-1}\, \sfh\big((j \ i)\big) \qquad \mbox{for} \quad i>1 \ ,
\eea
where $(i \ j)=\sigma_i\cdots \sigma_{j-2}\,\sigma_{j-1}\, \sigma_{j-2} \cdots \sigma_i$ for $j>i$ is the transposition which interchanges $i$ and $j$. For example, using the definition of the central elements $C_T$ from (\ref{eq:CTdef}), for $n=3$ we obtain $C_{(1,1,1)}=1$, $C_{(2,1)}=q\, (L_2+L_3)$ and $C_{(3)}=\frac{q^2}{2}\, (L_2\, L_3+L_3\, L_2)$ as homogeneous symmetric polynomials in Murphy operators. 

Given a symmetric polynomial $s(L_1,\dots,L_{n})$ in $L_i\in\sfH_q(\frS_{n})$, we need to embed it into $\sfH_q(\frS_{n+1})$, but a symmetric polynomial in $n$ variables is not necessarily a symmetric polynomial in $n+1$ variables so we need a non-trivial embedding. Because of (\ref{eq:CindlimitTr}) we require
\be \label{eq:phi_s_p}
\varphi_n\big(s(L_1,\dots,L_{n})\big) \, = \, s(L_1,\dots,L_{n}) \, + \, p(L_1,\dots,L_{n+1}) \ ,
\ee
where $p(L_1,\dots,L_{n+1})$ is not necessarily a symmetric polynomial. If $\widetilde{Z}\big(\sfH_q(\frS_{n})\big)$ is the space of homogeneous symmetric polynomials in Murphy operators, then $\varphi_n$ is unique and so there exists just one $p$ for every $s$ in (\ref{eq:phi_s_p}). For example one has
\bea
\varphi_1(1)& = & 1  \ , \nonumber \\[4pt]
\varphi_2(L_2) &=&L_2+L_3 \ , \nonumber \\[4pt] 
\varphi_3(L_2\,L_3+L_3\, L_2)&=& L_2\,L_3+L_3\,L_2+L_2\,L_4+L_4\,L_2+L_3\,L_4+L_4\,L_3 \ .
\eea

In the representation $R_{\omega_1}^{\otimes n}$, the Murphy operator $L_{n+1}$ is represented as $0$ if $\sfg_{n+i}\cdot( R_{\omega_1}\otimes\dots\otimes R_{\omega_1}) = 0$ for $i\geq 0$. Hence we get
\be
\varphi_n\big(s(L_1,\dots,L_{n})\big)\cdot(R_{\omega_1}\otimes\dots\otimes R_{\omega_1}) \, = \, s(L_1,\dots,L_{n})\cdot(R_{\omega_1}\otimes\dots\otimes R_{\omega_1}) \ .
\ee
All elements of $\widetilde{Z}\big(\sfH_q(\frS_{n+1})\big)$ either belong to the image of $\varphi_n$, or else all of their monomials contain at least one factor $L_{n+1}$ and so are represented as $0$ in $R_{\omega_1}^{\otimes n}$. 

Using this embedding, we can now take the inductive limit of $\widetilde{Z}\big(\sfH_q(\frS_{n})\big)$ as $n\to\infty$, which is given by the equivalence classes
\bea
\widetilde{Z}\big(\sfH_q(\frS_{\infty})\big) = \bigsqcup_{n=1}^\infty\, \widetilde{Z}\big(\sfH_q(\frS_{n})\big) \ \big/ \ \sim
\eea
where $x \sim  y$ if and only if $y=\varphi_{m-1}\circ\varphi_{m-2}\circ\cdots\circ\varphi_{n}(x)$ for $x\in \widetilde{Z}\big(\sfH_q(\frS_{n})\big)$, $y\in \widetilde{Z}\big(\sfH_q(\frS_{m})\big)$ and $m>n$; here the disjoint union over $\widetilde{Z}\big(\sfH_q(\frS_{n})\big)$ is factorized with a sequence of the embeddings. 

Now we can consider the transformation of a function $f(U)$ for $U\in T$ given by
\be
f(C) \, = \, \sum_{n=0}^\infty \ \int_T\, [\dd U]_{q,t} \ \Tr_{R_{\omega_1}^{\otimes n}}(\Phi_n \, y_n \, C \, U) \ f(U) \ ,
\ee
for $y_n\in\sfH_q(\frS_{n})$ and $C\in\widetilde{Z}\big(\sfH_q(\frS_{\infty})\big)$, provided the series converges. Here we defined the integration
measure
\be
[\dd U]_{q,t}:= \frac{(-1)^N}{N!}\, \prod_{i=1}^N\, \dd z_i \
\Delta_{q,t}(x) \, \Delta_{q,t}(x^{-1})
\label{eq:Macintmeasure}\ee
where
\be
\Delta_{q,t}(x):= t^{-\frac{N\, (N-1)}2}\, \prod_{i< j}\, \frac{\big(x_i\,x_j^{-1};
  q\big)_\infty}{\big(t\, x_i\, x_j^{-1}; q\big)_\infty} =
\prod_{m=0}^{\beta-1} \ \prod_{i< j}\, \big(q^{-m/2}\, \e^{(z_j-z_i)/2}-q^{m/2}\, \e^{(z_i-z_j)/2} \big)
\ee
for $\beta\in\IZ_{>0}$, with $U=\e^{(z,H)}$ and $x=\e^z$. In the unrefined limit $\beta=1$, the measure \eqref{eq:Macintmeasure} reduces to the usual Haar measure
\be 
[\dd U]_q=[\dd U]_{q,q} = \frac1{N!}\, \prod_{i=1}^N\, \dd z_i \
\Delta_q(x)^2
\ee
for integration over the maximal torus $T\subset G$, where
\be 
\Delta_q(x)=\Delta_{q,q}(x) = \prod_{i<j}\, 2\sinh\Big(\,
\frac{z_i-z_j}2\, \Big)
\ee
is the Weyl determinant for $G=U(N)$.


\bigskip

\begin{thebibliography}{99}

\bibitem{Aganagic2012ref}
  M.~Aganagic and K.~Schaeffer,
  ``Orientifolds and the refined topological string,''
  JHEP {\bf 1209} (2012) 084
  [arXiv:1202.4456 [hep-th]].

\bibitem{Aganagic2012}
  M.~Aganagic and K.~Schaeffer,
  ``Refined black hole ensembles and topological strings,''
  JHEP {\bf 1301} (2013) 060
  [arXiv:1210.1865 [hep-th]].

\bibitem{Aganagic2011}
  M.~Aganagic and S.~Shakirov,
  ``Knot homology and refined Chern-Simons index,''
  Commun. Math. Phys. {\bf 333} (2015) 187--228
  [arXiv:1105.5117 [hep-th]].
  
 \bibitem{Aganagic2005}
   M.~Aganagic, H.~Ooguri, N.~Saulina and C.~Vafa,
   ``Black holes, $q$-deformed $2D$ Yang-Mills and nonperturbative topological strings,''
   Nucl. Phys. B {\bf 715} (2005) 304--348
   [arXiv:hep-th/0411280].

\bibitem{Billo2001}
  M.~Bill\'o, A.~D'Adda and P.~Provero,
  ``Branched coverings and interacting matrix strings in two dimensions,''
  Nucl.\ Phys.\ B {\bf 616} (2001) 495--516
  [arXiv:hep-th/0103242].
  
\bibitem{Brini2010}
  A.~Brini, M.~Mari\~no and S.~Stevan,
  ``The uses of the refined matrix model recursion,''
  J.\ Math.\ Phys.\ {\bf 52} (2011) 052305
  [arXiv:1010.1210 [hep-th]].
  
\bibitem{Bryan2004}
  J.~Bryan and R.~Pandharipande,
  ``The local Gromov-Witten theory of curves,''
  J. Amer. Math. Soc. {\bf 21} (2008) 101--136 
  [arXiv:math.AG/0411037].
  
\bibitem{Caporaso2005a}
  N.~Caporaso, M.~Cirafici, L.~Griguolo, S.~Pasquetti, D.~Seminara and R.~J.~Szabo,
  ``Topological strings and large $N$ phase transitions I: Nonchiral expansion of $q$-deformed Yang-Mills theory,''
  JHEP {\bf 0601} (2006) 035
  [arXiv:hep-th/0509041].

\bibitem{Caporaso2005b}
  N.~Caporaso, M.~Cirafici, L.~Griguolo, S.~Pasquetti, D.~Seminara and R.~J.~Szabo,
  ``Topological strings and large $N$ phase transitions II: Chiral expansion of $q$-deformed Yang-Mills theory,''
  JHEP {\bf 0601} (2006) 036
  [arXiv:hep-th/0511043].
  
\bibitem{Caporaso2006}
  N.~Caporaso, M.~Cirafici, L.~Griguolo, S.~Pasquetti, D.~Seminara and R.~J.~Szabo,
  ``Topological strings, two-dimensional Yang-Mills theory and Chern-Simons theory on torus bundles,''
  Adv.\ Theor.\ Math.\ Phys.\  {\bf 12} (2008) 981--1058
  [arXiv:hep-th/0609129].
  
\bibitem{Chen2013}
  H.-Y.~Chen and A.~Sinkovics,
  ``On integrable structure and geometric transition in supersymmetric gauge theories,''
  JHEP {\bf 1305} (2013) 158
  [arXiv:1303.4237 [hep-th]].
  
\bibitem{Chuang2010}
  W.-y.~Chuang, D.-E.~Diaconescu and G.~Pan,
  ``Wall-crossing and cohomology of the moduli space of Hitchin pairs,''
  Commun.\ Num.\ Theor.\ Phys.\ {\bf 5} (2011) 1--56
  [arXiv:1004.4195 [math.AG]].
  
\bibitem{Cordes1994}
  S.~Cordes, G.~W.~Moore and S.~Ramgoolam,
  ``Large $N$ $2D$ Yang-Mills theory and topological string theory,''
  Commun.\ Math.\ Phys.\  {\bf 185} (1997) 543--619
  [arXiv:hep-th/9402107].
  
\bibitem{Cordes1995}
S.~Cordes, G.~W.~Moore and S.~Ramgoolam, 
``Lectures on $2D$ Yang-Mills theory, equivariant cohomology and topological field theories,'' 
Nucl. Phys. Proc. Suppl. {\bf 41} (1995) 184--244
[arXiv:hep-th/9411210].

\bibitem{Dijkgraaf1996}
  R.~Dijkgraaf and G.~W.~Moore,
  ``Balanced topological field theories,''
  Commun.\ Math.\ Phys.\  {\bf 185} (1997) 411--440
  [arXiv:hep-th/9608169].

\bibitem{DijkgraafVafa2009}
  R.~Dijkgraaf and C.~Vafa,
  ``Toda theories, matrix models, topological strings and $\cN=2$ gauge systems,''
  arXiv:0909.2453 [hep-th].

\bibitem{Dijkgraaf2008}
  R.~Dijkgraaf, L.~Hollands and P.~Sulkowski,
  ``Quantum curves and D-modules,''
  JHEP {\bf 0911} (2009) 047
  [arXiv:0810.4157 [hep-th]].
  
\bibitem{Dipper1987}
  R.~Dipper and G.~James,
  ``Blocks and idempotents of Hecke algebras of general linear groups,''
  Proc. London Math. Soc. {\bf 54} (1987) 57--82.

\bibitem{Do2015}
N.~Do and M.~Karev,
``Monotone orbifold Hurwitz numbers,''
arXiv:1505.06503 [math.GT].

\bibitem{Ekedahl2001}
T.~Ekedahl, S.~K.~Lando, M.~Shapiro and A.~Vainshtein,
``Hurwitz numbers and intersections on moduli spaces of
curves,''
Invent. Math. {\bf 146} (2001) 297--327
[arXiv:math.AG/0004096].

\bibitem{Etingof1993}
  P.~I.~Etingof and A.~A.~Kirillov, Jr.,
  ``Macdonald's polynomials and representations of quantum groups,''
  Math. Res. Lett. {\bf 1} (1994) 279--296
  [arXiv:hep-th/9312103].

\bibitem{Eynard2008}
  B.~Eynard and O.~Marchal,
  ``Topological expansion of the Bethe ansatz and noncommutative algebraic geometry,''
  JHEP {\bf 0903} (2009) 094
  [arXiv:0809.3367 [math-ph]].

\bibitem{Faddeev1989}
  L.~D. Faddeev, N.~Yu.~Reshetikhin and L.~A.~Takhtajan, ``Quantization of Lie groups and Lie algebras,'' 
Lengingrad Math. J. {\bf 1} (1990) 193--225.

\bibitem{Gadde2011}
  A.~Gadde, L.~Rastelli, S.~S.~Razamat and W.~Yan,
  ``Gauge theories and Macdonald polynomials,''
  Commun.\ Math.\ Phys.\ {\bf 319} (2013) 147--193
  [arXiv:1110.3740 [hep-th]].
  
\bibitem{Gaiotto2009}
  {D.~Gaiotto},
  ``{$\cN=2$ dualities},''
  JHEP {\bf 1208} (2012) 034
  [arXiv:0904.2715 [hep-th]].
  
\bibitem{GaiottoMoore2009}
  {D.~Gaiotto, G.~W.~Moore and A.~Neitzke},
  ``{Wall-crossing, Hitchin systems and the WKB approximation},''
  Adv. Math. {\bf 234} (2013) 239--403
  [arXiv:0907.3987 [hep-th]].
  
\bibitem{Gorsky2016}
  A.~Gorsky, A.~Milekhin and S.~Nechaev,
  ``Douglas-Kazakov on the road to superfluidity: From random walks to black holes,''
  arXiv:1604.06381 [hep-th].
  
\bibitem{Goulden1999}
I.~P.~Goulden, J.~L. Harer and D.~M.~Jackson,
``A geometric parameterization for the
virtual Euler characteristics of the moduli spaces of real and complex algebraic curves,''
Trans. Amer. Math. Soc. {\bf 353} (2001) 4405--4427
[arXiv:math.AG/9902044]

\bibitem{Green1955}
J.~A.~Green,
``The characters of the finite general linear groups,''
Trans. Amer. Math. Soc. {\bf 80} (1955) 402--447.

\bibitem{Gross1993}
  D.~J.~Gross, ``Two-dimensional QCD as a string theory,'' 
Nucl. Phys. B {\bf 400} (1993) 161--180
[arXiv:hep-th/9212149].

\bibitem{Gross1993a}
  D.~J.~Gross and W.~I.~Taylor, ``Two-dimensional QCD is a string theory,'' 
Nucl. Phys. B {\bf 400} (1993) 181--208
[arXiv:hep-th/9301068].

\bibitem{Gross1993b}
  D.~J.~Gross and W.~I.~Taylor, ``Twists and Wilson loops in the string theory of two-dimensional QCD,'' 
Nucl. Phys. B {\bf 403} (1993) 395--452
[arXiv:hep-th/9303046].

\bibitem{deHaro2006}
  S.~de Haro, S.~Ramgoolam and A.~Torrielli,
  ``Large $N$ expansion of $q$-deformed two-dimensional Yang-Mills theory and Hecke algebras,''
  Commun.\ Math.\ Phys.\ {\bf 273} (2007) 317--355
  [arXiv:hep-th/0603056].

\bibitem{Iqbal2011}
  A.~Iqbal and C.~Kozcaz,
  ``Refined Hopf link revisited,''
  JHEP {\bf 1204} (2012) 046
  [arXiv:1111.0525 [hep-th]].
  
\bibitem{Iqbal2015}
  A.~Iqbal, A.~Z.~Khan, B.~A.~Qureshi, K.~Shabbir and M.~A.~Shehper,
  ``Topological field theory amplitudes for $A_{N-1}$ fibration,''
  JHEP {\bf 1512} (2015) 017
  [arXiv:1507.02662 [hep-th]].

\bibitem{Jimbo1986}
  M.~Jimbo,
  ``A $q$-analog of $\cU(\frgl_{N+1})$, Hecke algebras and the Yang-Baxter equation,''
  Lett. Math. Phys. {\bf 11} (1986) 247--252.

\bibitem{King1992}
  R.~C.~King and B.~G.~Wybourne, 
  ``Representations and traces of the Hecke algebras $\sfH_n(q)$ of type $A_{n-1}$,''
 J. Math. Phys. {\bf 33} (1992) 4--14.
 
\bibitem{Klimcik1999}
  C.~Klimcik,
  ``The formulae of Kontsevich and Verlinde from the perspective of the Drinfeld double,''
  Commun.\ Math.\ Phys.\  {\bf 217} (2001) 203--228
  [arXiv:hep-th/9911239].

\bibitem{Kokenyesi2013}
  Z.~K\"ok\'enyesi, A.~Sinkovics and R.~J.~Szabo,
  ``Refined Chern-Simons theory and $(q, t)$-deformed Yang-Mills theory: Semi-classical expansion and planar limit,''
  JHEP {\bf 1310} (2013) 067
  [arXiv:1306.1707 [hep-th]].

\bibitem{Kostov1997}
  I.~K.~Kostov, M.~Staudacher and T.~Wynter,
  ``Complex matrix models and statistics of branched coverings of $2D$ surfaces,''
  Commun.\ Math.\ Phys.\  {\bf 191} (1998) 283--298
  [arXiv:hep-th/9703189].
 
\bibitem{Krefl2013}
  D.~Krefl and A.~Schwarz, 
  ``Refined Chern-Simons versus Vogel universality,''
 J. Geom. Phys. {\bf 74} (2013) 119--129
 [arXiv:1304.7873 [hep-th]].
  
\bibitem{Krefl2010}
  D.~Krefl and J.~Walcher,
  ``Extended holomorphic anomaly in gauge theory,''
  Lett.\ Math.\ Phys.\ {\bf 95} (2011) 67--88
  [arXiv:1007.0263 [hep-th]].

\bibitem{Lando2004}
S.~K.~Lando, A.~K.~Zvonkin and D.~Zagier,
\emph{Graphs on Surfaces and their Applications}
(Springer, New York, 2004).

\bibitem{Macdonald1995}
I.~G.~Macdonald,
  \emph{Symmetric Functions and Hall Polynomials}
 (Oxford University Press, Oxford, 1995).

\bibitem{Migdal1975}
  A.~A.~Migdal, 
  ``Recursion equations in gauge field theories,''
 Sov. Phys. JETP {\bf 42} (1975) 413--418.
 
\bibitem{Nakajima2003}
  H.~Nakajima and K.~Yoshioka,
  ``Lectures on instanton counting,''
  CRM Proc. Lect. Notes {\bf 38} (2004) 31--102
  [arXiv:math.AG/0311058].

\bibitem{Ooguri2002}
  H.~Ooguri and C.~Vafa,
  ``Worldsheet derivation of a large $N$ duality,''
  Nucl.\ Phys.\ B {\bf 641} (2002) 3--34
  [arXiv:hep-th/0205297].

\bibitem{OSV2004}
H.~Ooguri, A.~Strominger and C.~Vafa,
``Black hole attractors and the topological string,''
Phys. Rev. D {\bf 70 } (2004) 106007
[arXiv:hep-th/0405146].

\bibitem{Ramgoolam1995}
  S.~Ramgoolam,
  ``Wilson loops in $2D$ Yang-Mills: Euler characters and loop equations,''
  Int.\ J.\ Mod.\ Phys.\ A {\bf 11} (1996) 3885--3933
  [arXiv:hep-th/9412110].

\bibitem{Rusakov1990}
  B.~E.~Rusakov,
  ``Loop averages and partition functions in $U(N)$ gauge theory on two-dimensional manifolds,''
  Mod.\ Phys.\ Lett.\ A {\bf 5} (1990) 693--703.
  
\bibitem{Szabo2010}
  R.~J.~Szabo and M.~Tierz,
  ``Chern-Simons matrix models, two-dimensional Yang-Mills theory and the Sutherland model,''
  J.\ Phys.\ A {\bf 43} (2010) 265401
  [arXiv:1003.1228 [hep-th]].
  
\bibitem{Szabo2010a}
  R.~J.~Szabo and M.~Tierz,
  ``Matrix models and stochastic growth in Donaldson-Thomas theory,''
  J.\ Math.\ Phys.\  {\bf 53} (2012) 103502
  [arXiv:1005.5643 [hep-th]].
  
\bibitem{Szabo2013}
  R.~J.~Szabo and M.~Tierz,
  ``$q$-deformations of two-dimensional Yang-Mills theory: Classification, categorification and refinement,''
  Nucl.\ Phys.\ B {\bf 876} (2013) 234--308
  [arXiv:1305.1580 [hep-th]].

\bibitem{Zenkevich2015}
  Y.~Zenkevich,
  ``Quantum spectral curve for $(q,t)$-matrix model,''
  arXiv:1507.00519 [hep-th].
  
\end{thebibliography}
\end{document}